\documentclass[twocolumn]{aastex63}
\usepackage{multirow}
\usepackage{float}
\usepackage{xcolor}

\newcommand{\ang}{$\rm\AA$}
\newcommand{\msun}{M$_{\odot}$}
\newcommand{\myr}{M$_\odot$~yr$^{-1}$} 
\newcommand{\myrkpc}{M$_\odot$~yr$^{-1}$~kpc$^{-2}$} 
\newcommand{\mstar}{M$_{*}$}

\newcommand{\ha}{H$\alpha$}
\newcommand{\lha}{L$_{H\alpha}$}

\newcommand{\hb}{H$\beta$}

\newcommand{\kms}{km\,s$^{-1}$}

\newcommand{\ergs}{erg s$^{-1}$}

\newcommand{\sfrd}{$\Sigma_{SFR}$}

\newcommand{\hii}{H\,{\sc II} }
\newcommand{\hi}{H\,{\sc I} }
\newcommand{\oii}{[O{\sc II}]}
\newcommand{\oiii}{[O{\sc III}]}
\newcommand{\nii}{[N{\sc II}]}
\newcommand{\stromgren}{Str\"{o}mgren }

\received{October 14, 2021}
\revised{January 20, 2022}
\accepted{February 4, 2022}
\submitjournal{ApJ}

\shorttitle{IC\,10}
\shortauthors{Cosens et al.}

\begin{document}

\title{Kinematics and Feedback in \hii regions in the Dwarf Starburst Galaxy IC\,10}   

\correspondingauthor{Maren Cosens}
\email{mcosens@ucsd.edu}

\author[0000-0002-2248-6107]{Maren Cosens}
\affil{Physics Department, University of California, San Diego, 9500 Gilman Drive, La Jolla, CA 92093 USA} 
\affil{Center for Astrophysics and Space Sciences, University of California, San Diego, 9500 Gilman Drive, La Jolla, CA 92093 USA}

\author[0000-0003-1034-8054]{Shelley A. Wright}
\affil{Physics Department, University of California, San Diego, 9500 Gilman Drive, La Jolla, CA 92093 USA} 
\affil{Center for Astrophysics and Space Sciences, University of California, San Diego, 9500 Gilman Drive, La Jolla, CA 92093 USA}

\author{Norman Murray}
\affil{Canadian Institute for Theoretical Astrophysics, University of Toronto, 60 St. George Street, Toronto, ON M5S 3H8, Canada}
\altaffiliation{Canada Research Chair in Astrophysics}

\author[0000-0003-3498-2973]{Lee Armus}
\affil{Spitzer Science Center, California Institute of Technology, 1200 E. California Blvd., Pasadena, CA 91125 USA}

\author[0000-0002-4378-8534]{Karin Sandstrom}
\affil{Physics Department, University of California, San Diego, 9500 Gilman Drive, La Jolla, CA 92093 USA} 
\affil{Center for Astrophysics and Space Sciences, University of California, San Diego, 9500 Gilman Drive, La Jolla, CA 92093 USA}

\author[0000-0001-9554-6062]{Tuan Do}
\affil{UCLA Galactic Center Group, Department of Physics \& Astronomy, University of California, Los Angeles, Los Angeles, CA 90095, USA}

\author[0000-0003-3917-6460]{Kirsten Larson}
\affil{Spitzer Science Center, California Institute of Technology, 1200 E. California Blvd., Pasadena, CA 91125 USA}

\author{Gregory Martinez}
\affil{UCLA Galactic Center Group, Department of Physics \& Astronomy, University of California, Los Angeles, Los Angeles, CA 90095, USA}

\author[0000-0002-8780-8226]{Sanchit Sabhlok}
\affil{Physics Department, University of California, San Diego, 9500 Gilman Drive, La Jolla, CA 92093 USA} 
\affil{Center for Astrophysics and Space Sciences, University of California, San Diego, 9500 Gilman Drive, La Jolla, CA 92093 USA}

\author[0000-0002-0710-3729]{Andrey Vayner}
\affil{Department of Physics and Astronomy, Johns Hopkins University, Baltimore, MD 21218, USA}

\author[0000-0003-2687-9618]{James Wiley}
\affil{Physics Department, University of California, San Diego, 9500 Gilman Drive, La Jolla, CA 92093 USA} 
\affil{Center for Astrophysics and Space Sciences, University of California, San Diego, 9500 Gilman Drive, La Jolla, CA 92093 USA}

\begin{abstract}
We present a survey of the central region of the nearest starburst galaxy, IC\,10, using the W. M. Keck Observatory Keck Cosmic Web Imager (KCWI) at high spectral and spatial resolution. We map the central starburst of IC\,10 to sample the kinematic and ionization properties of the individual star-forming regions. Using the low spectral resolution mode of KCWI we map the oxygen abundance and with the high spectral resolution mode we identify 46 individual \hii regions. These \hii regions have an average radius of 4.0\,pc, star formation rate $\sim1.3\times10^{-4}$ \myr, and velocity dispersion $\sim$16\, \kms. None of the \hii regions appear to be virialized ($\rm \alpha_{vir}>>1$), and, on average, they show evidence of ongoing expansion. IC\,10's \hii regions are offset from the star forming region size-luminosity scaling relationships, as well as Larson's Law that relates size and velocity dispersion. We investigate the balance of inward and outward pressure, $\rm P_{in}$ and $\rm P_{out}$, finding $\rm P_{out}>P_{in}$ in 89\% of \hii regions, indicating feedback driven expansion even in these low mass \hii regions. We find warm gas pressure ($\rm P_{gas}$) provides the dominant contribution to the outward pressure ($\rm P_{out}$). This counteracts the inward pressure which is dominated by turbulence in the surrounding gas rather than self-gravity. Five \hii regions show evidence of outflows which are most likely supported by either stellar winds (2 regions) or champagne flows (3 regions). These observations provide new insights into the state of the star-forming regions in IC\,10 and negative feedback from low mass clusters.
\end{abstract}


\section{Introduction} \label{sec:intro}
\hii regions are formed when UV photons from young stars and clusters ionize the surrounding gas cloud. The ionization within these regions is typically dominated by the most massive and luminous stars. This can be due to just a single O or B star \citep[e.g.,][]{Armentrout2021} or a cluster of massive stars. \hii regions are observed to have typical lifetimes $\lesssim10$Myr, starting out spatially compact ($<$pc) and expanding as they age \citep[e.g.,][]{Spitzer1978, Zamora-Aviles2019} before the \hii region dissipates. As the \hii regions expand they interact with and influence the surrounding gas. As the sites of recent massive star formation, \hii regions are intrinsically linked to the efficiency of star formation in the larger molecular cloud, the properties of the Interstellar Medium (ISM), and the evolution of galaxies. The photometric and kinematic properties of \hii regions are therefore of great interest for studying and understanding the progression of star formation.

There are a number of surveys studying ionized and diffuse gas in and around the \hii regions of nearby galaxies. Some of the first studies of extragalactic \hii regions used \ha \, imaging with photographic plates and CCD's to map the ionized gas in the Small Magellanic Cloud (SMC) and Large Magellanic Cloud (LMC) \citep[e.g.,][]{DEM1976, Kennicutt1986} as well as other nearby galaxies \citep[e.g. NGC 6822,][]{Hodge1989}. With \ha \, imaging they were able to measure the size and flux of star forming regions and explore their size distributions and luminosity functions. As instrumentation improved, Fabry-Perot mapping added measurements of the ionized gas kinematics both within the \hii regions and in the diffuse gas component, finding $\rm \sim4\times$ higher velocity dispersions ($\sigma$) in the diffuse gas \citep[e.g.,][]{Valdez-Gutierrez2002}. With the advent of integral field spectrographs (IFS), surveys could map the resolved gas properties and kinematics at $\sim$kpc scale. Large surveys such as CALIFA \citep{Sanchez2012a}, SAMI \citep{Croom2012}, and MaNGA\citep{Bundy2015}, studied hundreds to thousands of star forming galaxies and their resolved properties. These surveys have resulted in numerous publications including studies of galaxy dynamical scaling relations \citep{Cortese2014}, measurement of a ``resolved" star forming main sequence \citep{Ellison2018}, and the fundamental metallicity relation \citep{Cresci2019}. The CALIFA survey also made the important characterization of the Diffuse Ionized Gas (DIG) showing a trend in the \ha \, equivalent width with both the position on the BPT diagram as well as galaxy morphological type for this large sample of galaxies \citep{Espinosa-Ponce2020}. Using the MaNGA survey, \citet{RodriguezdelPino2019} identify ionized gas outflows in 7\% of the studied \ha \, emitting galaxies, finding evidence of shocks in most of the outflows with larger velocities associated with more massive galaxies.

State of the art IFS's operating at visible wavelengths such as the Multi-Unit Spectroscopic Explorer \citep[MUSE,][]{Bacon2010}, and the Keck Cosmic Web Imager \citep[KCWI,][]{Morrissey2018}, have begun to allow incredibly high spatial and spectral resolution mapping of nearby \hii regions. For example, \citet{Castro2018} used MUSE to map the giant \hii region 30 Doradus, generating resolved maps of the ionization state and revealing bi-modal gas velocities surrounding the star cluster R136. In another study, \citet{McLeod2019} use MUSE to map two LMC \hii region complexes and characterize the role of stellar feedback mechanisms, finding stellar winds and thermal gas pressure to be dominant. Studying two giant \hii regions in M101 with KCWI, \citet{Bresolin2020} find evidence of expanding shells and an underlying broad emission component potentially attributable to stellar winds interacting with cold gas.

The process of energy from star formation being injected into and influencing the surrounding gas through feedback can be caused by a variety of mechanisms. These mechanisms and their impact are typically discussed in the context of the larger molecular clouds surrounding the compact \hii regions where the pressure originates. An important form of feedback is radiation pressure which occurs when stellar photons interact with dust grains in the surrounding molecular cloud thereby transferring both energy and momentum. The energy imparted to the molecular gas may be radiated away, but the momentum cannot and therefore may be able to more effectively cause expansion of the gas \citep{Krumholz2014}.

The ionizing photons produced in a star cluster also act to heat the surrounding \hii region gas to typical temperatures of $\rm \sim10^4 K$. This warm gas generates an important source of outward pressure that may cause the expansion and eventual disruption of the region and surrounding gas. An additional source of thermal pressure comes from hot gas heated by shocks from stellar winds. The bubbles of hot gas produced can be observed via emitted X-rays. However, this hot gas is less effective in disrupting the region since it is limited by leakage through low density regions and turbulent mixing occurring with the neighboring cold gas resulting in enhanced thermal emission \citep{Krumholz2019}. Supernovae explosions also produce shocked winds in a short burst that can disrupt star-forming regions, but these don't occur until a few million years after the formation of the first massive stars by which point the molecular gas may be significantly disrupted or cleared already.

Which of these feedback mechanisms is dominant in different star forming environments is still a matter of some debate. Observational studies seek to determine the relative impact of each of these forms of pressure by estimating the energy input to the ISM or the pressure produced by each component. For instance, the giant \hii region, 30 Doradus, has been one target of such pressure studies investigating the physical processes leading to the complex structure in this single region \citep{Lopez2011, Pellegrini2011}. These types of studies are often limited to just a few \hii regions or a single giant \hii region, making it difficult to form conclusions about the general population of star forming regions. IC\,10 provides a unique laboratory to study the effectiveness of many of these forms of feedback in a statistically significant sample. Thanks to the $>$100 previously identified \hii regions, a large sample of compact regions can be studied simultaneously, and the recent nature of the starburst allows the effectiveness of pre-SNe feedback to be investigated.

IC\,10 is the nearest starburst galaxy at a distance of $715$kpc \citep{Kim2009}, and the only one in the Local Group. It is also a dwarf galaxy with low metallicity; approximately 0.25$\times$ solar \citep{Magrini2009, Skillman1989}. IC\,10 has a higher density of Wolf-Rayet stars than both the SMC and LMC \citep[e.g.][]{Tehrani2017}, indicating that the current observed burst of star formation is relatively recent. These unique characteristics as well as it's close proximity have made IC\,10 the subject of numerous studies. 

Studies of the gas in IC\,10 have shown an expansive H {\sc I} component stretching $\sim$7$\times$ larger than the optical component of the galaxy \citep[e.g.,][]{Huchtmeier1979, Namumba2019}. The central region of the H {\sc I} gas has been observed to have a regularly rotating disk structure with an extended counter-rotating component beyond that \citep[e.g.,][]{Shostak1989, Wilcots1998, Ashley2014}. These studies also find kinematically distinct ``spurs" and ``plumes" that do not follow the main H {\sc I} disk. The origin of these features is currently not well known but possible explanations presented include an as of yet undetected companion galaxy or a late stage merger \citep[e.g.,][]{Ashley2014}, ongoing accretion of primordial gas onto the main body of IC\,10 \citep[e.g.,][]{Wilcots1998}, past interaction with a body such as another dwarf galaxy \citep[e.g.,][]{Nidever2013}, or some combination of these mechanisms.

Narrow-band \ha \, imaging by \citet{HL1990} was used to identify 144 individual \hii regions and complexes throughout IC\,10 and measure their characteristic properties such as size and SFR. The majority of the identified star-forming regions lie in the central 2.5$'\times$2.5$\arcmin$ of the irregular galaxy. \citet{Thurow2005} studied the ionized gas kinematics in a portion of this field with a fiber-fed IFS utilizing 3$''$ fibers and achieving a maximum resolution of 23 \kms. Interestingly, they find larger line widths in the diffuse gas than in the compact \hii regions which they attribute to a superposition of components with different velocities. They find that stellar winds are likely to have shaped much of the ionized and neutral gas in this region. \citet{Polles2019} model fine structure cooling lines observed in IC\,10 with the photoionization code Cloudy \citep{Ferland2017} finding relatively uniform properties between the five regions studied, which match the characteristics of matter-bounded regions allowing photons to escape and ionize the diffuse gas. These unique characteristics of IC\,10 make it an ideal target to study a large sample of young, evolving star-forming regions.

In order to better understand the conditions of star formation and its impact on galaxy properties it is important to study not only local \hii regions, but also the sites of star formation throughout cosmic time. The \hii regions of IC\,10 provide an important tool for comparison with the $\sim$kpc scale star-forming ``clumps" found at $\rm z\gtrsim1$ \citep[e.g.,][]{Livermore2012, Mieda2016}. These clumps are found to have high velocity dispersions \citep[$\sigma \gtrsim 50$\kms, e.g., ][]{Genzel2011, Mieda2016} indicating strong energetics and significant amounts of turbulence present. In an effort to understand these massive star-forming regions and compare them to their more compact local counterparts, the scaling relationships between clump properties such as size, luminosity, and velocity dispersion are explored in order to provide insight into the process driving clump formation. However, these studies have yielded some conflicting results, in particular regarding whether high-z clumps are offset to higher luminosities than local \hii regions for a given size \citep[e.g.,][]{Wisnioski2012, Livermore2015}. This was initially proposed as a possible redshift evolution in the size-luminosity scaling relationship \citep{Livermore2012, Livermore2015}, but later studies by \citet{Wisnioski2012} and \citet{Mieda2016} did not find evidence of such an evolution. In order to investigate this discrepancy, our team compiled a comprehensive sample of high-z and local star-forming regions and developed a Bayesian Markov Chain Monte Carlo (MCMC) fitting framework to investigate these scaling relations in detail \citep{Cosens2018}. We did not find any clear evidence of redshift evolution with this expansive sample, nor a definite selection affect between lensed and field galaxies at high-z. Instead, we found evidence that there may be a break in the size-luminosity relationship based on the star formation rate (SFR) surface density, \sfrd.

A key area of parameter space missing in these scaling relationship investigations are compact ($\lesssim$50pc), low-luminosity ($10^{34-36}$\ergs) star-forming regions studied with the same methodology as at high-z. These compact regions set the constraint on the intercept of the relationship; a critical component in interpreting changes in slope or offsets between samples. Observing the \hii regions of IC\,10 with the KCWI IFS provides an ideal target to study a large sample of \hii regions at unprecedented angular resolution, probing this missing parameter space and allowing for an improved comparison of local and high-z star-forming regions.

Despite the extensive study of IC\,10, the exact distance to the dwarf galaxy remains rather uncertain since it lies close to the plane of the Milky Way. Measured distances have ranged between 500kpc \citep{Sakai1999} and $>$2Mpc \citep{Bottinelli1984}, with distances $\sim$700kpc being used more recently \citep[e.g.,][]{Ashley2014, Polles2019}. Throughout this paper we will use the distance of 715$\pm$60 kpc measured by \citet{Kim2009} but we will also report the angular size of all measured structures. For the systemic velocity of IC\,10 we will use the value of $\rm -348\pm10$\kms \, determined from the 21 cm line \citep{Tifft1988}.

In Section \ref{sec:obs_red} we describe the observations carried out on the central starburst of IC\,10 with KCWI and the method of reduction. Then we identify \hii regions in our observations (Section \ref{sec:astrodendro}), extract \hii region spectra (Section \ref{sec:spectra}), and determine SFRs (Section \ref{sec:SFR}) and masses (Section \ref{sec:mass}). In Section \ref{sec:kinematics} we investigate the kinematics of the field and \hii regions, virialization and energetics. We estimate the metallicity throughout the field (Section \ref{sec:metallicity}), investigate the Diffuse Ionized Gas (Section \ref{sec:DIG}), and study the star formation scaling relations (Section \ref{sec:scaling}). In Section \ref{sec:discussion} we discuss how these results inform a picture of young \hii regions still evolving. Lastly, in Section \ref{sec:conclusion} we summarize our results.

\section{Observations \& Data Reduction} \label{sec:obs_red}
\subsection{KCWI Observations} \label{sec:obs}
We used KCWI \citep{Morrissey2018} at the W.M. Keck Observatory to observe the star forming \hii regions of IC\,10. These observations tile a combined $\sim$1.25 sq. deg. field of view (FoV) in the central region of the galaxy with the highest concentration of \hii regions. We use a low resolution mode making use of KCWI's large slicer and BL grating which we will refer to as the ``large slicer, R$\sim$900" mode, as well as a high resolution mode which uses the small slicer and BH3 grating which we will refer to as the ``small slicer, R$\sim$18,000" mode. Observations in these two modes cover approximately the same FoV in order to combine the exceptional spatial sampling (0.35$\arcsec$/pixel) and spectral resolution (0.125\ang/channel) of the small slicer, with the wavelength coverage afforded with the low resolution grating (3500-5500\ang). We determine our achieved resolution by measuring the point-spread-function of the observed standard stars giving an average FWHM$\sim1"$ across the observing nights in both modes. In this paper we will limit our discussion primarily to the high resolution ``small slicer, R$\sim$18,000" observations with the ``large slicer, R$\sim$900" mode providing the extinction correction and metallicity diagnostics in Section \ref{sec:metallicity}.

\begin{figure*}
    \centering
    \includegraphics[width=\textwidth]{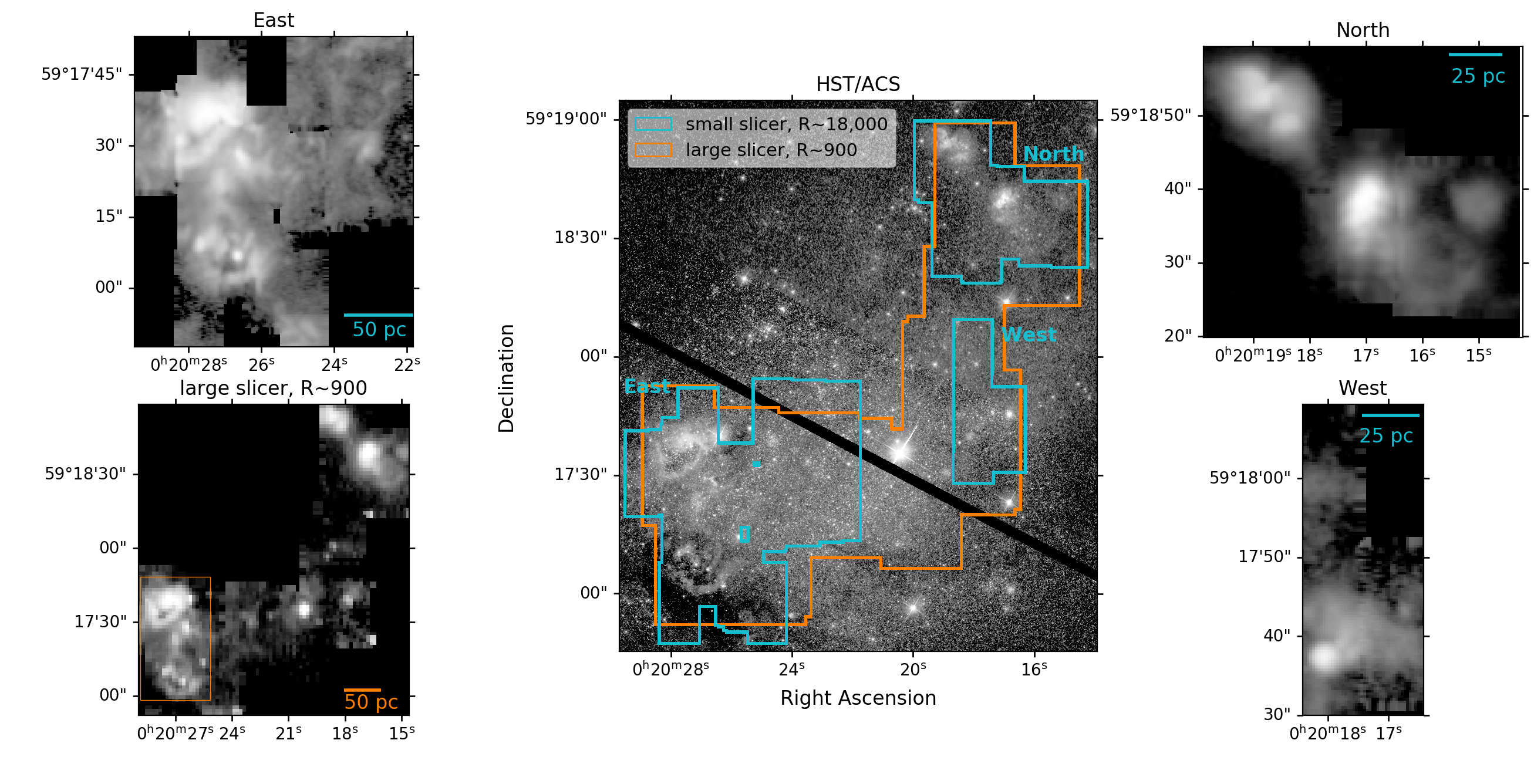}
    \caption{(Middle): HST/ACS image of IC\,10's central 2$'$x2$'$ showing outlined field coverage of our completed and in progress KCWI observations in two modes: the high resolution small slicer, R$\sim$18,000 mode (blue), and the coarse resolution large slicer, R$\sim$900 mode (orange). (Top Left): Integrated \oiii5007\ang \, flux map from the East region of our small slicer, R$\sim$18,000 observations. (Bottom Left): Large slicer, R$\sim$900 observations. The extinction correction (Section \ref{sec:spectra}) is determined from this full field, but the metallicity analysis of Section \ref{sec:metallicity} is limited to the field inside the orange outline due to low SNR outside of the \hii regions in the remainder of the field. (Top Right): North region of the small slicer, R$\sim$18,000 observations. (Bottom Right): West region of the small slicer, R$\sim$18,000 observations.}
    \label{fig:HST_KCWIobs}
    
\end{figure*}

In the high resolution mode we obtain a wavelength range of 4700-5200\ang \, and a FoV of 8.4$\arcsec$ $\times$ 20.4$\arcsec$ for each exposure, providing coverage of \hb, \oiii4959\ang, and \oiii5007\ang. A typical pointing consists of three 120s exposures with a dither pattern of 0, -1.5, +2 slices\footnote{(-): left; (+): right} perpendicular to the slices to improve sampling and avoid saturation of the bright \oiii5007\ang \, line. In the large slicer, R$\sim$900 mode we obtain wider FoV and spectral coverage in each exposure (3500-5500\ang \, and 33$\arcsec$ $\times$ 20.4$\arcsec$) with lower spatial resolution of 1.35$\arcsec$/pixel. We limit these exposure times to 6s to avoid saturation at \oiii5007\ang \, and complete 5 exposures per pointing with a dither pattern of 0, -0.5, -1.5, -2.5, -0.5, 0 slices. Due to the nature of the extended diffuse emission in the FoV of our science observations we took standalone sky frames approximately once every hour for each mode at an exposure time of 120s in the small slicer mode and 6s in the larger slicer mode. With each exposure we saved the associated guide camera image to be used in correcting WCS errors. The observation details are summarized in Table \ref{tbl:KCWI_observation_dates} with the total FoV of each observing mode illustrated in Figure \ref{fig:HST_KCWIobs}.

\begin{deluxetable*}{llccccc}
\tablecaption{KCWI Observational Summary \label{tbl:KCWI_observation_dates}}
\tablehead{\colhead{Date (mm-dd-yyyy)} & \colhead{Time (UT)} & \colhead{Airmass} & \colhead{Pointings} & \colhead{Exposures/Pointing} & \colhead{Exposure Time(s)} & \colhead{Standard Star}}
\startdata
\multicolumn{6}{c}{small slicer, R$\sim$18,000} \\
11-22-2017 & 4:30 - 5:40 & 1.41-1.36 & 1 & 3 & 900 & L870-2\\
11-23-2017 & 4:17-5:50 & 1.43-1.31 & 3 & 3 & 300 & Feige 24\\
08-16-2018 & 11:06 - 13:23 & 1.41-1.30 & 8 & 3 & 120 & NGC7293/Feige 24\\
09-03-2018 & 11:20 - 14:23\tablenotemark{a} & 1.3-1.5 & 5 & 3 & 120 & NGC7293 \\
08-16-2020 & 10:57 - 14:22 & 1.30-1.41 & 11 & 3 & 120-360 & NGC7293/Feige24 \\
\multicolumn{6}{c}{large slicer, R$\sim$900} \\
08-16-2018 & 14:00 - 14:37 & 1.32-1.37 & 5 & 5 & 6 & Feige24\\
08-16-2020 & 14:41 - 15:20 & 1.38-1.46 & 7 & 5 & 6 & Feige24 \\
\enddata
\tablecomments{Summary of Keck/KCWI observations of IC\,10's \hii regions. Approximately the same fields are observed in both modes for complimentary observations. There is some overlap in pointings between nights to increase SNR in fainter areas of IC\,10 so the total number of pointings is not the sum of each night. Approximately 1/2 of the small slicer R$\sim$900 pointings are still in progress with more exposure time needed to achieve sufficient SNR at \oii3727\ang \, to determine metallicities, but the SNR is sufficient throughout the field for determining an extinction correction at the \hii regions.
\tablenotetext{a}{There was an observational gap from 12:31-13:51 UT due to inclement weather.}}
\end{deluxetable*}

\subsection{Data Reduction} \label{sec:reduction}
Raw data were reduced using the IDL version of the public KCWI Data Reduction Pipeline (DRP) version 1.1.0 \citep{KCWIDRP} with modifications for our data set described here.

The first stage of the DRP consists of bias subtraction, gain correction, and cosmic ray removal procedures. In the default pipeline, the overscan region is used to perform a secondary bias subtraction after removal of the master bias to account for variation in the read noise between the calibration and science frames. We take this a step further to modify the bias subtraction used in our reduction to include a \textit{scaled} bias subtraction. Before the bias subtraction occurs we take the ratio of the overscan regions in the science and master bias frame for each row of the detector and multiply the master bias row by this ratio before subtracting it. This gives us a better match to the readnoise throughout the night and between the two distinct chips of the detector (with distinct amplifiers and readnoise).

Stages 2-4 of the DRP perform scattered light subtraction, determine transformations to 3D data cubes (used later), and flat field correction, respectively. We skip stage 5 of the DRP which performs sky subtraction in favor of using our own scaled sky subtraction routine on the reduced data cubes. Before this subtraction is performed, we run DRP stage 6 generating data cubes with the geometric solutions of stage 3, stage 7 to perform a correction for differential atmospheric refraction, and stage 8 which uses observations of standard stars to flux calibrate the cubes. The final data products from this pipeline are flux calibrated data cubes for both sky and science observations with associated variance cubes.

After completion of the standard pipeline steps we run a custom scaled sky subtraction on the data. This routine takes an average spectrum over the entire cube for the science frame and associated sky observation. The ratio of these two spectra are computed away from any known emission lines. This ratio is then used to scale the average sky spectrum, which is then subtracted from every spaxel of the science data cube. Errors are propagated in this step using the associated object and sky variance cubes along with the computed scale factor.

WCS offsets between individual frames in the ``small slicer, R$\sim$18,000" mode were corrected by matching stars in the guide camera images to HST/ACS imaging of IC\,10. The average offset of the measured and expected coordinates of stars in the field was used to shift the associated KCWI frame. On average, a $0.7\arcsec$ offset in declination and a $1.2\arcsec$ offset in right ascension were found for the ``small slicer, R$\sim$18,000" frames. We did not find WCS offsets in the ``large slicer, R$\sim$900 observations, and thus do not apply this step for that mode. All observations from each of the two observing modes were mosaicked with the rectangular KCWI pixels binned to square using the Python package \texttt{reproject} \citep{reproject} before analysis.

Since the observations were spread over a multi-year period, we compared the calibration frames to ensure consistency in these steps. The master bias frames produced in DRP stage 1 for each night show a standard deviation in median flux of $<0.5$\% in the small slicer, R$\sim$18,000 mode and $<0.2$\% in the large slicer, R$\sim$900 mode. Similarly, the master flats produced in DRP stage 4 show only a standard deviation of $<1$\% in the small slicer, R$\sim$18,000 mode and $<0.1$\% in the large slicer, R$\sim$900 mode.

\section{Analysis} \label{sec:analysis}

\subsection{Identifying star-forming regions}\label{sec:astrodendro}
Preliminary flux maps are generated for each emission line by summing over 15 channels (1.875\ang) centered at the systemic velocity of IC\,10 \citep[-348\kms,][]{Tifft1988}. More robust flux maps are generated later from spectral fitting, but these preliminary maps are used so as not to introduce boundary effects from low SNR regions in the \hii region identification routine. We use the python package, \texttt{astrodendro} \citep{astrodendro}, to find the locations and extent of star forming \hii regions in our \oiii \, and \hb \, flux maps. \texttt{Astrodendro} finds hierarchical structure in data sets by starting at the pixels with the highest flux and progressing to lower flux pixels surrounding those. If a local maximum is found \texttt{astrodendro} creates a new structure with that as the peak when the local maximum is above a user defined threshold. As the algorithm progresses to lower flux values a system of leaves, branches, and trunks are defined relating these local maxima \citep[see][for an illustration of this method]{Rosolowsky2008}. In this system the leaves are the most compact structures, the individual \hii regions in this study, while the branches connect the larger \hii region complexes. The trunks are the bottom level of the hierarchical structure identified by \texttt{astrodendro} and show the extent of the ionized gas emission in our KCWI observations. If these observations covered the full optical extent of IC\,10, we would expect the trunks to identify distinct areas of star-formation in the galaxy. However, since these observations are focused only on areas of high star formation activity the trunks fill the majority of the field and are therefore not physically significant in this study.

In order for a local flux maximum to be considered a real structure we have set a series of constraints to be applied by \texttt{astrodendro} - some of which are standard parameters in the package and some that are required routines unique to our data set. Standard parameters that we constrain with \texttt{astrodendro} are the minimum peak value for a structure, the minimum flux a pixel can contain in order to be added to any structure, and the minimum step size between independent structures (leaves) derived from the variance cubes resulting from the KCWI DRP. We have also written custom routines to set the minimum radius and minor axis length required for a real structure defined by the point spread function (PSF) of standard star observations. The values provided for these constraints are given in Table \ref{tbl:Astrodendro_params} for the \oiii5007\ang \, flux maps; the analysis with the \hb \, maps uses the values obtained with the same $\sigma$ requirements.

\begin{deluxetable}{cc}
    \tablecaption{Clump Identification Constraints \label{tbl:Astrodendro_params}}
    \tablehead{\colhead{Type} & \colhead{Value}}
    \startdata
    Minimum radius & 1.4px \\
        (HWHM) & $\sim0.5"$ \\
        \hline
        Minimum peak flux & $3.7\times10^{-17}$ \\
        $[ \rm erg \, s^{-1} \, cm^{-2} \, \AA^{-1}]$ & ($5\sigma$) \\
        \hline
        Minimum peak delta & $7.3\times10^{-18}$ \\
        $[ \rm erg \, s^{-1} \, cm^{-2} \, \AA^{-1}]$ & ($1\sigma$) \\
        \hline
        Minimum flux & $2.2\times10^{-17}$ \\
        $[ \rm erg \, s^{-1} \, cm^{-2} \, \AA^{-1}]$ & ($3\sigma$) \\
    \enddata
    \tablecomments{Parameters used to constrain the \texttt{astrodendro} clump fitting procedure after correcting the KCWI small slicer, R$\sim$18,000 flux maps for the background DIG. The values listed for each constraint are for the \oiii5007\ang \, flux maps, but the same constraint method (e.g., 5$\sigma$ minimum peak flux) is used for all emission lines.}
\end{deluxetable}

\begin{figure*}[ht!]
\includegraphics[width=0.95\textwidth]{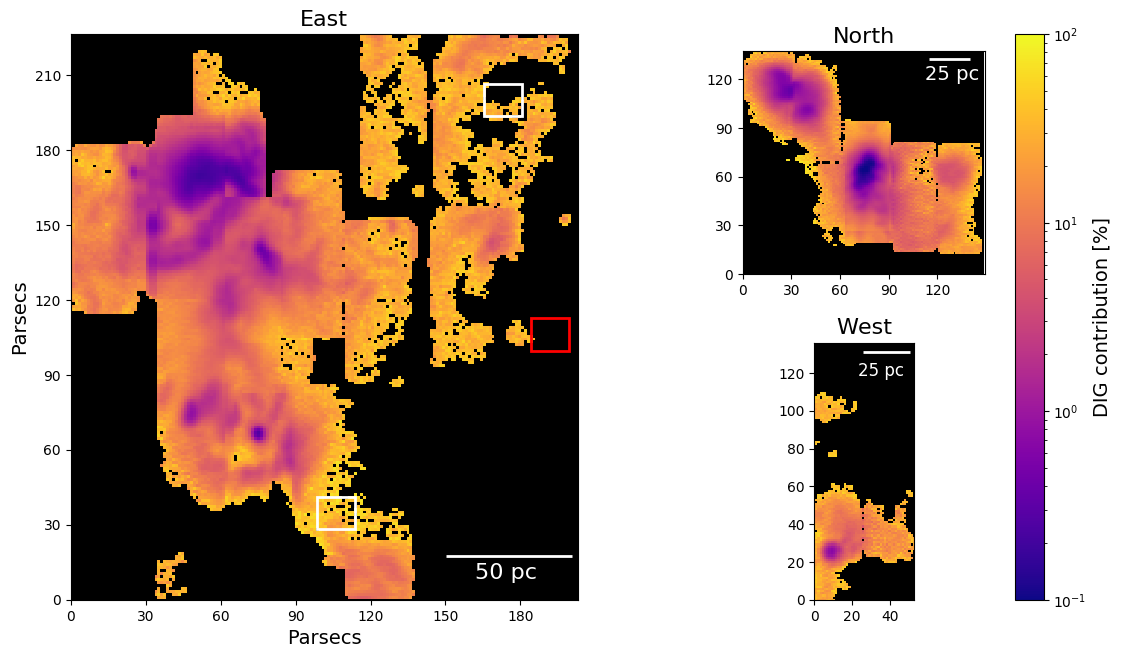}
\caption{Map of the Diffuse Ionized Gas (DIG) contribution to the total \oiii5007\ang \, flux before subtraction. The locations of HII regions are identified after this DIG contribution is removed from the flux maps. Spaxels with SNR$<2$ are masked in this figure. The DIG flux per pixel was evaluated in the three outlined boxes, with the region outlined in red chosen for the DIG subtraction as it contained the minimum flux per pixel.\label{fig:DIG_frac}}
\end{figure*}

To accurately identify \hii regions and determine SFRs, we first remove the background of Diffuse Ionized Gas (DIG). It is not possible to do this spectrally on individual spaxels since some locations have too low of SNR to determine the velocity shift (particularly in locations of greater DIG contribution). We identified three regions where there are no known \hii regions in our observed field (identified by the white and red boxes in Figure \ref{fig:DIG_frac}). We find the mean flux per pixel in each of these regions, and take the minimum value to be our DIG contribution so as not to over-subtract flux at this stage (the red box). The average DIG contribution in all three cases is on the order of a percent in pixels associated with \hii regions, so the DIG region chosen should not make a significant difference in the subsequent analysis. Maps of the DIG contribution to the integrated \oiii5007\ang \, flux are shown in Figure \ref{fig:DIG_frac}. Finally, we subtract the mean DIG flux/pixel from every location in our flux maps. For the small slicer, R$\sim$18,000 observations, this removes on average 21\% of the \oiii5007$\rm\AA$ flux per spaxel, with significantly lower contribution at the \hii regions. We repeat this subtraction on integrated flux maps for each emission line using the same DIG region throughout. We also extract a mean DIG spectrum over this same region for use in correcting \hii region spectra to be discussed in Section \ref{sec:spectra}.

The \texttt{astrodendro} package has the option to compute dendrograms on either 2D (position-position) or 3D (position-position-velocity) data. We chose to use our flux maps for \hii region identification due to the need to subtract the DIG contribution to identify fainter regions. To test whether any major differences in identified \hii regions were produced by the choice of using flux maps, the dendrograms constructed from un-subtracted flux maps and data cubes were compared. For the latter, a defined range of spectral channels is used to limit the analysis to a single emission line and a requirement is added that the spectral extent of each structure exceeds $\rm 0.38$\ang, the average width of arc lines determined from Gaussian fits to lines in the calibration frames separated every 100\ang. There was not a significant deviation in either the size or number of \hii regions identified with these two methods and therefore we proceed in the analysis of regions identified from the DIG subtracted flux maps.

\begin{figure*}[h]
    \centering
    \gridline{\fig{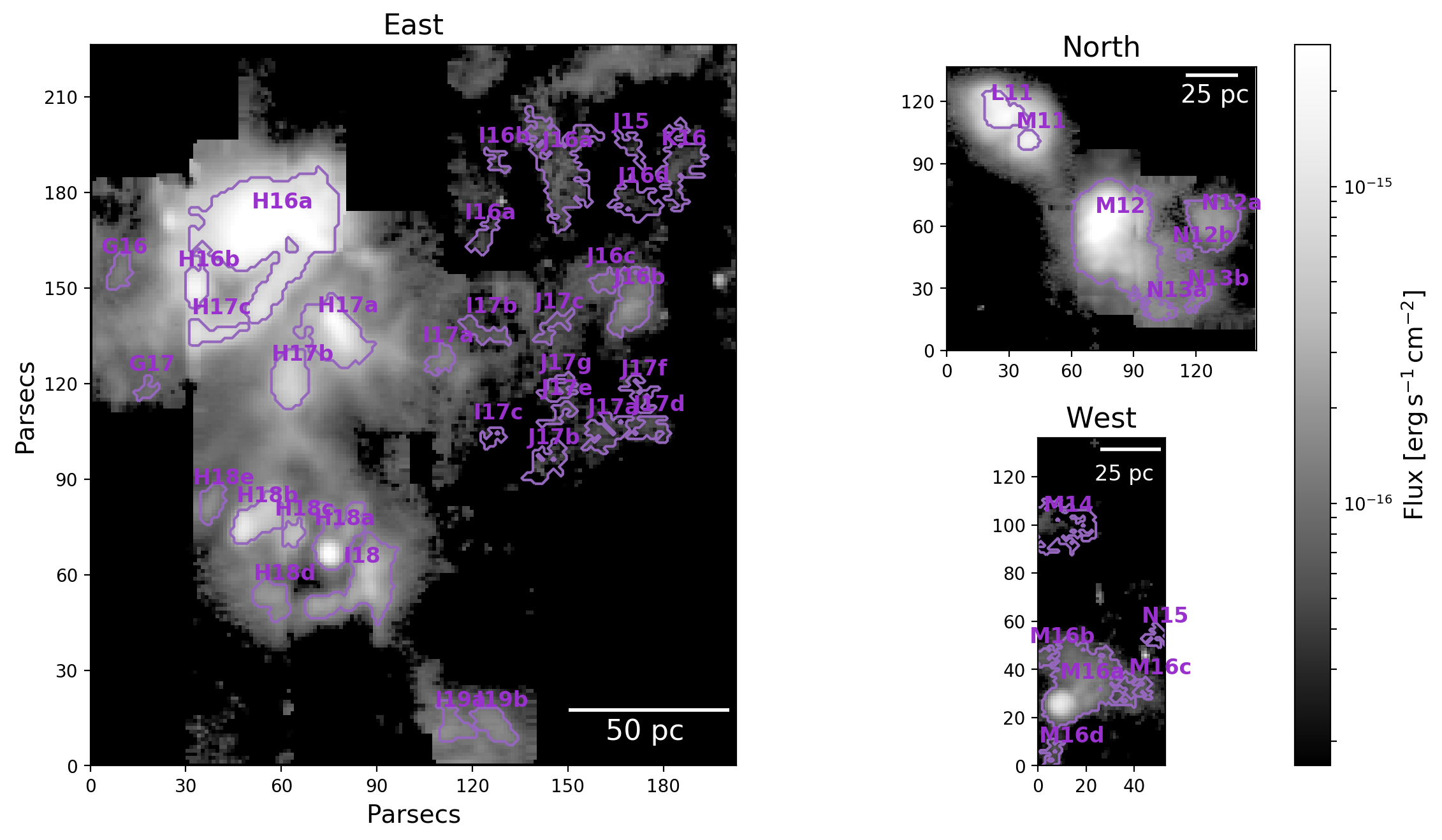}{.95\textwidth}{(a) \oiii5007\ang}}
    \gridline{\fig{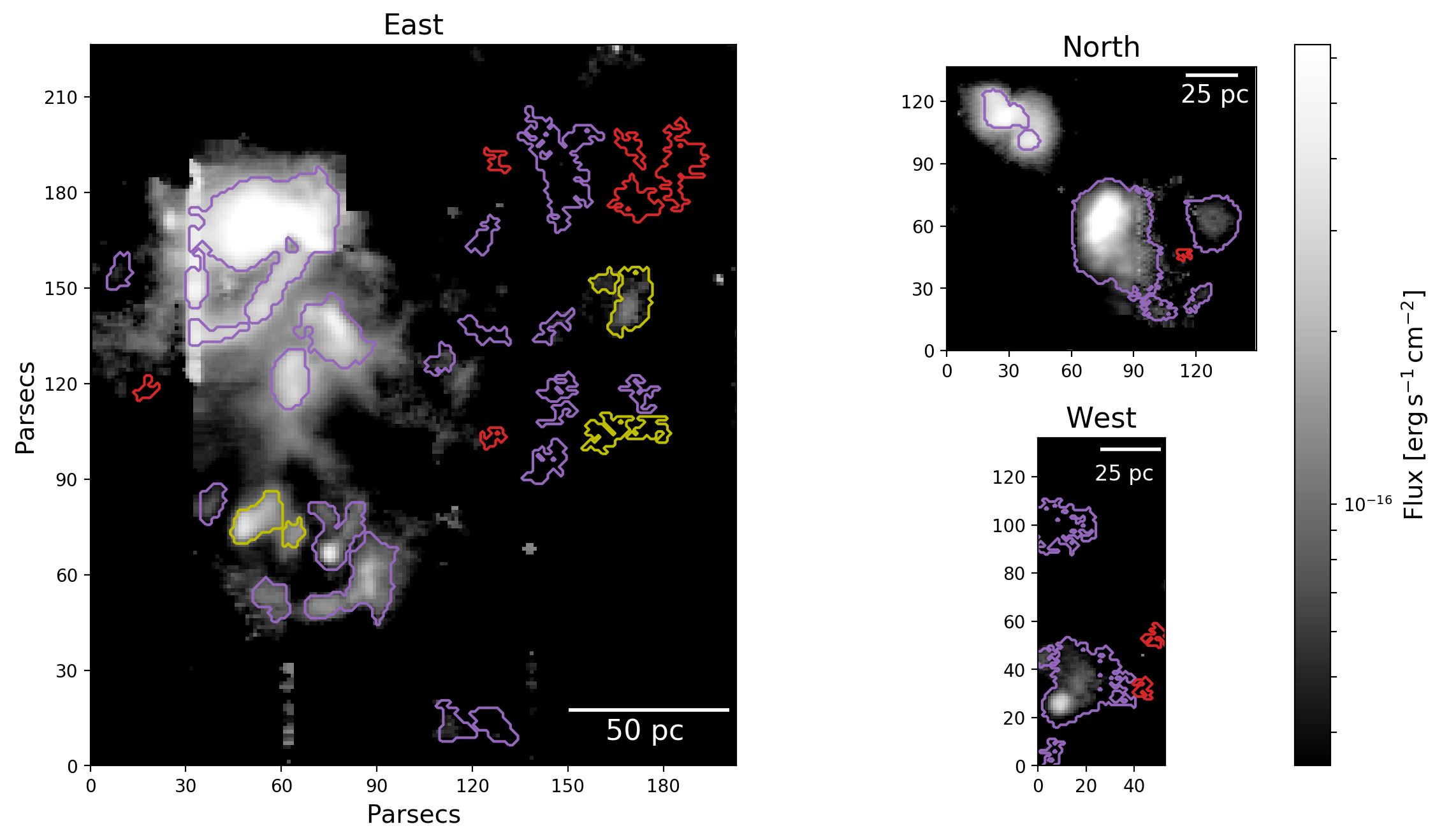}{.95\textwidth}{(b) \hb}}
    \caption{Maps of integrated flux of the ionized gas in IC\,10 measured from Gaussian fits to the \oiii5007\ang \, (a) and \hb \, (b) lines at each spaxel for the high resolution ``small slicer, R$\sim$18,000" observing mode. Each map is divided into the ``North" portion of the FoV in the upper right, the lower right shows the ``West" field, and the ``East" region of the observed field on the left. The purple contours mark the edges of the most compact \hii region structures found by \texttt{astrodendro} in the \oiii5007\ang \, flux map. The \hii regions outlined in red in the \hb\, map were not found by \texttt{astrodendro} on this lower SNR map, and the \hii regions outlined in yellow were blended into single regions. SNR $>$2 is required for these maps, with more spaxels falling below this cut in \hb. It should be noted that the \hii region contours are determined from integrated flux maps at \oiii5007\ang \, rather than these maps generated from Gaussian fits at each spaxel due to the regions with too low of SNR to perform DIG subtraction at the individual spaxel.\label{fig:flux_maps}}
\end{figure*}

Running \texttt{astrodendro} on the 2D \oiii5007\ang \, flux map and manually removing narrow filaments or regions truncated by the edge of the mosaic results in the identification of 46 \hii regions. The average radius is 4.0\,pc with a \sfrd \, of 0.20\,\myrkpc. Each of these identified regions are listed in Table \ref{tbl:regions_BH3} with the contours shown in the \oiii5007\ang \, and \hb \, flux maps of Figure \ref{fig:flux_maps}. Using the 2D \hb \, flux map produces the same \hii region locations in areas of high SNR, but the lower SNR of the \hb \, line means that faint regions cannot be identified in this map that can in \oiii5007\ang. Since the results are consistent in areas where the SNR is high for both lines, we proceed with the region identification from the \oiii5007\ang \, flux map.

It should be noted that there are more compact structures within the identified \hii regions which can be seen visually (Figure \ref{fig:flux_maps}) but do not result in unique structures detected by \texttt{astrodendro}. In some cases these structures do not meet the resolution requirements, but in others it is a result of using the same identification criteria across the entire field. A larger number of \hii regions can be identified using location dependent criteria, but that introduces an extra element of uncertainty in requiring manual tuning of the parameters. It is more prudent to maintain consistent requirements for \hii region identification across the study even if it does not result in perfect separation of compact structures.

\begin{deluxetable*}{lllrrcc}
	 \tablecaption{\hii Region Catalogue (small slicer, R$\sim$18,000) \label{tbl:regions_BH3}} 
\tablehead{\colhead{Region ID} & \colhead{Complex} & \colhead{HL90 ID} & \colhead{RA} & \colhead{Dec} & \multicolumn{2}{c}{Radius} \\ 
	 \colhead{} & \colhead{} & \colhead{} & \colhead{(J2000)} & \colhead{(J2000)} & \colhead{(pc)} & \colhead{(arcseconds)}} 
\startdata 
G16 & \nodata & 125 & 00h20m29.14s & +59d17m32.37s & 2.60 $\pm$ 0.22 & 0.75 $\pm$ 0.06 \\ 
G17 & \nodata & \nodata & 00h20m28.81s & +59d17m21.76s & 2.11 $\pm$ 0.18 & 0.61 $\pm$ 0.05 \\ 
H16a & c\_H16 & 111c/111e & 00h20m27.35s & +59d17m36.49s & 10.32 $\pm$ 0.87 & 2.98 $\pm$ 0.25 \\ 
H16b & c\_H16 & 111d & 00h20m28.23s & +59d17m31.24s & 2.95 $\pm$ 0.25 & 0.85 $\pm$ 0.07 \\ 
H17a & c\_H17 & 111a & 00h20m26.57s & +59d17m27.05s & 5.08 $\pm$ 0.43 & 1.47 $\pm$ 0.12 \\ 
H17b & c\_H17 & 111b & 00h20m27.12s & +59d17m22.66s & 4.14 $\pm$ 0.35 & 1.20 $\pm$ 0.10 \\ 
H17c & c\_H16 & 111e & 00h20m28.06s & +59d17m26.90s & 3.54 $\pm$ 0.30 & 1.02 $\pm$ 0.09 \\ 
H18a & c\_I18 & 106a & 00h20m26.61s & +59d17m07.72s & 4.74 $\pm$ 0.40 & 1.37 $\pm$ 0.12 \\ 
H18b & c\_H18 & 106b & 00h20m27.53s & +59d17m09.82s & 4.08 $\pm$ 0.34 & 1.18 $\pm$ 0.10 \\ 
H18c & c\_H18 & 106 & 00h20m27.08s & +59d17m08.61s & 2.19 $\pm$ 0.18 & 0.63 $\pm$ 0.05 \\ 
H18d & \nodata & 106 & 00h20m27.33s & +59d17m02.74s & 3.38 $\pm$ 0.28 & 0.98 $\pm$ 0.08 \\ 
H18e & \nodata & 115 & 00h20m28.05s & +59d17m11.42s & 2.72 $\pm$ 0.23 & 0.79 $\pm$ 0.07 \\ 
I16a & \nodata & \nodata & 00h20m24.82s & +59d17m35.54s & 2.60 $\pm$ 0.22 & 0.75 $\pm$ 0.06 \\
I16b & \nodata & \nodata & 00h20m24.66s & +59d17m42.49s & 2.31 $\pm$ 0.19 & 0.67 $\pm$ 0.05 \\
I17a & \nodata & 100 & 00h20m25.33s & +59d17m24.36s & 2.56 $\pm$ 0.21 & 0.74 $\pm$ 0.06 \\ 
I17b & \nodata & \nodata & 00h20m24.82s & +59d17m26.99s & 3.02 $\pm$ 0.25 & 0.87 $\pm$ 0.07 \\
I17c & \nodata & \nodata & 00h20m24.72s & +59d17m17.36s & 2.10 $\pm$ 0.18 & 0.61 $\pm$ 0.05 \\
I18 & c\_I18 & 106 & 00h20m26.26s & +59d17m04.28s & 7.02 $\pm$ 0.59 & 2.03 $\pm$ 0.17 \\ 
I19a & c\_I19 & 97 & 00h20m25.18s & +59d16m51.19s & 3.51 $\pm$ 0.30 & 1.01 $\pm$ 0.09 \\ 
I19b & c\_I19 & 91 & 00h20m24.69s & +59d16m51.24s & 3.30 $\pm$ 0.28 & 0.95 $\pm$ 0.08 \\ 
J15 & \nodata & \nodata & 00h20m23.07s & +59d17m43.78s & 2.71 $\pm$ 0.23 & 0.78 $\pm$ 0.07 \\ 
J16a & \nodata & 86/87 & 00h20m23.91s & +59d17m42.08s & 7.72 $\pm$ 0.65 & 2.23 $\pm$ 0.19 \\ 
J16b & c\_J16 & 73 & 00h20m23.06s & +59d17m29.61s & 4.46 $\pm$ 0.37 & 1.29 $\pm$ 0.11 \\ 
J16c & c\_J16 & 77 & 00h20m23.38s & +59d17m31.51s & 2.24 $\pm$ 0.19 & 0.65 $\pm$ 0.05 \\ 
J16d & \nodata & \nodata & 00h20m23.01s & +59d17m38.83s & 3.74 $\pm$ 0.31 & 1.08 $\pm$ 0.09 \\
J17a & \nodata & 74 & 00h20m23.37s & +59d17m17.83s & 3.76 $\pm$ 0.32 & 1.09 $\pm$ 0.09 \\ 
J17b & \nodata & \nodata & 00h20m24.08s & +59d17m15.08s & 3.50 $\pm$ 0.29 & 1.01 $\pm$ 0.08 \\
J17c & \nodata & 85 & 00h20m23.99s & +59d17m27.41s & 2.94 $\pm$ 0.25 & 0.85 $\pm$ 0.07 \\ 
J17d & \nodata & 74a & 00h20m22.83s & +59d17m18.09s & 3.09 $\pm$ 0.26 & 0.89 $\pm$ 0.08 \\ 
J17e & c\_J17 & 83 & 00h20m23.92s & +59d17m19.54s & 2.61 $\pm$ 0.22 & 0.75 $\pm$ 0.06 \\ 
J17f & \nodata & 74 & 00h20m22.97s & +59d17m21.34s & 3.18 $\pm$ 0.27 & 0.92 $\pm$ 0.08 \\ 
J17g & c\_J17 & 84 & 00h20m23.93s & +59d17m21.84s & 2.91 $\pm$ 0.24 & 0.84 $\pm$ 0.07 \\ 
K16 & \nodata & \nodata & 00h20m22.50s & +59d17m42.22s & 5.10 $\pm$ 0.43 & 1.47 $\pm$ 0.12 \\ 
L11 & c\_L11 & 50b/50c & 00h20m18.85s & +59d18m53.14s & 5.32 $\pm$ 0.45 & 1.54 $\pm$ 0.13 \\ 
M11 & c\_L11 & 50a & 00h20m18.38s & +59d18m49.17s & 2.77 $\pm$ 0.23 & 0.80 $\pm$ 0.07 \\ 
M12 & c\_M12 & 45 & 00h20m16.95s & +59d18m37.43s & 9.24 $\pm$ 0.78 & 2.67 $\pm$ 0.23 \\ 
M14 & \nodata & 49 & 00h20m18.10s & +59d17m58.82s & 6.74 $\pm$ 0.57 & 1.95 $\pm$ 0.16 \\ 
M16a & c\_M16 & 46a/46b & 00h20m17.84s & +59d17m38.78s & 8.45 $\pm$ 0.71 & 2.44 $\pm$ 0.21 \\ 
M16b & c\_M16 & 46c & 00h20m18.32s & +59d17m43.01s & 2.62 $\pm$ 0.22 & 0.76 $\pm$ 0.06 \\ 
M16c & c\_M16 & 44 & 00h20m16.76s & +59d17m39.25s & 2.61 $\pm$ 0.22 & 0.75 $\pm$ 0.06 \\ 
M16d & \nodata & 48 & 00h20m18.17s & +59d17m31.05s & 3.33 $\pm$ 0.28 & 0.96 $\pm$ 0.08 \\ 
N12a & \nodata & 36 & 00h20m15.03s & +59d18m37.82s & 6.79 $\pm$ 0.57 & 1.96 $\pm$ 0.16 \\ 
N12b & \nodata & 37 & 00h20m15.56s & +59d18m33.30s & 1.98 $\pm$ 0.17 & 0.57 $\pm$ 0.05 \\ 
N13a & c\_M12 & 41 & 00h20m16.02s & +59d18m25.67s & 3.91 $\pm$ 0.33 & 1.13 $\pm$ 0.10 \\ 
N13b & \nodata & 35 & 00h20m15.28s & +59d18m27.28s & 3.26 $\pm$ 0.27 & 0.94 $\pm$ 0.08 \\ 
N15 & \nodata & 43 & 00h20m16.56s & +59d17m45.47s & 2.85 $\pm$ 0.24 & 0.82 $\pm$ 0.07 \\ 
\enddata 
\tablecomments{\hii regions identified in small slicer, R$\sim$18,000 observations of IC\,10 following the new region identification scheme laid out in Section \ref{sec:naming} and the radius definition adopted in Section  \ref{sec:radius}.} 
\end{deluxetable*} 

\subsubsection{\hii Region Naming}\label{sec:naming}
\begin{figure*}[ht!]
    \centering
    \includegraphics[width=0.9\textwidth]{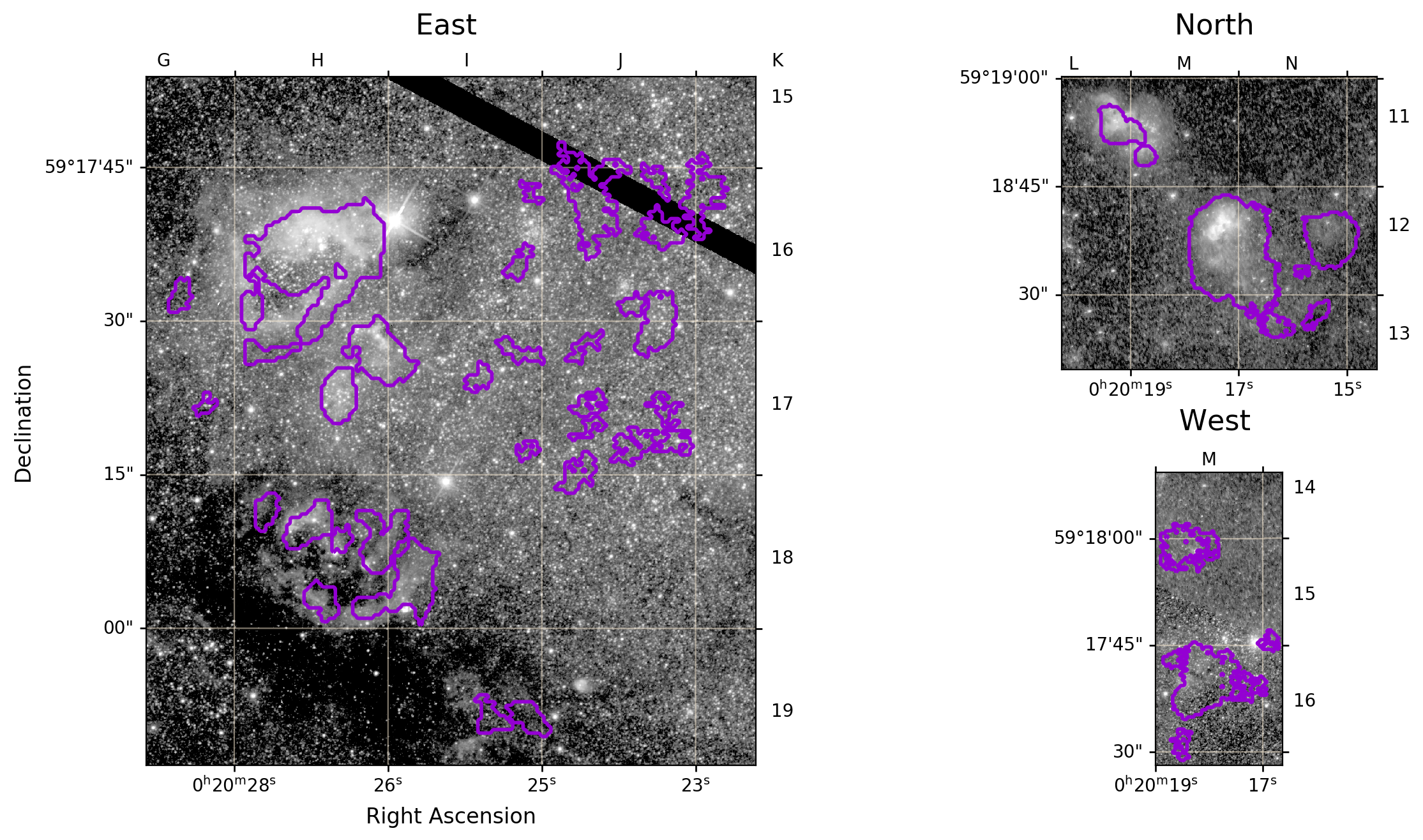}
    \caption{Coordinate grid for proposed \hii region naming scheme overlaid on an HST/ACS image of the galaxy. The column letter is shown along the top of each image with the row number along the right hand side. \hii regions are named based on the column and row corresponding to their center. The HST image is divided into the ``East", ``West", and ``North" sections as in the presentation of the KCWI maps throughout this paper (e.g., Figure \ref{fig:flux_maps}) with the identified \hii regions outlined in purple. \label{fig:coord_regs}}
\end{figure*}

Throughout this paper the identified \hii regions will be referred to with a naming convention based on a simple grid divided into 15$\arcsec\times$15$\arcsec$ regions spanning the optical extent of IC\,10. This grid based naming convention provides a simpler method of comparison between studies and extension to a larger FoV than the initial \hii region naming developed in the study by \citet{HL1990}, as discussed in more detail in Appendix \ref{app:region_naming}. The grid rows are numbered from 0 - 24 with columns designated A - X as illustrated in Figure \ref{fig:coord_regs} and Appendix Figure \ref{fig:coord_grid}. \hii regions will be assigned a name consisting of their column followed by their row (e.g., J16). In the case of multiple \hii regions falling into the same square of this grid, they are assigned an additional letter, ``a,b,c,etc.", in order of decreasing luminosity (e.g., J16a). Each knot belonging to a larger complex will be given its own designation based on the knot center with the parent complex listed in column 2 of Table \ref{tbl:regions_BH3} along with the designation from \citet{HL1990} if there is a corresponding one in column 3.

\subsubsection{Defining the Radius}\label{sec:radius}
There are a number of ways in which to define the size of star-forming regions, and the use of these methods is not always consistent between studies, particularly when comparing local and high-redshift samples with widely varying resolution. One method is to assume a spherically symmetric region (such as a \stromgren \, sphere) and fit the flux profile with a 2D elliptical Gaussian \citep[e.g.,][]{Wisnioski2012}. The half-width at half-maximum (HWHM) then gives an estimate of the half-light radius, $r_{1/2}$. A second method is to fit contours to the flux profile at a defined level above the noise and then sum the pixels contained within the contour to determine the area, $A$, of the region. This area can then be used to define the effective radius, $r_{\rm{eff}}=\sqrt{A/\pi}$ \citep[e.g.,][]{Larson2020}. This total area is produced by the \texttt{astrodendro} algorithm, from which we calculate $r_{\rm{eff}}$ of the \hii regions in IC\,10. A third method, also produced by \texttt{astrodendro}, takes an approach that compromises between the two previous methods. Rather than summing the pixels within the defined contour the second moment of the structure is determined along the direction of greatest elongation and the direction perpendicular to that. The HWHM determined by these second moments is then used to define an ellipse centered at the region peak and calculate its area. We use this area to define a pseudo half-light radius, $r_{1/2}^* = \sqrt{A_{\rm{ellipse}}/\pi}$ which we will use as the characteristic size of the \hii regions throughout this paper. We compare the result of using each of these three methods to define the extent of the \hii regions in IC\,10 in Appendix \ref{app:radius_MCMC} along with a discussion of the definitions used in similar studies.

Our choice of determining region sizes from the second moments will be most directly comparable to high-redshift studies which use $r_{1/2}$, but we do not expect significant biases from including studies using $r_{\rm{eff}}$ in our investigation of the scaling relationships (Section \ref{sec:scaling}) due to the larger impact of the PSF in high-redshift observations and the systematically larger uncertainties on measured sizes.

\subsection{\hii Region Spectra}\label{sec:spectra}
Spectra are extracted for each identified \hii region by integrating the flux over a circular aperture defined by the region center position and $r_{1/2}^*$ at each wavelength channel using the \texttt{aperture\_photometry} function in the Python package \texttt{photutils}. With this we produce an integrated flux and error spectrum for each region. The use of a circular aperture will necessarily exclude the edges of asymmetric \hii regions, but it will capture the core which provides the dominant contribution to the flux. The use of a circular aperture has also been found to include less bias from background emission \citep{Wisnioski2012}. Since the DIG subtraction prior to the construction of the dendrogram uses a lower limit on this emission source, the exact boundaries may be biased by this estimation while the circular aperture is less impacted. To ensure the choice of aperture does not bias the conclusions of this paper key analysis was carried out with both integration methods. While the size, luminosity, and dependent properties do increase slightly for integration over the exact area, the trends observed and conclusions reached in the following analysis do not change. Therefore we proceed with the analysis and results using the circular aperture with $r_{1/2}^*$. The integrated spectra are corrected for the underlying DIG contribution. To do this a Gaussian profile is fit to the \oiii5007\ang \, line in both the region and mean DIG spectrum. Any wavelength shift between these Gaussian centers is corrected and the DIG spectrum is subtracted from the integrated region spectrum with a scaling factor for the number of spaxels in the integrated region. For the remainder of our analysis we use this DIG subtracted spectrum.

\begin{figure}[h]
    \centering
    \includegraphics{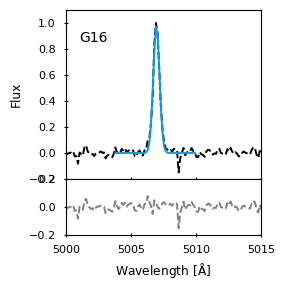}
    \caption{Thumbnail of region G16 integrated spectrum at \oiii5007\ang\, in the small slicer, R$\sim$18,000 observing mode. The complete figure set including spectra from all identified \hii regions (46 images) are available in the online journal and as thumbnails in Appendix \ref{app:spec_thumbs}. The spectral lines are fit by a single Gaussian profile (except for J16a with a double Gaussian profile) shown in cyan with the residuals shown in grey below the associated spectrum. The spectra have been normalized to the peak of the \oiii5007\ang \, line.}
    \label{fig:spec_thumb}
\end{figure}

We fit each spectrum with a continuum and a Gaussian profile at each emission line, weighted by the associated error spectrum. An example of these fits is shown in Figure \ref{fig:spec_thumb} for region G16 with the full figure set of fits for the remainder of the \hii regions available in Appendix \ref{app:spec_thumbs}. The velocity shift of the region is determined from the mean of the Gaussian fit to \oiii5007\ang \, relative to the proper motion of IC\,10 \citep[$\rm -348\pm1 \, km \, s^{-1}$;][]{Tifft1988}. The velocity dispersion is determined from the standard deviation of the Gaussian fit with the instrumental width subtracted in quadrature. The total flux of the emission line is determined by integrating over the Gaussian profile and is converted to luminosity using a distance to IC\,10 of $\rm 715\pm60 \, kpc$ \citep{Kim2009}. For nearly all cases the emission is well represented by a single Gaussian, but one region, J16a, exhibits a double peaked \oiii5007\ang \, line. For this special case integrating over the single Gaussian profile underestimates the flux by $\sim$30\% compared to a pure sum over the emission line. For this region we instead fit a double Gaussian, using the primary component to derive the velocity shift and dispersion, and integrating over both peaks to determine the flux.

Emission line luminosities are corrected for extinction determined from the ratio of the \hb \, and H$\gamma$ Balmer lines in the large slicer, R$\sim$900 \hii region spectra. We use the ratio of the integrated flux of each of these lines along with the theoretical line ratio of $0.47$ assuming Case B recombination \citep{OsterbrockFerland2006} to determine the $\rm E(B-V)$ reddening in each of the identified \hii regions following \citet{Momcheva2013}'s Equation A10:
\begin{equation}
    E(B-V) = \frac{-2.5}{\kappa(H\beta)-\kappa(H\gamma)}\times log_{10}\left(\frac{0.47}{\left(H\gamma / H\beta)\right )_{\rm{obs}}}\right)
\end{equation}
where $\kappa(H\beta)=4.6$ and $\kappa(H\gamma)=5.12$.

For the \hii regions identified, we determine an average total reddening value of $\rm E(B-V)=0.67\pm0.10$ with a higher nominal reddening $\rm E(B-V)=0.97\pm0.24$ in a stacked spectrum of spaxels outside the \hii regions, though these values overlap when incorporating the measurement uncertainties. The reddening in the stacked spectrum outside the \hii regions had to be determined using the peak flux of the \hb \, and H$\gamma$ lines due to lower SNR at H$\gamma$ resulting in a poor fit. We expect this to be a reasonable approximation as the difference introduced by this method inside the \hii regions is less than 1/3 of the uncertainty on the associated $\rm E(B-V)$. The larger uncertainty on the estimated reddening outside the \hii regions is likely to encompass the possible source of error introduced by the use of peak flux.

The reddening determined in our \hii region spectra are lower than the values found by \citet{Kim2009} via NIR colors of RGB stars ($\rm E(B-V)=1.01\pm0.03$) and from UBV photometry of early type stars ($\rm E(B-V)=0.95\pm0.06$), but the estimated reddening for IC\,10 varies widely throughout the literature. These estimates have ranged from $\rm E(B-V)=0.47$ \citep{Lequeux1979} to upwards of 1.7 \citep{YangSKillman1993} with a variety of methods used. These estimates are of the total reddening, including the foreground reddening from the Milky Way. IC\,10 is at a low galactic latitude, so estimates of even the foreground reddening show large variation. The estimate from the commonly used survey by \citet{SFD1998} gives $\rm E(B-V)\sim1.6$, larger than many of the estimates for the total reddening in IC\,10. \citet{Kim2009} notes the large uncertainties in the \citet{SFD1998} maps at low galactic latitude and finds a foreground extinction of $\rm E(B-V)=0.52$ towards IC\,10. Given the wide variation in reddening determined for IC\,10 we therefore proceed with the values determined from our spectra as this is likely to provide the most accurate measure of the extinction in our FoV. 

For the remainder of the analysis all spectra will be extinction corrected according to the associated $\rm E(B-V)$ determined from the appropriate stacked spectrum of spaxels either inside or outside of \hii regions and a \citet{Cardelli1989} extinction law. These two different measurements of extinction are particularly important in accurately correcting the extinction of these two distinct gas regions prior to estimates of the metallicity in Section \ref{sec:metallicity} using the $\rm R_{23}$ parameter with emission lines covering a wide wavelength range \citep[e.g.,][]{Kewley2019}. Ideally each spaxel would be extinction corrected individually, but the SNR was not sufficient to reliably measure H$\gamma$ at each spaxel in order to determine the localized extinction. However, due to the high foreground extinction the uncertainty in using global average extinction corrections is reduced compared to environments with a high amount of internal extinction. 

\subsection{Star Formation Rate Indicators} \label{sec:SFR}
Our high resolution observations cover the \hb \, and \oiii5007\ang \, emission lines, both of which can be used as SFR tracers. \citet{Kennicutt1992} and \citet{Moustakas2006} investigate the accuracy of these emission lines as SFR diagnostics and provide a detailed description of their advantages and disadvantages. Both determine the \ha \, luminosity to be the more reliable SFR indicator, but unfortunately this lies beyond the wavelength coverage of KCWI. We provide here a brief description of the method of calculating SFR from each of the observed emission lines.

\subsubsection{\hb}
In order to determine the SFR from the extinction corrected \hb \, flux we first convert to equivalent H$\alpha$ luminosity based on the Balmer decrement and then use the calibration of \citet{Murphy2011}:
\begin{equation}
    SFR = 5.37\times10^{-42}L_{\rm{H\alpha}} \label{eq:Ha_SFR}
\end{equation}
This calibration is based solar metallicity and a \citet{Kroupa2001} IMF. It is updated from the calibration of \citet{Kennicutt1998, Moustakas2006} which make use of a \citet{Salpeter1955} IMF.

\subsubsection{\oiii5007$\AA$}
\citet{Moustakas2006} uses their sample of SDSS star-forming galaxy spectra to explore the uncertainties present when using \oiii5007\ang \, to calculate SFR. They find a significant amount of scatter when comparing the luminosity of \oiii\, to the \ha \, derived SFR, resulting in a factor of 3-4 uncertainty which they attribute to variation in chemical abundance and excitation. However, for a single galaxy like IC\,10, and particularly for \hii regions in the central starburst, we would not expect the same level of variation in these parameters as would be present in a sample of unique galaxies. Further, previous studies have used the \oiii5007\ang, flux to estimate SFR when Balmer line measurements were not present by assuming an \oiii5007\ang/\ha \, ratio of unity \citep{Teplitz2000}.

\begin{figure}[h]
    \centering
    \includegraphics[width=0.44\textwidth]{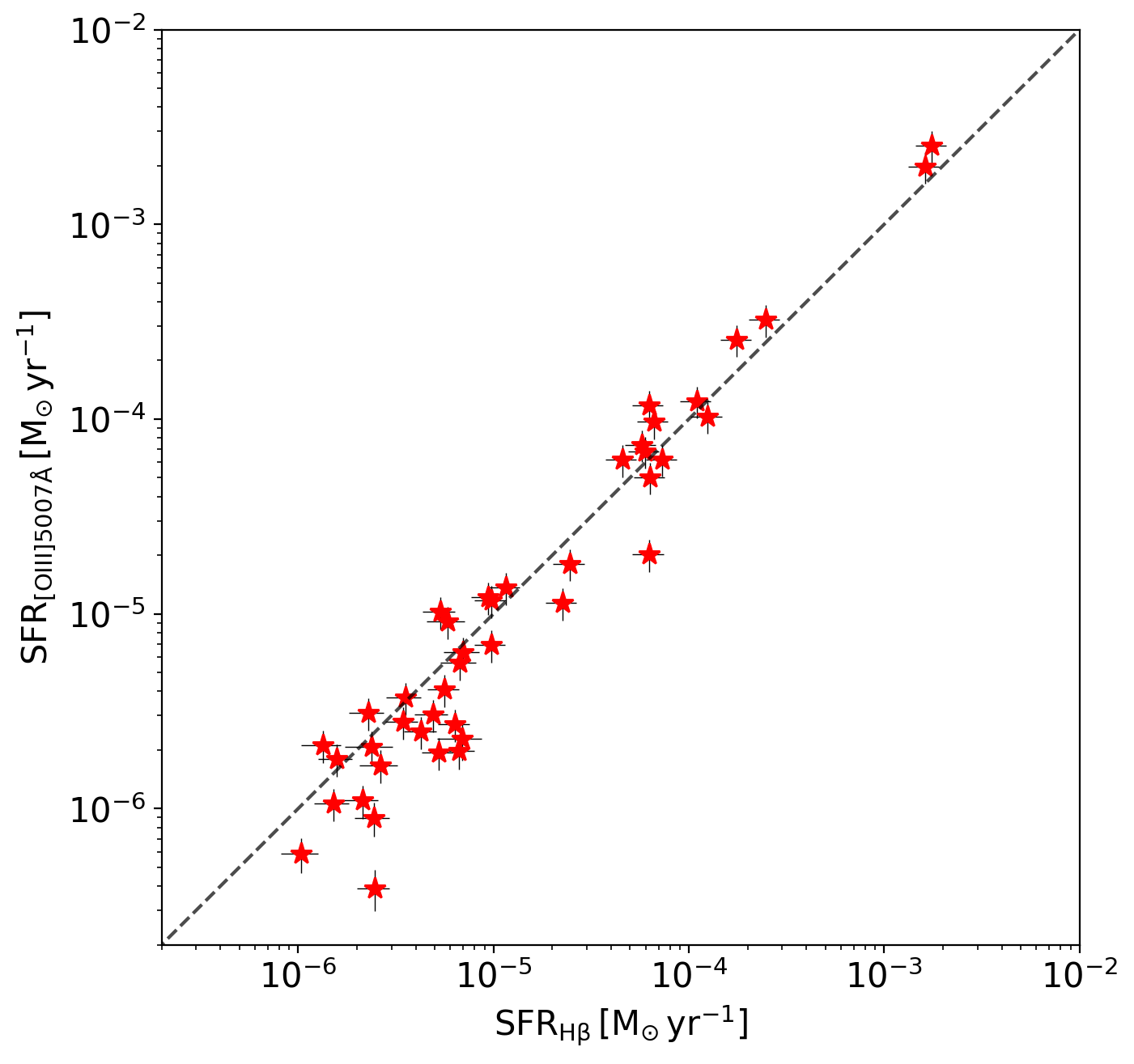}
    \caption{Comparison of the SFR determined from \hb \, and \oiii5007\ang. The dashed black line denotes the one-to-one line. Given an average $\rm SFR_{\oiii5007}/SFR_{H\beta} = 0.9$, the $\rm SFR_{\oiii5007}$ does, on average, provide a good match to the more commonly used $\rm SFR_{H\beta}$.}
    \label{fig:BH3_SFR_comp}
\end{figure}

We investigate substituting the extinction corrected \oiii5007\ang \, luminosity for $L_{\rm{H\alpha}}$ in Equation \ref{eq:Ha_SFR} to estimate the SFR in the \hii regions of IC\,10, beginning with the same factor of $\rm L_{\oiii}/L_{H\alpha}=1$ as \citet{Teplitz2000}. This is compared to the \hb \, derived SFR in Figure \ref{fig:BH3_SFR_comp} to determine if the two SFR estimates are correlated and if the coefficient of 1 between $L_{\rm{\oiii}}$ and $L_{\rm{H\alpha}}$ provides the best match. We find that there is good agreement between the two methods with an average $\rm SFR_{\oiii5007}/SFR_{H\beta} = 0.9$, though this ratio drops to $\rm SFR_{\oiii5007}/SFR_{H\beta} = 0.63$ in the faintest $1/3$ of the \hii regions. Since these are well matched overall in this sample, the higher signal-to-noise ratio \oiii5007\ang \, line will be used in the remainder of the analysis allowing for the identification of fainter star-forming regions. We checked whether the apparent lower $\rm SFR_{\oiii}/SFR_{H\beta}$ ratio in the faintest regions would introduce a bias in the results of the following sections by underestimating $\rm L_{H\alpha}$. This was not found to have a significant impact on the results and conclusions in the remainder of this paper with the potential differences being captured in the existing uncertainties. We therefore report only measurement uncertainties on quantities like SFR which rely on $\rm L_{\oiii}$ as a proxy for $\rm L_{H\alpha}$ and proceed with $\rm L_{\oiii}/L_{H\alpha}=1$ for the full sample of IC\,10 \hii regions for consistency with previous studies.

\subsubsection{Low SFR \hii Regions} \label{sec:SFR_limit}
As can be seen in Figure \ref{fig:BH3_SFR_comp}, many of the identified \hii regions in IC\,10 have very low SFR. The rate of production of ionizing photons for a single 18$M_{\odot}$ O9 star is $\rm Q=10^{47.90} \, s^{-1}$ \citep{Martins2005}. Assuming Case B recombination this results in an expected $\rm L_{H\alpha} \approx 10^{36} \, erg \, s^{-1}$. In this sample of IC\,10 \hii regions, 23/46 of the identified regions have luminosities less than this, indicating that the primary ionization source is a less massive star or cluster. There are a limited number of extragalactic surveys which to compare to in this low luminosity regime as the spatial resolution needed to differentiate these compact sources necessitates nearby objects as well as the sensitivity to detect low luminosities. The identification of IC\,10's many \hii regions by \citet{HL1990} in \ha \, imaging survey does get to this regime as well. They also find some regions in which they attribute the ionization to a single late B or early A type star. The same regime of single star ionization is reached by \citet{Hodge1989} in NGC6822 though they note that some of the regions with low surface brightness may instead be diffuse emission rather than the \hii regions they are identified as.

Within the Milky Way a number of studies of compact and ultra-compact \hii regions have been conducted which fall into this category of ionization by a single intermediate mass star at radio and infrared wavelengths. \citet{Lundquist2014} studied four intermediate mass star forming regions identified in the \textit{Infrared Astronomical Satellite (IRAS)} Point Source Catalog and \textit{Wide-field Infrared Survey Explorer (WISE)} images which they determined were powered by low mass clusters with mid-B stars as the most massive components and therefore the dominant source of ionizing photons.
Over 900 so-called ``yellowball" regions, compact \hii regions sometimes ionized by a single B type star were identified by citizen scientists as part of the Milky Way Project \citep{Kerton2015}. The yellowballs regions show spatially coincident 8 and 24$\mu$m fluxes,which the authors attribute to the early stages of the \hii region evolution. As part of HRDS, \citet{Armentrout2021} identify single star \hii regions with \textit{WISE} and the Very Large Array (VLA). They find that these single star regions have similar morphologies as their more luminous counterparts and can be powered by a single B2 or earlier spectral type star.

\begin{figure}[h]
    \centering
    \includegraphics[width=0.44\textwidth]{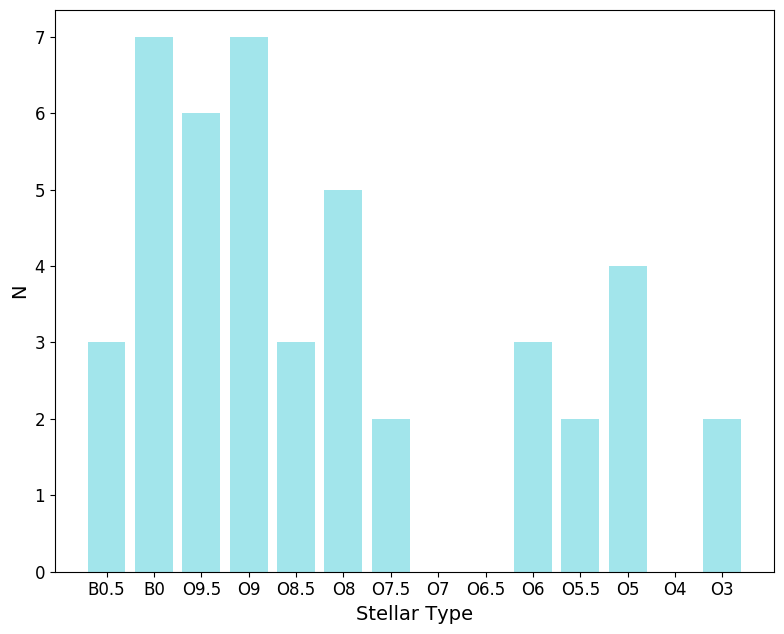}
    \caption{Distribution of the potential central ionizing source for each \hii region. The stellar type identified here is determined by the minimum number of ionizing photons which can produce the observed luminosity. In the case of a single star providing the dominant source of ionization, each IC\,10 \hii region requires at least a B0.5 star.}
    \label{fig:ionizing_source}
\end{figure}

To determine the likely stellar type for the ionizing star in IC\,10's \hii regions, we compare the observed luminosity to models of the number of Lyman continuum photons produced by a given stellar type. For consistent comparison with the \citet{Armentrout2021} Milky Way study we use the same stellar models for the number of ionizing photons produced by a single star, \citet{Martins2005} for O type stars and \citet{Smith2002} for B type stars. The minimum number of ionizing photons required to power a given \hii region is given by:
\begin{equation}
    Q_{\rm{req}} = 7.31\times10^{11} L_{\rm{H\alpha}} \label{eqn:Qreg}
\end{equation}
under case B recombination \citep{OsterbrockFerland2006}. The lowest luminosity region identified in IC\,10 can be produced by ionization from a B0.5 star, with 10 \hii regions potentially being produced by ionization from a single B star. Forty-four of the identified \hii regions have luminosities which can be produced by a single star, while the remaining two would require an ionizing photon production rate equivalent to at least six O3 stars. The number of IC\,10 \hii regions that could be produced by ionization from a single star of each spectral type is shown in Figure \ref{fig:ionizing_source}.

\subsection{Mass} \label{sec:mass}
The mass for each \hii region is estimated based on the assumption that the \hii regions are approximately spherical making the gas mass simply a function of the volume and density:
\begin{equation}
    M_{\rm{HII}} = \frac{4}{3} \pi r^3 n_{\rm{h}} m_{\rm{h}} \label{eqn:M_HII}
\end{equation}
where $\rm r$ is the \hii region radius and $\rm m_h$ is the mass of a hydrogen atom. We cannot directly measure the number density of hydrogen, $\rm n_h$, in our KCWI spectra, but we can estimate it via the \stromgren sphere approximation; the assumption that the gas is fully ionized and therefore the measured luminosity directly measures the amount of hydrogen gas present:
\begin{equation}
    n_{\rm{h}} = \sqrt{\frac{3L_{\rm{H\alpha}}\lambda_{\rm{H\alpha}}}{4\pi h c \alpha_{\rm{B}} r^3}} \label{eqn:stromgren_n}
\end{equation}
where $\lambda_{\rm{H\alpha}}$ is the wavelength of \ha, $h$ is Planck's constant, $c$ is the speed of light, and $\alpha_B$ is the Case B recombination coefficient.

\begin{figure}[h]
    \centering
    \includegraphics[width=0.48\textwidth]{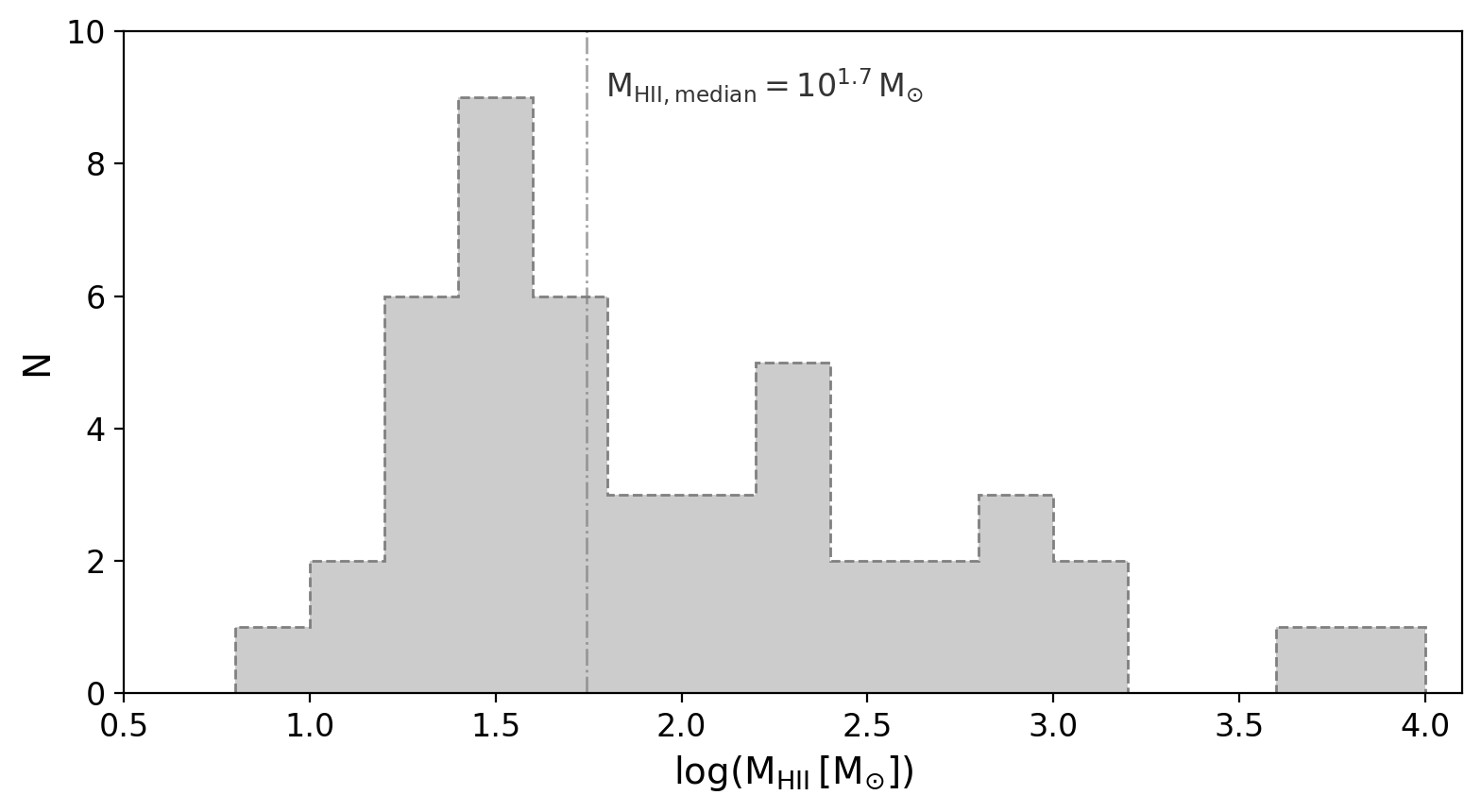}
    \caption{Histogram of the \hii region ionized gas masses, $\rm M_{HII}$, estimated from the \stromgren sphere approximation and the measured size and luminosity.}
    \label{fig:HII_mass_hist}
\end{figure}

Combining these two equations we can solve for the mass of ionized hydrogen gas in each \hii region. The resulting masses are shown in the histogram of Figure \ref{fig:HII_mass_hist} with a median region mass for the sample $\rm \sim 56 \, M_{\odot}$ and a total $\rm M_{\hii}\sim2\times10^4 \, M_{\odot}$ residing in the identified \hii regions. We will compare this to estimates of the mass based on the measured \hii region kinematics in Section \ref{sec:virialization}.

The stellar mass of each region is estimated by integrating the IMF over the full mass range from $\rm 0.1M_{\odot}$ to $\rm 100M_{\odot}$ following the method laid out in \citet{Relano2005}. We summarize the method here but refer the reader to \citet{Relano2005} for a more detailed discussion.

The total stellar mass with the previously stated upper and lower bounds is defined as:
\begin{equation}
    M_* = \int_{0.1M_{\odot}}^{100M_{\odot}} m\Phi(m)dm \label{eqn:Mstar_IMF}
\end{equation}
where $\Phi(m)=Am^{-2.3}$ for a \citet{Kroupa2001} IMF. In order to perform the integration, the normalization factor, $\rm A$, needs to be determined. This is done by estimating the stellar mass over a smaller range of stellar types. A first order estimate of $\rm M_*$ is determined by calculating a required number of O5 stars needed to produce the measured \hii region luminosity and multiplying by the mass of an O5 star ($\rm 49.53 \, M_{\odot}$). This is then substituted into Equation \ref{eqn:Mstar_IMF} with the upper and lower limits of integration replaced by the masses of an O3 ($\rm 55.3 \, M_{\odot}$) and O9 ($\rm 22.1 \, M_{\odot}$) star, respectively. The normalization factor, $\rm A$, can then be solved for and the IMF can be integrated over the full mass range to give a more accurate estimate of $\rm M_*$. For the low mass \hii regions of IC\,10 where we find luminosities consistent with ionization by a single early B star, integrating over the full IMF is somewhat uncertain. To address this, we include an estimated 30\% systematic uncertainty on $\rm M_*$ in addition to the propagated uncertainty from the measured luminosity. 

\begin{figure*}[h]
    \centering
    \includegraphics[width=0.98\textwidth]{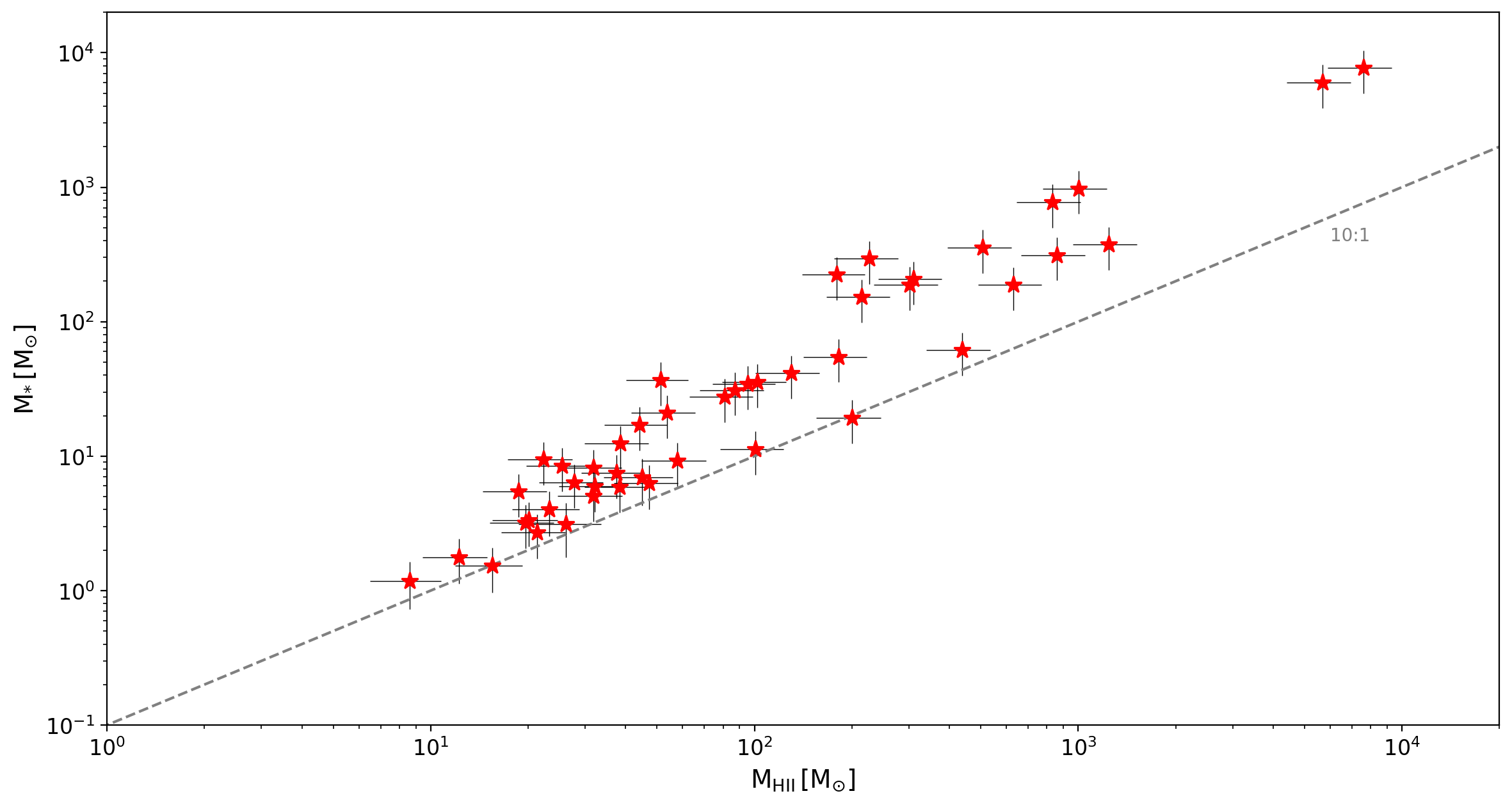}
    \caption{Comparison of the estimated $\rm M_{HII}$ and $\rm M_*$ for each IC\,10 \hii region identified in the KCWI data. On average $\rm M_* \sim 30\% \, M_{HII}$.}
    \label{fig:MHII_M*}
\end{figure*}

For IC\,10's \hii regions we find a mean $\rm M_* \sim 400 \, M_{\odot}$ from this method with a significantly lower median $\rm M_* \sim 18 \, M_{\odot}$. The large difference in the mean and median masses is due in part to a number of very low $\rm M_*$ estimates for low luminosity regions in which the assumption of a fully sampled IMF is less reliable. The estimates of $\rm M_{HII}$ and $\rm M_*$ for each \hii region are shown in Figure \ref{fig:MHII_M*}, with $\rm M_* \sim 0.3\times M_{HII}$ on average.

\subsection{Kinematics}\label{sec:kinematics} 
As discussed in Section \ref{sec:spectra}, we fit a Gaussian to each emission line in the integrated region spectra to determine the velocity shift and dispersion in each detected \hii region. Due to the higher SNR of the \oiii5007\ang \, line, we will report the values from this fit here, though the trends are the same regardless of emission line used. Previous studies of ionized gas in giant \hii regions find a systematic difference in linewidths measured from \ha \, or \hb \, and \oiii, with \oiii \, measurements giving a $\sim2$\kms \, underestimate of the dispersion, $\sigma$ \citep[e.g.,][]{Bresolin2020}. After correcting for thermal broadening of both lines, we do not observe such an offset in the spectra of IC\,10's \hii regions indicating that turbulence is likely the dominant source of the line broadening \citep{ODell2017}. We will therefore proceed with the \oiii5007\ang \, measurements to trace the kinematics of the ionized gas. The fitted kinematic properties for \oiii5007\ang \, are shown in Table \ref{tab:oiii_BH3} with properties for \oiii4959\ang \, and \hb \, included in a machine readable version of the table online. The average velocity shift is -12$\pm$12 \kms relative to the systemic velocity of IC\,10 with typical velocity dispersions of 16$\pm$8 \kms. 

\begin{figure}[h]
    \centering
    \includegraphics[width=0.45\textwidth]{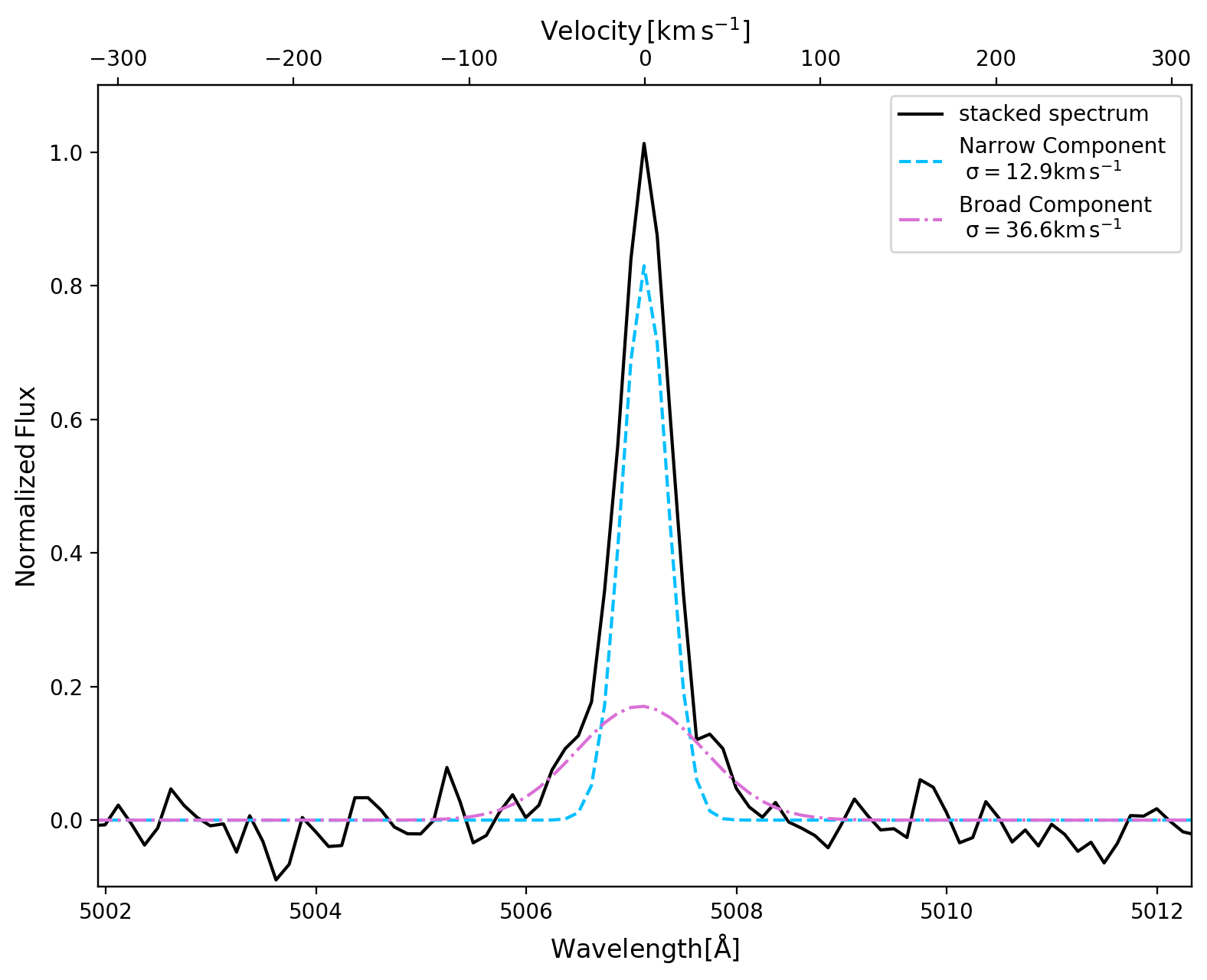}
    \caption{Stacked spectrum from each identified \hii region (black). Each individual spectrum is normalized to the \oiii5007\ang \, flux before stacking to investigate the shape of the emission lines. This reveals a faint broad component at \oiii5007\ang\, that could not be detected in the individual regions. Fits to the broad and narrow components of the \oiii5007\ang \, line are shown in magenta and cyan, respectively.}
    \label{fig:stacked_spec}
\end{figure}

Each of the \hii region DIG subtracted spectra are shifted to a common velocity based on the mean of the Gaussian fit to \oiii5007\ang, normalized to the \oiii5007\ang \, peak flux, and stacked to generate a composite spectrum as shown in Figure \ref{fig:stacked_spec}. Other than removal of the DIG, no other corrections are applied to the spectra before stacking. This reveals a lower luminosity broad component to the \oiii5007\ang \, line too faint to be detected in the individual \hii region spectra or the lower SNR \hb \, line. The broad component has a velocity dispersion $\rm \sigma = 36.6$ \kms and shows only a 2.4\kms \, velocity shift relative to the narrow component. Its peak is only 21\% of the narrow component peak, but contributes $\sim$37\% of the integrated flux. The double Gaussian fit is shown in Figure \ref{fig:stacked_spec} with the stacked spectrum.

\movetabledown=.3in
\begin{longrotatetable} 
\begin{deluxetable*}{cccccccchhhhhhh}
	 \tablecaption{\hii Region Properties (small slicer, R$\sim$18,000) \label{tab:oiii_BH3}} 
\tablehead{\colhead{Region} & \colhead{radius} & \colhead{$\Delta V_{\rm{[OIII]5007}}$} & \colhead{$\sigma_{\rm{[OIII]5007}}$} & \colhead{$L_{\rm{[OIII]5007}}$} & \colhead{SFR} & \colhead{$M_{\rm{HII}}$} & \colhead{$\tau_{\rm{cr}}$} & \nocolhead{$\Delta V_{[OIII]4959}$} & \nocolhead{$\sigma_{[OIII]4959}$} & \nocolhead{$L_{[OIII]4959}$} & \nocolhead{$\Delta V_{H\beta}$} & \nocolhead{$\sigma_{H\beta}$} & \nocolhead{$L_{H\beta}$} & \nocolhead{SFR_{H\beta}} \\ 
	 \colhead{} & \colhead{(pc)} & \colhead{($\rm km \, s^{-1}$)} & \colhead{($\rm km \, s^{-1}$)} & \colhead{($\rm 10^{35} \,erg \, s^{-1}$)} &\colhead{($\rm10^{-6}$ \myr)} & \colhead{(\msun)} & \colhead{$\rm(10^6 \ yrs)$} & \nocolhead{($\rm km \, s^{-1}$)} & \nocolhead{($\rm km \, s^{-1}$)} & \nocolhead{($\rm 10^{35} \,erg \, s^{-1}$)} & \nocolhead{($\rm km \, s^{-1}$)} & \nocolhead{($\rm km \, s^{-1}$)} & \nocolhead{($\rm 10^{35} \,erg \, s^{-1}$)} &\nocolhead{($\rm10^{-6}$ \myr)}}
\startdata 
G16 & 2.60 $\pm$ 0.22 & -8.73 $\pm$ 1.77 & 14.52 $\pm$ 0.18 & 7.62 $\pm$ 1.43 & 4.09 $\pm$ 0.77 & 38.48 $\pm$ 8.59 & 0.18 $\pm$ 0.05 & -5.58 $\pm$ 3.31 & 16.62 $\pm$ 2.93 & 1.78 $\pm$ 0.43 & 17.92 $\pm$ 2.19 & 15.97 $\pm$ 1.53 & 3.66 $\pm$ 0.67 & 5.62 $\pm$ 1.03 \\ 
G17 & 2.11 $\pm$ 0.18 & 5.62 $\pm$ 1.57 & 14.52 $\pm$ 0.28 & 3.35 $\pm$ 0.63 & 1.80 $\pm$ 0.34 & 18.65 $\pm$ 4.19 & 0.14 $\pm$ 0.04 & 8.93 $\pm$ 2.00 & 16.06 $\pm$ 1.30 & 2.14 $\pm$ 0.42 & 29.23 $\pm$ 1.99 & 16.81 $\pm$ 1.24 & 1.03 $\pm$ 0.20 & 1.58 $\pm$ 0.31 \\ 
H16a$^*$ & 10.32 $\pm$ 0.87 & -11.98 $\pm$ 1.61 & 17.02 $\pm$ 0.00 & 4741.00 $\pm$ 889.10 & 2546.00 $\pm$ 477.40 & 7591.80 $\pm$ 1690.19 & 0.59 $\pm$ 0.17 & -9.93 $\pm$ 2.24 & 17.82 $\pm$ 1.64 & 3.83 $\pm$ 0.77 & 10.40 $\pm$ 2.95 & 20.68 $\pm$ 2.46 & 1144.00 $\pm$ 205.80 & 1757.00 $\pm$ 316.10 \\ 
H16b$^*$ & 2.95 $\pm$ 0.25 & -11.52 $\pm$ 1.50 & 15.15 $\pm$ 0.08 & 181.00 $\pm$ 33.95 & 97.20 $\pm$ 18.23 & 226.71 $\pm$ 50.65 & 0.19 $\pm$ 0.06 & -8.43 $\pm$ 1.51 & 13.90 $\pm$ 0.08 & 27.58 $\pm$ 5.10 & 14.96 $\pm$ 1.55 & 16.42 $\pm$ 0.11 & 43.12 $\pm$ 7.81 & 66.22 $\pm$ 11.99 \\ 
H17a$^*$ & 5.08 $\pm$ 0.43 & -4.43 $\pm$ 2.18 & 15.85 $\pm$ 0.01 & 476.90 $\pm$ 89.41 & 256.10 $\pm$ 48.01 & 831.57 $\pm$ 185.60 & 0.31 $\pm$ 0.09 & -13.03 $\pm$ 2.85 & 15.04 $\pm$ 2.41 & 0.64 $\pm$ 0.15 & 17.68 $\pm$ 1.91 & 11.24 $\pm$ 1.12 & 114.60 $\pm$ 20.61 & 176.00 $\pm$ 31.65 \\ 
H17b$^*$ & 4.14 $\pm$ 0.35 & -0.44 $\pm$ 1.98 & 11.83 $\pm$ 0.01 & 115.60 $\pm$ 21.68 & 62.08 $\pm$ 11.64 & 301.21 $\pm$ 67.18 & 0.34 $\pm$ 0.10 & -11.19 $\pm$ 2.82 & 11.25 $\pm$ 2.37 & 1.42 $\pm$ 0.37 & 27.19 $\pm$ 4.09 & 21.70 $\pm$ 3.78 & 47.68 $\pm$ 8.58 & 73.23 $\pm$ 13.17 \\ 
H17c$^*$ & 3.54 $\pm$ 0.30 & -3.46 $\pm$ 5.64 & 12.95 $\pm$ 0.03 & 93.73 $\pm$ 17.58 & 50.33 $\pm$ 9.44 & 214.45 $\pm$ 47.91 & 0.27 $\pm$ 0.08 & 114.11 $\pm$ 7.08 & 6.29 $\pm$ 6.84 & 0.00 $\pm$ 0.00 & -9751.25 $\pm$ 0.00 & 0.00 $\pm$ 0.00 & 41.29 $\pm$ 7.43 & 63.41 $\pm$ 11.41 \\ 
H18a$^*$ & 4.74 $\pm$ 0.40 & -14.98 $\pm$ 1.52 & 16.84 $\pm$ 0.03 & 218.70 $\pm$ 41.01 & 117.40 $\pm$ 22.02 & 507.55 $\pm$ 113.07 & 0.28 $\pm$ 0.08 & -11.98 $\pm$ 1.62 & 27.70 $\pm$ 0.57 & 6.86 $\pm$ 1.27 & 12.89 $\pm$ 1.78 & 29.62 $\pm$ 0.88 & 40.64 $\pm$ 7.32 & 62.42 $\pm$ 11.25 \\ 
H18b & 4.08 $\pm$ 0.34 & -38.16 $\pm$ 1.67 & 11.83 $\pm$ 0.01 & 127.50 $\pm$ 23.90 & 68.47 $\pm$ 12.83 & 309.48 $\pm$ 68.38 & 0.34 $\pm$ 0.10 & -139.53 $\pm$ 3.71 & 13.74 $\pm$ 3.40 & 0.12 $\pm$ 0.04 & -2.11 $\pm$ 1.72 & 3.70 $\pm$ 1.11 & 38.82 $\pm$ 6.98 & 59.62 $\pm$ 10.73 \\ 
H18c & 2.19 $\pm$ 0.18 & -31.40 $\pm$ 1.72 & 11.75 $\pm$ 0.03 & 22.63 $\pm$ 4.24 & 12.15 $\pm$ 2.28 & 51.27 $\pm$ 11.23 & 0.18 $\pm$ 0.05 & 0.06 $\pm$ 8.43 & 40.66 $\pm$ 8.33 & 0.71 $\pm$ 0.18 & 162.40 $\pm$ 2.09 & 4.12 $\pm$ 1.06 & 6.11 $\pm$ 1.10 & 9.39 $\pm$ 1.69 \\ 
H18d & 3.38 $\pm$ 0.28 & -18.39 $\pm$ 1.50 & 13.70 $\pm$ 0.04 & 21.16 $\pm$ 3.97 & 11.36 $\pm$ 2.13 & 95.07 $\pm$ 20.93 & 0.24 $\pm$ 0.07 & -16.77 $\pm$ 1.56 & 19.40 $\pm$ 0.37 & 6.29 $\pm$ 1.17 & 15.09 $\pm$ 1.96 & 23.50 $\pm$ 1.21 & 14.68 $\pm$ 2.64 & 22.55 $\pm$ 4.06 \\ 
H18e & 2.72 $\pm$ 0.23 & -15.82 $\pm$ 1.50 & 11.93 $\pm$ 0.08 & 12.89 $\pm$ 2.42 & 6.92 $\pm$ 1.30 & 53.56 $\pm$ 11.95 & 0.22 $\pm$ 0.06 & -13.15 $\pm$ 1.52 & 15.40 $\pm$ 0.16 & 63.51 $\pm$ 11.76 & 10.59 $\pm$ 1.56 & 18.70 $\pm$ 0.23 & 6.34 $\pm$ 1.15 & 9.73 $\pm$ 1.76 \\ 
I16a & 2.60 $\pm$ 0.22 & -6.80 $\pm$ 1.51 & 12.18 $\pm$ 0.55 & 1.97 $\pm$ 0.38 & 1.06 $\pm$ 0.20 & 19.59 $\pm$ 4.40 & 0.21 $\pm$ 0.06 & -4.43 $\pm$ 1.58 & 10.02 $\pm$ 0.45 & 12.36 $\pm$ 2.34 & 16.77 $\pm$ 1.73 & 14.57 $\pm$ 0.76 & 0.98 $\pm$ 0.20 & 1.51 $\pm$ 0.31 \\ 
I16b & 2.31 $\pm$ 0.19 & -1.66 $\pm$ 1.51 & 16.64 $\pm$ 1.53 & 1.09 $\pm$ 0.22 & 0.59 $\pm$ 0.12 & 12.20 $\pm$ 2.76 & 0.14 $\pm$ 0.04 & 1.11 $\pm$ 1.60 & 11.96 $\pm$ 0.51 & 1.83 $\pm$ 0.35 & 21.68 $\pm$ 1.67 & 14.42 $\pm$ 0.63 & 0.68 $\pm$ 0.15 & 1.04 $\pm$ 0.23 \\ 
I17a$^*$ & 2.56 $\pm$ 0.21 & 7.85 $\pm$ 1.68 & 17.05 $\pm$ 0.57 & 10.48 $\pm$ 1.99 & 5.63 $\pm$ 1.07 & 44.10 $\pm$ 9.69 & 0.15 $\pm$ 0.04 & 10.58 $\pm$ 3.07 & 17.33 $\pm$ 2.64 & 1.64 $\pm$ 0.37 & 37.91 $\pm$ 2.54 & 14.63 $\pm$ 2.01 & 4.37 $\pm$ 0.90 & 6.71 $\pm$ 1.38 \\ 
I17b & 3.02 $\pm$ 0.25 & -3.88 $\pm$ 1.50 & 15.08 $\pm$ 0.43 & 4.62 $\pm$ 0.87 & 2.48 $\pm$ 0.47 & 37.50 $\pm$ 8.28 & 0.20 $\pm$ 0.06 & -0.75 $\pm$ 1.51 & 12.05 $\pm$ 0.03 & 33.67 $\pm$ 6.23 & 23.04 $\pm$ 1.54 & 14.88 $\pm$ 0.04 & 2.78 $\pm$ 0.52 & 4.27 $\pm$ 0.81 \\ 
I17c$^*$ & 2.10 $\pm$ 0.18 & -2.29 $\pm$ 1.50 & 36.14 $\pm$ 6.23 & 0.73 $\pm$ 0.17 & 0.39 $\pm$ 0.09 & 8.63 $\pm$ 2.14 & 0.06 $\pm$ 0.02 & 0.62 $\pm$ 1.51 & 11.96 $\pm$ 0.08 & 7.59 $\pm$ 1.41 & 23.40 $\pm$ 1.55 & 15.23 $\pm$ 0.14 & 1.60 $\pm$ 0.31 & 2.46 $\pm$ 0.47 \\ 
I18$^*$ & 7.02 $\pm$ 0.59 & -63.67 $\pm$ 1.65 & 15.16 $\pm$ 0.06 & 192.30 $\pm$ 36.07 & 103.30 $\pm$ 19.37 & 857.80 $\pm$ 190.61 & 0.45 $\pm$ 0.13 & 9.91 $\pm$ 2.77 & 16.96 $\pm$ 2.32 & 1.85 $\pm$ 0.41 & 44.85 $\pm$ 1.90 & 3.56 $\pm$ 1.23 & 81.26 $\pm$ 14.64 & 124.80 $\pm$ 22.48 \\ 
I19a & 3.51 $\pm$ 0.30 & -22.49 $\pm$ 1.51 & 32.39 $\pm$ 0.23 & 21.85 $\pm$ 4.10 & 11.73 $\pm$ 2.20 & 102.23 $\pm$ 22.97 & 0.11 $\pm$ 0.03 & -18.99 $\pm$ 1.59 & 23.72 $\pm$ 0.50 & 7.44 $\pm$ 1.38 & 27.09 $\pm$ 1.99 & 21.20 $\pm$ 1.26 & 6.36 $\pm$ 1.16 & 9.77 $\pm$ 1.77 \\ 
I19b$^*$ & 3.30 $\pm$ 0.28 & -29.09 $\pm$ 1.64 & 19.26 $\pm$ 0.13 & 19.07 $\pm$ 3.58 & 10.24 $\pm$ 1.92 & 87.06 $\pm$ 19.47 & 0.17 $\pm$ 0.05 & 94.05 $\pm$ 3.17 & 10.23 $\pm$ 2.78 & 0.44 $\pm$ 0.13 & 185.28 $\pm$ nan & 0.91 $\pm$ nan & 3.48 $\pm$ 0.65 & 5.34 $\pm$ 0.99 \\ 
J15 & 2.71 $\pm$ 0.23 & -1.79 $\pm$ 1.51 & 53.08 $\pm$ 6.37 & 2.46 $\pm$ 0.52 & 1.32 $\pm$ 0.28 & 23.27 $\pm$ 5.46 & 0.05 $\pm$ 0.02 & -0.23 $\pm$ 1.59 & 14.13 $\pm$ 0.47 & 2.32 $\pm$ 0.44 & 23.10 $\pm$ 1.71 & 17.59 $\pm$ 0.74 & 0.00 $\pm$ 0.00 & 0.00 $\pm$ 0.00 \\ 
J16a & 7.72 $\pm$ 0.65 & -6.02 $\pm$ 1.50 & 11.91 $\pm$ 0.47 & 37.66 $\pm$ 7.13 & 20.22 $\pm$ 3.83 & 437.78 $\pm$ 97.70 & 0.63 $\pm$ 0.19 & -2.57 $\pm$ 1.51 & 17.19 $\pm$ 0.01 & 1575.00 $\pm$ 291.40 & 19.31 $\pm$ 1.54 & 19.47 $\pm$ 0.02 & 40.97 $\pm$ 7.44 & 62.92 $\pm$ 11.43 \\ 
J16b & 4.46 $\pm$ 0.37 & -9.92 $\pm$ 1.76 & 9.88 $\pm$ 0.17 & 33.71 $\pm$ 6.35 & 18.10 $\pm$ 3.41 & 181.87 $\pm$ 40.13 & 0.44 $\pm$ 0.13 & -15.57 $\pm$ 3.22 & 13.99 $\pm$ 2.83 & 0.55 $\pm$ 0.15 & 14.83 $\pm$ 1.90 & 11.49 $\pm$ 1.10 & 16.01 $\pm$ 2.97 & 24.59 $\pm$ 4.56 \\ 
J16c & 2.24 $\pm$ 0.19 & -20.49 $\pm$ 1.56 & 10.21 $\pm$ 0.16 & 5.21 $\pm$ 0.98 & 2.80 $\pm$ 0.53 & 25.45 $\pm$ 5.69 & 0.21 $\pm$ 0.06 & -20.61 $\pm$ 1.82 & 13.70 $\pm$ 1.01 & 1.37 $\pm$ 0.27 & 8.23 $\pm$ 1.89 & 16.91 $\pm$ 1.08 & 2.24 $\pm$ 0.41 & 3.44 $\pm$ 0.63 \\ 
J16d & 3.74 $\pm$ 0.31 & -15.46 $\pm$ 1.53 & 22.29 $\pm$ 1.51 & 3.87 $\pm$ 0.76 & 2.08 $\pm$ 0.41 & 47.33 $\pm$ 10.59 & 0.16 $\pm$ 0.05 & -13.21 $\pm$ 1.78 & 14.29 $\pm$ 0.94 & 2.91 $\pm$ 0.56 & 12.63 $\pm$ 1.69 & 14.14 $\pm$ 0.69 & 1.55 $\pm$ 0.42 & 2.38 $\pm$ 0.65 \\ 
J17a & 3.76 $\pm$ 0.32 & -12.90 $\pm$ 1.50 & 15.70 $\pm$ 0.47 & 5.67 $\pm$ 1.07 & 3.05 $\pm$ 0.58 & 57.76 $\pm$ 12.98 & 0.23 $\pm$ 0.07 & -10.16 $\pm$ 1.51 & 17.10 $\pm$ 0.09 & 71.35 $\pm$ 13.20 & 13.15 $\pm$ 1.56 & 18.80 $\pm$ 0.23 & 3.18 $\pm$ 0.61 & 4.89 $\pm$ 0.94 \\ 
J17b & 3.50 $\pm$ 0.29 & 0.80 $\pm$ 1.50 & 21.47 $\pm$ 3.60 & 4.27 $\pm$ 1.00 & 2.29 $\pm$ 0.53 & 44.98 $\pm$ 10.84 & 0.16 $\pm$ 0.05 & 3.42 $\pm$ 1.51 & 13.35 $\pm$ 0.10 & 7.21 $\pm$ 1.33 & 25.87 $\pm$ 1.55 & 18.16 $\pm$ 0.14 & 4.49 $\pm$ 1.15 & 6.90 $\pm$ 1.76 \\ 
J17c & 2.94 $\pm$ 0.25 & 1.89 $\pm$ 1.52 & 14.66 $\pm$ 0.92 & 3.68 $\pm$ 0.72 & 1.98 $\pm$ 0.39 & 32.16 $\pm$ 7.32 & 0.20 $\pm$ 0.06 & 4.91 $\pm$ 1.69 & 14.73 $\pm$ 0.75 & 1.27 $\pm$ 0.24 & 28.13 $\pm$ 1.92 & 13.38 $\pm$ 1.14 & 4.32 $\pm$ 0.86 & 6.63 $\pm$ 1.32 \\ 
J17d$^*$ & 3.09 $\pm$ 0.26 & -2.98 $\pm$ 1.50 & 13.66 $\pm$ 0.76 & 3.11 $\pm$ 0.60 & 1.67 $\pm$ 0.32 & 31.85 $\pm$ 7.18 & 0.22 $\pm$ 0.07 & -0.09 $\pm$ 1.51 & 11.75 $\pm$ 0.03 & 36.51 $\pm$ 6.75 & 22.76 $\pm$ 1.54 & 14.28 $\pm$ 0.04 & 1.72 $\pm$ 0.38 & 2.64 $\pm$ 0.58 \\ 
J17e & 2.61 $\pm$ 0.22 & 4.98 $\pm$ 1.50 & 22.06 $\pm$ 1.57 & 2.05 $\pm$ 0.40 & 1.10 $\pm$ 0.22 & 20.06 $\pm$ 4.55 & 0.12 $\pm$ 0.03 & 8.05 $\pm$ 1.52 & 12.34 $\pm$ 0.19 & 4.55 $\pm$ 0.84 & 32.93 $\pm$ 1.57 & 15.87 $\pm$ 0.30 & 1.39 $\pm$ 0.28 & 2.13 $\pm$ 0.42 \\ 
J17f & 3.18 $\pm$ 0.27 & -16.69 $\pm$ 3.92 & 18.31 $\pm$ 5.39 & 1.92 $\pm$ 0.60 & 1.03 $\pm$ 0.32 & 26.13 $\pm$ 7.45 & 0.17 $\pm$ 0.07 & 7.52 $\pm$ 12.41 & 30.73 $\pm$ 12.27 & 2.29 $\pm$ 0.94 & 14.11 $\pm$ 3.93 & 17.39 $\pm$ 3.59 & 0.00 $\pm$ 0.00 & 0.00 $\pm$ 0.00 \\ 
J17g & 2.91 $\pm$ 0.24 & -33.59 $\pm$ 6.47 & 12.12 $\pm$ 0.91 & 1.66 $\pm$ 0.33 & 0.89 $\pm$ 0.18 & 21.28 $\pm$ 4.77 & 0.23 $\pm$ 0.07 & -17.71 $\pm$ 2.03 & 6.48 $\pm$ 1.36 & 0.00 $\pm$ 0.00 & 19.14 $\pm$ 2.09 & 18.42 $\pm$ 1.40 & 1.59 $\pm$ 0.32 & 2.44 $\pm$ 0.49 \\ 
K16 & 5.10 $\pm$ 0.43 & -26.38 $\pm$ 1.55 & 12.79 $\pm$ 0.32 & 6.91 $\pm$ 1.31 & 3.71 $\pm$ 0.70 & 100.70 $\pm$ 22.48 & 0.39 $\pm$ 0.11 & -25.07 $\pm$ 1.86 & 17.76 $\pm$ 1.08 & 2.15 $\pm$ 0.41 & 3.06 $\pm$ 1.77 & 17.26 $\pm$ 0.86 & 2.30 $\pm$ 0.47 & 3.54 $\pm$ 0.71 \\ 
L11 & 5.32 $\pm$ 0.45 & -14.92 $\pm$ 3.02 & 10.21 $\pm$ 0.01 & 601.40 $\pm$ 112.80 & 323.00 $\pm$ 60.57 & 1000.78 $\pm$ 223.30 & 0.51 $\pm$ 0.15 & -2.98 $\pm$ 3.67 & 33.66 $\pm$ 3.36 & 2.22 $\pm$ 0.45 & 120.89 $\pm$ nan & 0.00 $\pm$ nan & 160.50 $\pm$ 28.87 & 246.50 $\pm$ 44.34 \\ 
M11$^*$ & 2.77 $\pm$ 0.23 & -19.93 $\pm$ 1.51 & 11.42 $\pm$ 0.02 & 137.30 $\pm$ 25.75 & 73.73 $\pm$ 13.83 & 179.66 $\pm$ 39.61 & 0.24 $\pm$ 0.07 & -17.36 $\pm$ 1.56 & 7.70 $\pm$ 0.40 & 1.87 $\pm$ 0.36 & -3.30 $\pm$ 2.52 & 12.71 $\pm$ 1.98 & 37.43 $\pm$ 6.74 & 57.49 $\pm$ 10.34 \\ 
M12$^*$ & 9.24 $\pm$ 0.78 & -45.03 $\pm$ 1.57 & 14.88 $\pm$ 0.00 & 3693.00 $\pm$ 692.40 & 1983.00 $\pm$ 371.80 & 5676.59 $\pm$ 1264.79 & 0.61 $\pm$ 0.18 & -45.03 $\pm$ 1.57 & 11.91 $\pm$ 0.47 & 38.69 $\pm$ 7.22 & -45.03 $\pm$ 1.57 & 11.91 $\pm$ 0.47 & 1053.00 $\pm$ 189.40 & 1617.00 $\pm$ 290.90 \\ 
M14 & 6.74 $\pm$ 0.57 & -17.98 $\pm$ 1.60 & 25.23 $\pm$ 0.95 & 11.84 $\pm$ 2.25 & 6.36 $\pm$ 1.21 & 200.24 $\pm$ 44.91 & 0.26 $\pm$ 0.08 & -15.74 $\pm$ 1.99 & 14.06 $\pm$ 1.29 & 0.84 $\pm$ 0.17 & 15.33 $\pm$ 1.98 & 12.16 $\pm$ 1.24 & 4.56 $\pm$ 0.93 & 7.00 $\pm$ 1.42 \\ 
M16a$^*$ & 8.45 $\pm$ 0.71 & -12.73 $\pm$ 1.66 & 10.06 $\pm$ 0.02 & 230.30 $\pm$ 43.19 & 123.70 $\pm$ 23.19 & 1239.72 $\pm$ 275.42 & 0.82 $\pm$ 0.24 & -18.22 $\pm$ 1.94 & 18.99 $\pm$ 1.20 & 5.37 $\pm$ 1.03 & 16.24 $\pm$ 2.48 & 27.71 $\pm$ 1.94 & 71.69 $\pm$ 12.90 & 110.10 $\pm$ 19.81 \\ 
M16b & 2.62 $\pm$ 0.22 & -9.04 $\pm$ 1.50 & 9.98 $\pm$ 0.15 & 5.06 $\pm$ 0.95 & 2.72 $\pm$ 0.51 & 31.73 $\pm$ 7.05 & 0.26 $\pm$ 0.07 & -5.80 $\pm$ 1.53 & 15.02 $\pm$ 0.22 & 15.42 $\pm$ 2.86 & 16.99 $\pm$ 1.58 & 16.10 $\pm$ 0.33 & 4.14 $\pm$ 0.75 & 6.35 $\pm$ 1.15 \\ 
M16c & 2.61 $\pm$ 0.22 & -13.06 $\pm$ 1.50 & 8.91 $\pm$ 0.21 & 3.92 $\pm$ 0.74 & 2.11 $\pm$ 0.40 & 27.78 $\pm$ 6.20 & 0.29 $\pm$ 0.08 & -9.71 $\pm$ 1.53 & 16.50 $\pm$ 0.23 & 63.82 $\pm$ 11.83 & 13.23 $\pm$ 1.61 & 19.09 $\pm$ 0.45 & 0.87 $\pm$ 0.20 & 1.34 $\pm$ 0.30 \\ 
M16d & 3.33 $\pm$ 0.28 & -3.86 $\pm$ 1.50 & 14.12 $\pm$ 0.38 & 3.61 $\pm$ 0.68 & 1.94 $\pm$ 0.37 & 38.39 $\pm$ 8.55 & 0.23 $\pm$ 0.07 & -0.74 $\pm$ 1.51 & 15.85 $\pm$ 0.02 & 152.90 $\pm$ 28.28 & 20.07 $\pm$ 1.54 & 17.55 $\pm$ 0.03 & 3.42 $\pm$ 0.63 & 5.25 $\pm$ 0.97 \\ 
N12a$^*$ & 6.79 $\pm$ 0.57 & -27.95 $\pm$ 1.53 & 12.18 $\pm$ 0.04 & 115.20 $\pm$ 21.61 & 61.86 $\pm$ 11.60 & 631.57 $\pm$ 140.24 & 0.55 $\pm$ 0.16 & -28.78 $\pm$ 1.74 & 13.44 $\pm$ 0.85 & 3.01 $\pm$ 0.58 & 2.10 $\pm$ 2.05 & 12.41 $\pm$ 1.35 & 29.75 $\pm$ 5.36 & 45.69 $\pm$ 8.23 \\ 
N12b$^*$ & 1.98 $\pm$ 0.17 & -24.54 $\pm$ 2.13 & 15.93 $\pm$ 0.30 & 5.78 $\pm$ 1.09 & 3.10 $\pm$ 0.58 & 22.28 $\pm$ 5.03 & 0.12 $\pm$ 0.04 & -18.42 $\pm$ 4.73 & 27.99 $\pm$ 4.51 & 2.01 $\pm$ 0.46 & 2.73 $\pm$ 5.30 & 20.49 $\pm$ 5.07 & 1.48 $\pm$ 0.30 & 2.28 $\pm$ 0.46 \\ 
N13a & 3.91 $\pm$ 0.33 & -21.74 $\pm$ 1.50 & 11.97 $\pm$ 0.08 & 25.42 $\pm$ 4.77 & 13.65 $\pm$ 2.56 & 129.64 $\pm$ 28.89 & 0.32 $\pm$ 0.09 & -18.81 $\pm$ 1.51 & 10.10 $\pm$ 0.05 & 79.24 $\pm$ 14.66 & 4.87 $\pm$ 1.54 & 13.21 $\pm$ 0.09 & 7.51 $\pm$ 1.36 & 11.53 $\pm$ 2.09 \\ 
N13b$^*$ & 3.26 $\pm$ 0.27 & -20.63 $\pm$ 1.65 & 12.80 $\pm$ 0.28 & 17.06 $\pm$ 3.22 & 9.16 $\pm$ 1.73 & 80.85 $\pm$ 17.84 & 0.25 $\pm$ 0.07 & -20.45 $\pm$ 2.46 & 18.18 $\pm$ 1.94 & 4.25 $\pm$ 0.88 & 10.81 $\pm$ 2.10 & 19.56 $\pm$ 1.44 & 3.80 $\pm$ 0.84 & 5.83 $\pm$ 1.29 \\ 
N15 & 2.85 $\pm$ 0.24 & -6.88 $\pm$ 1.58 & 14.88 $\pm$ 1.77 & 0.94 $\pm$ 0.20 & 0.50 $\pm$ 0.11 & 15.52 $\pm$ 3.64 & 0.19 $\pm$ 0.06 & -5.27 $\pm$ 2.31 & 31.11 $\pm$ 1.75 & 4.18 $\pm$ 0.80 & 15.79 $\pm$ 2.03 & 19.66 $\pm$ 1.30 & 0.00 $\pm$ 0.00 & 0.00 $\pm$ 0.00 \\ 
\enddata 
\tablecomments{Properties of the \oiii5007\ang \, emission line for \hii regions in the small slicer, R$\sim$18,000 observing mode. Spectral properties are determined by fitting a single Gaussian model to the emission line. Regions with elevated $\sigma$ in the surrounding gas are identified with a $^*$. The properties of the \oiii4959\ang \, and \hb \, lines are available in the machine-readable version of this table online.}
\end{deluxetable*} 
\end{longrotatetable}

\begin{figure*}
    \centering
    \begin{interactive}{animation}{3dstructrure2.mp4}
    \includegraphics[width=0.98\textwidth]{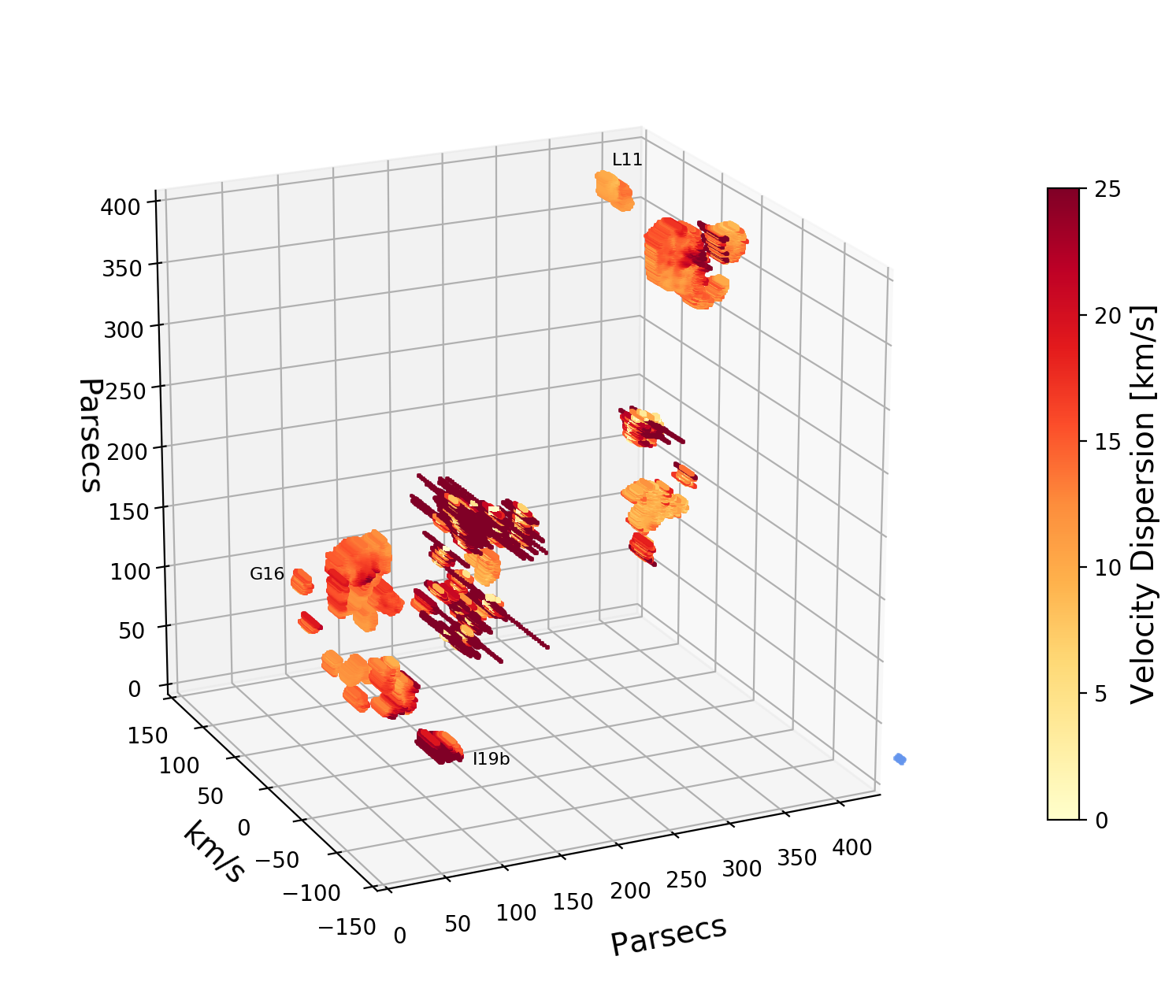}
    \end{interactive}
    \caption{Three dimensional view of the 46 identified \hii regions in the center of IC\,10 based on Gaussian fits to \oiii5007\ang \, at each spaxel within the region contours. For each spaxel belonging to an identified \hii region, the central velocity is set by the velocity shift relative to the systemic velocity of IC\,10 and the depth by the velocity dispersion. The colormap also illustrates the measured velocity dispersion at each spaxel. The blue cylinder at the lower right represents a single spectral (depth) and spatial (width/height) resolution element. This figure is available online as an animation which rotates about the field of view.} \label{fig:3d_map}
\end{figure*}

\subsubsection{Resolved Kinematics}\label{sec:resolved_kinematics}
\begin{figure*}
    \centering
    \gridline{\fig{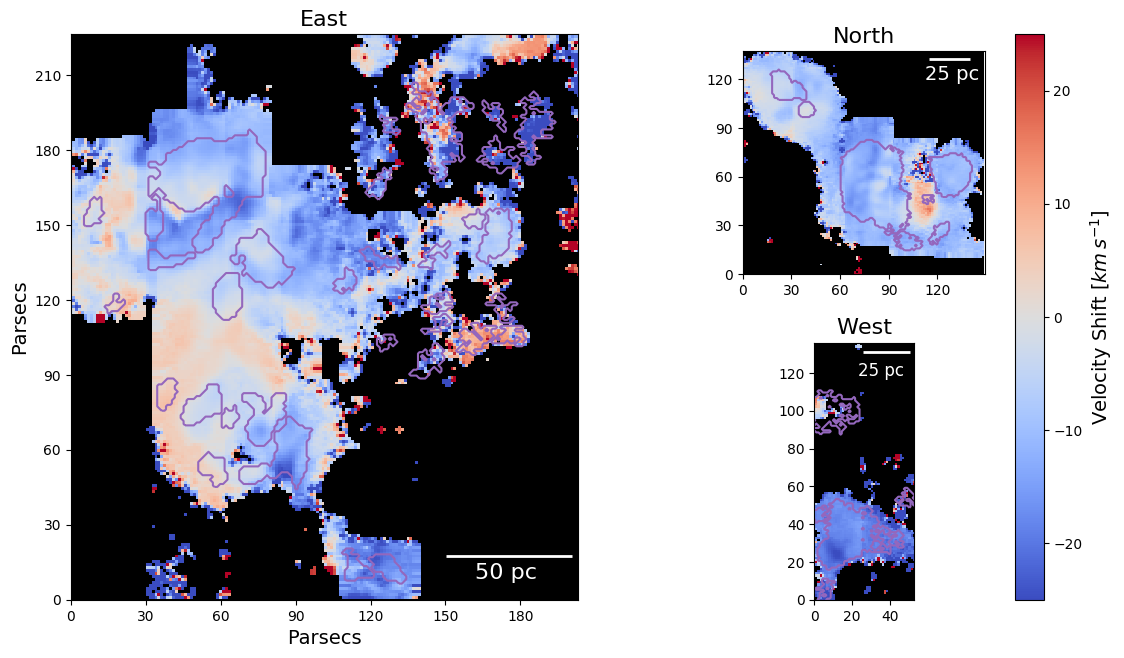}{.95\textwidth}{(a) \oiii5007\ang}}
    \gridline{\fig{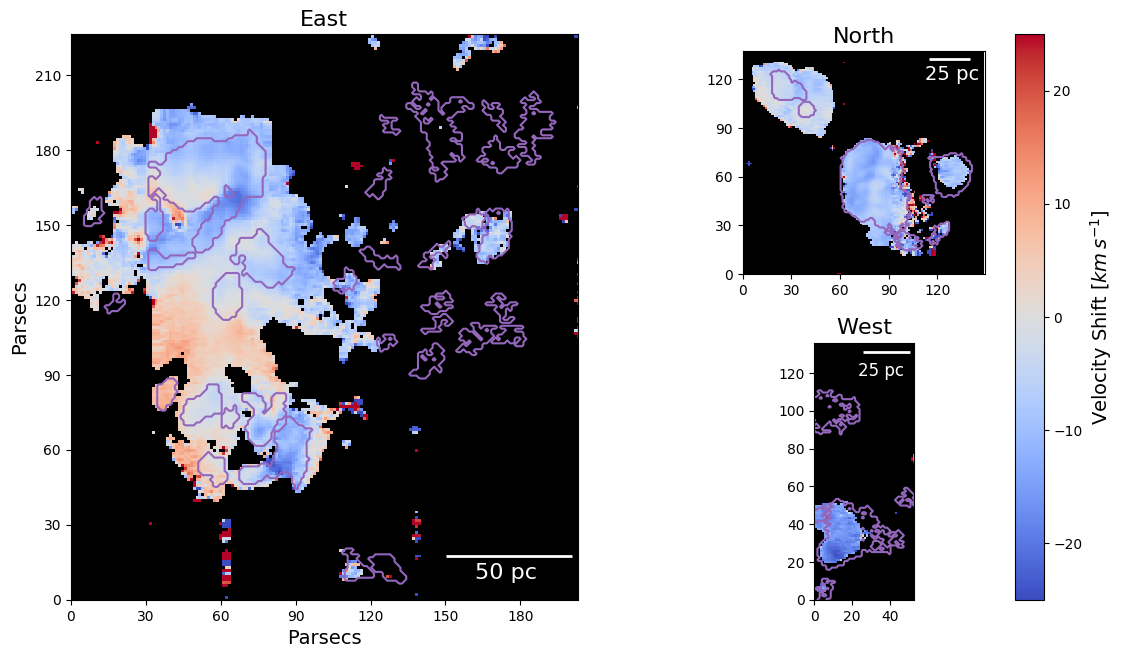}{.95\textwidth}{(b) \hb}}
    \caption{Maps of velocity shift of the ionized gas in IC\,10 measured from Gaussian fits to the \oiii5007\ang \, (a) and \hb \, (b) lines at each spaxel for the high resolution ``small slicer, R$\sim$18,000" observing mode. Each map is divided into the ``North" , ``West", and ``East" regions of the observed field. The purple contours mark the edges of the most compact \hii region structures found by \texttt{astrodendro}. Interestingly, these \hii regions predominantly show gas which is blue-shifted relative to the systemic velocity of IC\,10 while the surrounding diffuse gas more often shows red-shifted components. SNR $>$2 is required for these maps, with more spaxels falling below this cut in \hb, but with the velocity fields matching when measured from either emission line where the SNR is high enough for comparison. \label{fig:vel_maps}}
\end{figure*}

\begin{figure*}[h]
\includegraphics[width=\textwidth]{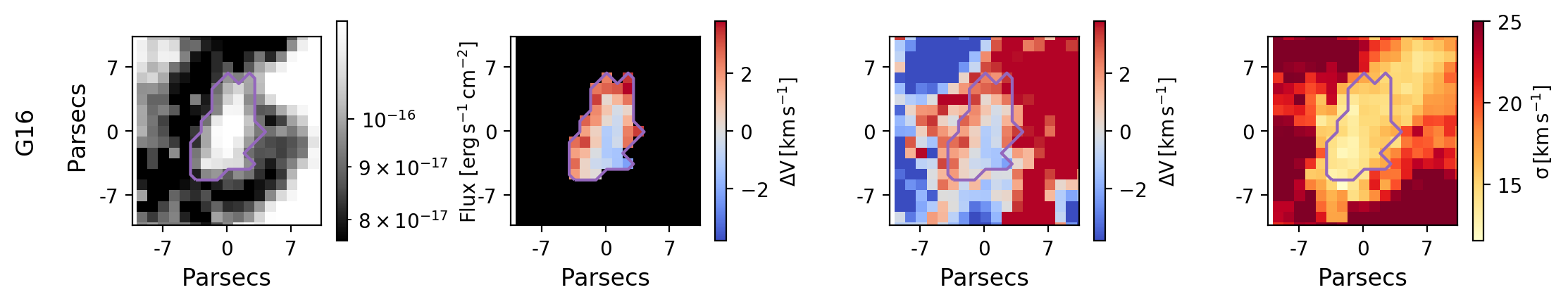}
\caption{Thumbnail map of \hii region G16 identified in our KCWI observations. The complete figure set (46 images) is available in the online journal. (Left): Flux maps of the surrounding area. (Center Left): Velocity shift of spaxels within the \hii region relative to the systemic velocity of the region. (Center Right): Velocity shift of the \hii region and the surrounding gas. (Left): Velocity dispersion within the \hii region and the surrounding gas. Regions of elevated velocity dispersion may be indicative of outflowing gas, particularly when correlated with a velocity shift relative to the surrounding gas.}
\label{fig:BH3_mapthumbs}
\end{figure*}

\begin{figure*}
    \centering
    \gridline{\fig{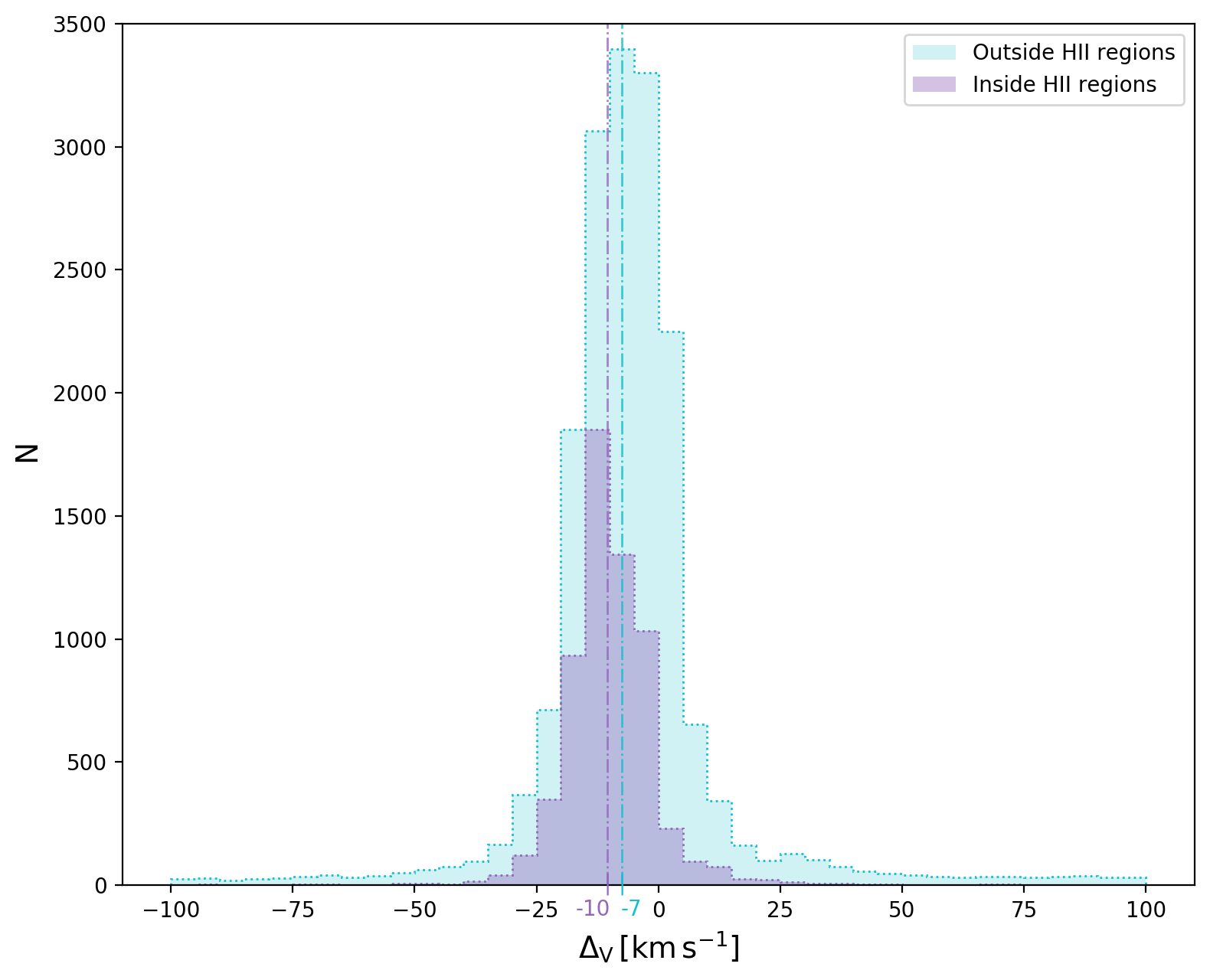}{0.48\textwidth}{(a)}
            \fig{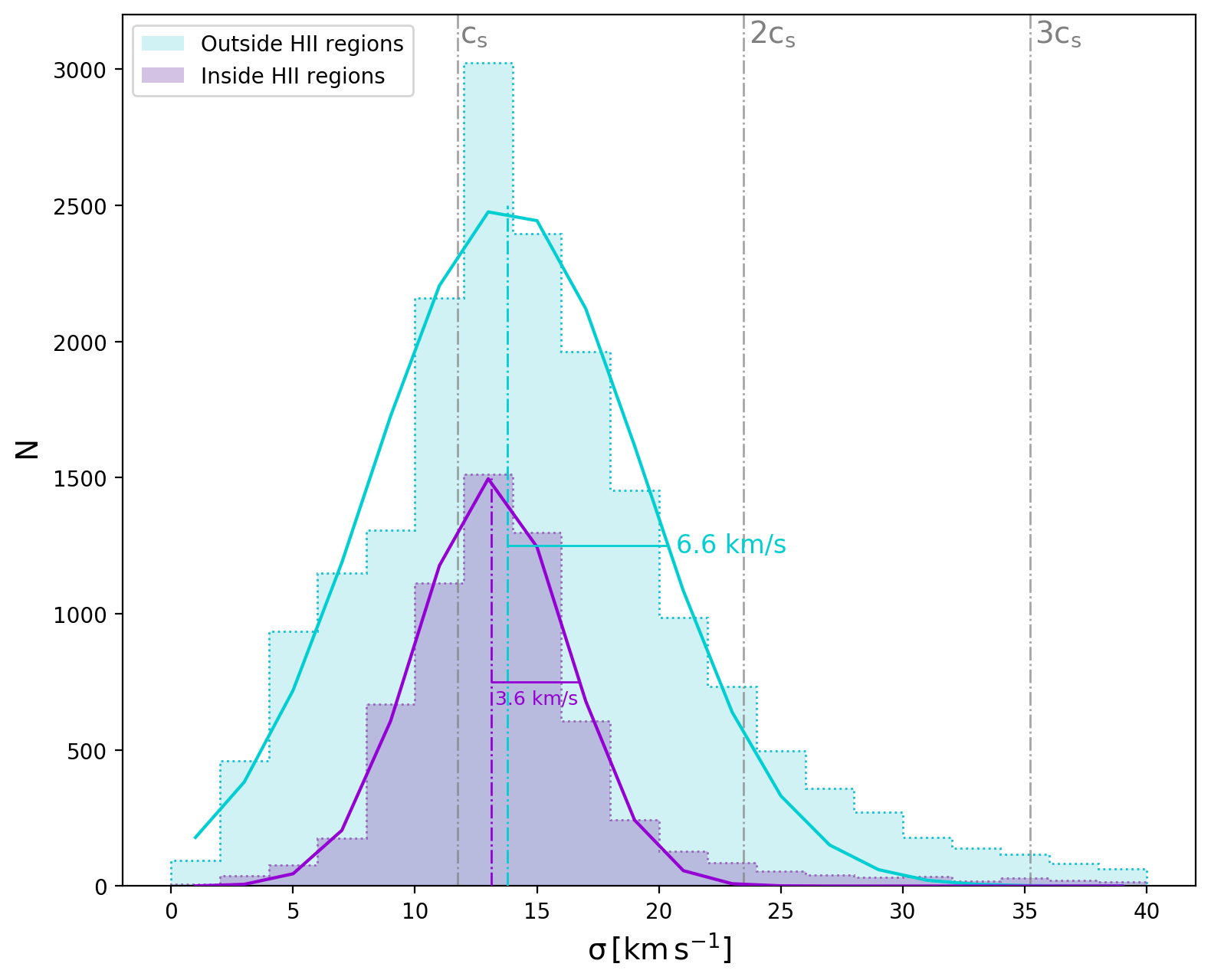}{0.48\textwidth}{(b)}}
    \caption{Distributions of kinematic properties derived from Gaussian fits to the \oiii5007\ang \, line at each spaxel. These distributions are seperated into spaxels which lie outside the \hii region boundaries (purple) and inside (cyan). (a): Distribution of $\Delta V$ determined from the Gaussian mean relative to the systemic velocity of IC\,10 (-348\kms). The median of each distribution is denoted with a dashed line, but it should be noted that the distributions are quite broad at the base, in particular for those spaxels which are not associated with an \hii region. (b): Distribution of $\sigma$, the Gaussian standard deviation after subtraction of instrumental width. These distributions are more regularly distributed about their peaks than their $\Delta V$ counterparts and are therefore fit with Gaussians (solid lines). The mean of the Gaussian fits to the distributions are marked with dashed lines and are quite similar (13\kms \, for spaxels inside \hii regions and 14\kms \, outside), but the width of the distribution outside the \hii regions is twice as large as can be seen in the annotated standard deviation of the fitted distribution.}
    \label{fig:kinematic_histograms}
\end{figure*}

In addition to spectra integrated over IC\,10's \hii regions, we investigate the resolved kinematic properties of the ionized gas at each spaxel. The velocity shift relative to the systemic velocity ($\rm v_{sys}=-348$\kms), $\Delta V$, and dispersion, $\sigma$, are measured from single Gaussian fits to \oiii5007\ang \, and \hb \, emission lines. The properties determined from the \oiii5007\ang \, line in spaxels inside \hii regions are used to illustrate the three dimensional structure of the identified \hii regions in Figure \ref{fig:3d_map}. Additionally, maps of $\Delta V$ and $\sigma$ at each spaxel in the FoV with SNR$>$2 for both the \oiii5007\ang \, and \hb \, fits are shown in Figures \ref{fig:vel_maps} \& \ref{fig:disp_maps}. Thumbnail flux and kinematic maps for each \hii region are included in Appendix \ref{app:map_thumbs} with an example for region G16 in Figure \ref{fig:BH3_mapthumbs}. These maps show interesting kinematic trends between gas inside and outside the \hii regions. The velocity shift shows clear differences between gas residing in the population of \hii regions (blue-shifted) and the diffuse gas not associated with a particular region (red-shifted). This is illustrated in both the map of Figure \ref{fig:vel_maps} and the distributions of Figure \ref{fig:kinematic_histograms}a which shows a much broader distribution of $\Delta V$ in spaxels outside the \hii regions. This is not surprising as less structure would be expected in the kinematics of this gas than that which is associated with a coherent \hii region.

\begin{figure*}
    \centering
    \gridline{\fig{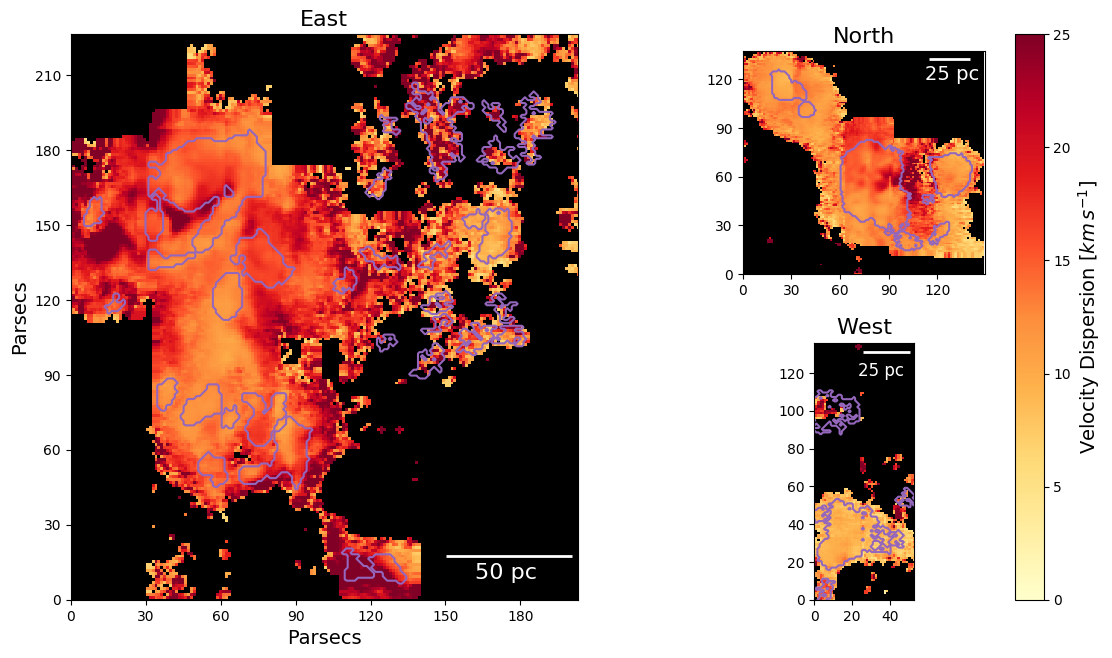}{.95\textwidth}{(a) \oiii5007\ang}}
    \gridline{\fig{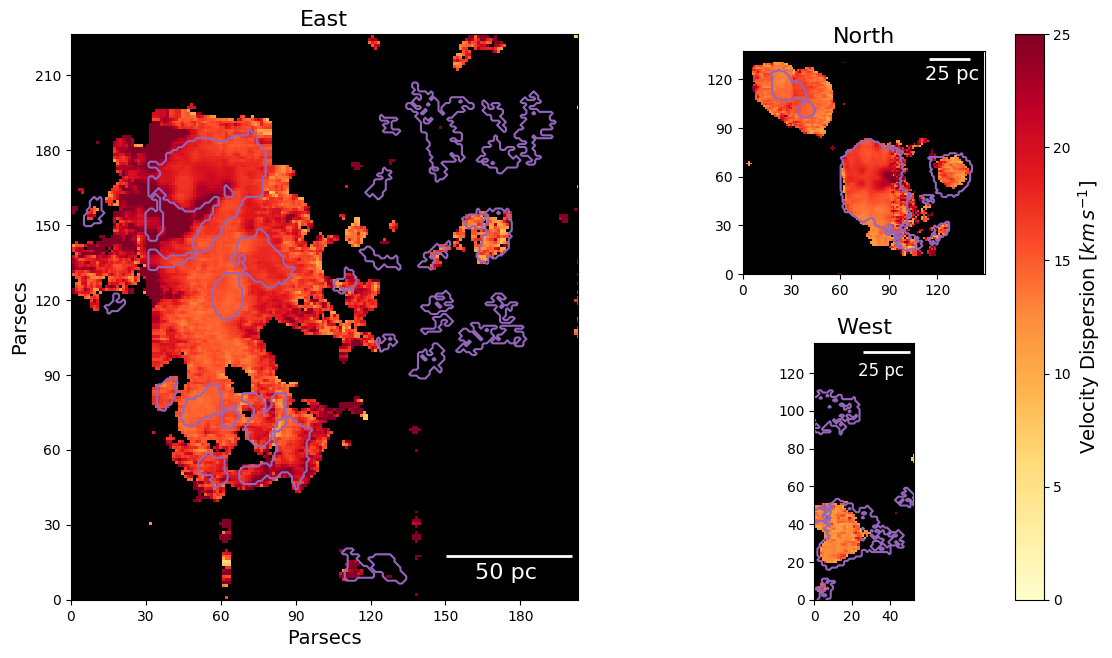}{.95\textwidth}{(b) \hb}}
    \caption{Maps of velocity dispersion of the ionized gas in IC\,10 measured from Gaussian fits to the \oiii5007\ang \, (a) and \hb \, (b) lines at each spaxel for the high resolution ``small slicer, R$\sim$18,000" observing mode. Each map is divided into the ``North" , ``West", and ``East" regions of the observed field. The purple contours mark the edges of the most compact \hii region structures found by \texttt{astrodendro}. SNR $>$2 is required for these maps, with more spaxels falling below this cut in \hb, but with the velocity fields matching when measured from either emission line where the SNR is high enough for comparison. In particular note the elevated dispersion along the outer edges of some \hii regions.} \label{fig:disp_maps}
\end{figure*}

\begin{figure*}[ht!]
    \centering
    \gridline{\fig{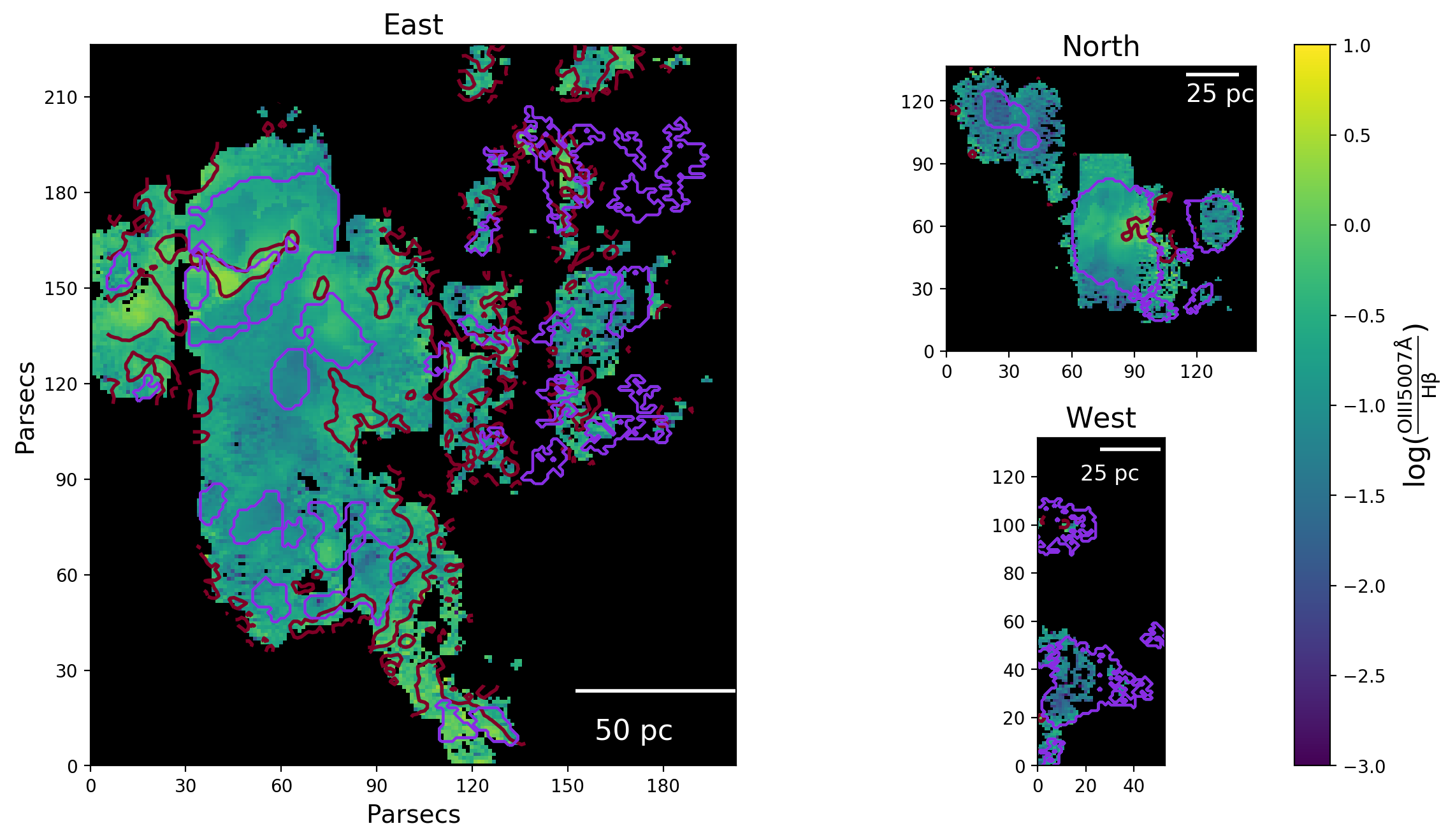}{0.6\textwidth}{(a)}
        \fig{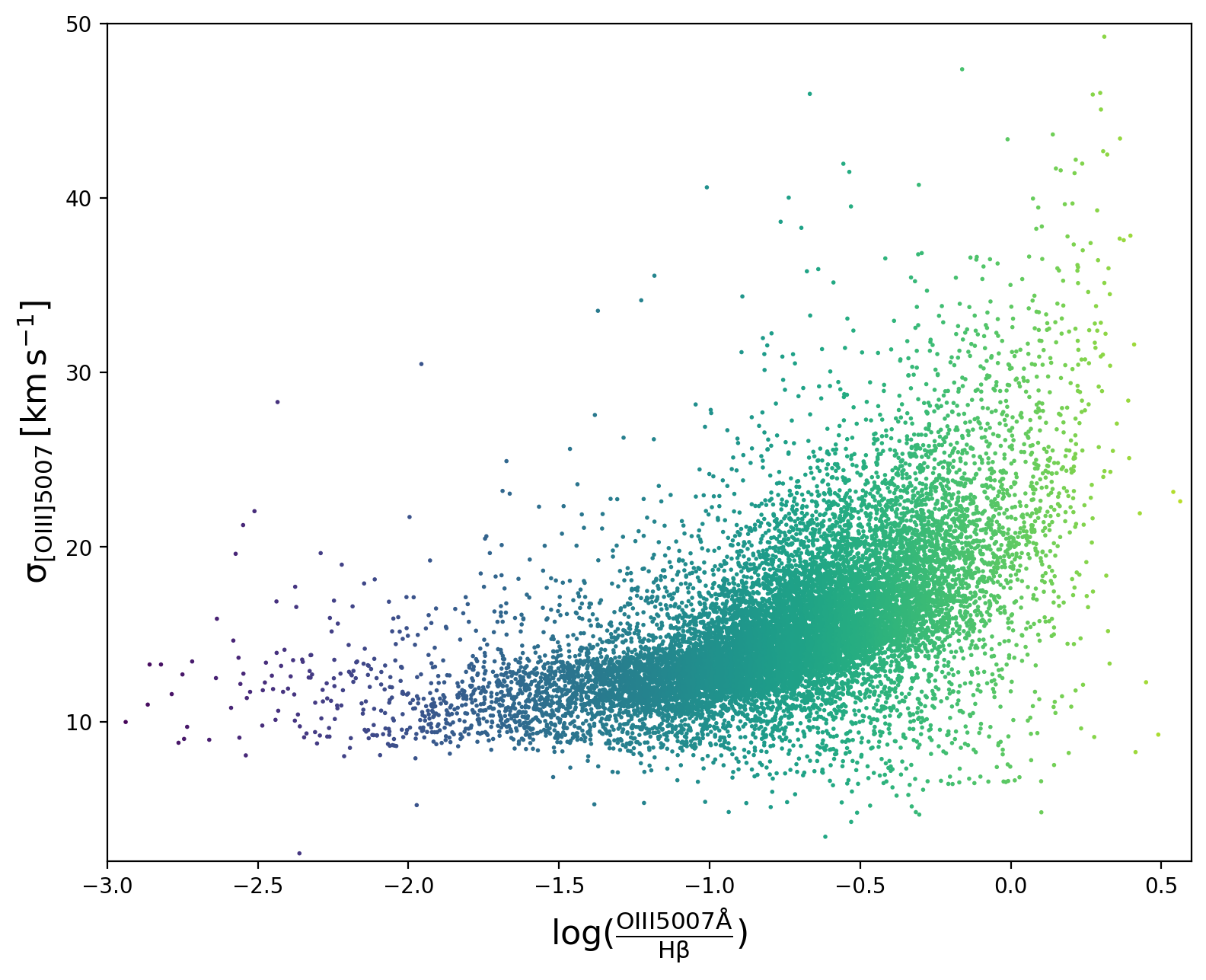}{0.38\textwidth}{(b)}}
    \caption{(a): Map of the \oiii5007\ang/\hb \, ratio for spaxels with SNR$>$2 at the weaker \hb \, line. As in Figure \ref{fig:flux_maps}, the purple contours show the edges of the identified \hii regions. The dark red contours show areas with velocity dispersion $>20$ \kms \, as measured from the \oiii5007\ang \, line. (b): The \oiii5007\ang/\hb \, ratio plotted against velocity dispersion for each spaxel with the same requirement that $\rm SNR_{H\beta}>2$. There is a significant amount of scatter here, but the general trend is for increased velocity dispersion with higher \oiii5007\ang/\hb, which largely coincides with the edges of the identified \hii regions in (a).}
    \label{fig:O3_Hb_ratio}
\end{figure*}

Furthermore, it can be seen in the maps of Figure \ref{fig:BH3_mapthumbs} and Appendix \ref{app:map_thumbs} that the velocity dispersion is highest at many of the \hii region borders, particularly those that reside in larger complexes. In fact, 37\% of the identified \hii regions, identified with a $^*$ in Table \ref{tab:oiii_BH3}, show elevated velocity dispersions at one or more of their edges which could be indicative of outflowing gas. The distributions of measured $\sigma$ at each spaxel inside and outside of the \hii region boundaries are shown in Figure \ref{fig:kinematic_histograms}b for further comparison. The mean $\sigma$ for spaxels inside and outside \hii regions is quite similar (13 and 14 \kms, respectively), but the width of the distribution is twice as large for spaxels outside the region boundaries. We investigate the possibility of shocked gas at the \hii region boundaries by evaluating the $\rm \oiii5007\AA/H\beta$ ratio. An area with an elevated line ratio may indicate the presence of shocked gas. To ensure this ratio is evaluated over the same physical gas column, we define a fixed velocity shift and width for \oiii \, and \hb \, in each spaxel determined from the Gaussian fit to the lower SNR \hb \, line. Figure \ref{fig:O3_Hb_ratio} shows the map of this line ratio along with contours of elevated velocity dispersion, and as can be seen, the spaxels which show the highest $\rm log \left(\oiii5007\AA/H\beta\right )$ also correspond to the areas of elevated velocity dispersion. This result indicates that the elevated velocity dispersion observed at the \hii region borders is due to shocked gas.

\subsubsection{Region Rotation}\label{sec:rotation}
In order to classify the kinematic structure of the individual \hii regions we generate $\Delta V$ maps centered on each region adjusted so the systemic shift of the \hii region is 0 \kms (based on the mean of the Gaussian fit to the integrated region spectrum). For each region we designate whether it is rotating both visually and quantitatively. In order for a region to be considered rotating by eye it must have a bimodality of the velocity shift relative to the region's systemic velocity shift; meaning it must have one region of negative shift and one of positive shift. If for example, there is positively shifted gas surrounded by negatively shifted gas  or vice-versa we do not consider that to be ordered rotation. With this method we find 35\% of regions to be rotating, which is likely a conservative estimate due to the strict visual criteria.

To assign a quantitative rotation criteria we fit the velocity gradient across the region at different position angles. For each position angle we generate a ``fit quality factor" based on the reduced $\chi^2$ combined with the steepness of the slope. The maximum value of the fit quality factor is taken to be the most probable rotation direction and magnitude for that region. If the slope is greater than a threshold value of 0.57 \kms $\rm pc^{-1}$ then the region is considered rotating. This threshold was determined by assuming an electron density, $\rm n_e=100\, cm^{-3}$ and determining the mass contained within a region of radius 1pc. Assuming the gas is virialized we then determined the expected value of velocity dispersion and range of velocity shifts\footnote{For an \hii region with the average radius of this sample,$\rm r=4 \, pc$ at $\rm n_e=100 \, cm^{-3}$, the resulting mass $\rm M_{\hii} ~500 M_{\odot}$. This is larger than the average mass estimated in the previous section for IC\,10's \hii regions, but that simply results in a conservative threshold for rotation.}. With this criteria we find 65\% of \hii regions to be rotating. More regions are classified as rotating using this method than the visual classification which in part is due to restricting our visual classification to regions which only have one transition between positive and negative velocity shifts rather than based only on the overall gradient. This restriction is then susceptible to bias from velocity shifts in a small number of pixels that may be outliers for the \hii region, classifying regions as ``not rotating" when they do in fact have underlying rotation. The quantitative method is also free from the inherent bias with all visual classification and is more easily extended to other samples and studies. Therefore we proceed with the rotation classification of the quantitative method in further analysis.

\subsubsection{Virialization}\label{sec:virialization}
In addition to $\rm M_{HII}$, we can also estimate the \hii region masses based on the measured kinematics of the ionized gas by calculating the virial, $\rm M_{vir}$, and enclosed, $\rm M_{encl}$, masses. However, these rely on the assumption that the motion of the ionized gas is dominated by the self-gravity of the region.

For $\rm M_{vir}$ the region is assumed to be bound and we apply the virial theorem:
\begin{equation}
    M_{\rm vir} = \frac{5\sigma^2r}{G}
\end{equation}
where the factor of 5 is a geometric factor representing the shape of the potential well for a spherical region. This results in a median $\rm M_{vir}= 7.6\times10^5 \, M_{\odot}$, orders of magnitude greater than $\rm M_{HII}$.

As a second method of estimating the \hii region masses kinematically we calculate the enclosed mass, $\rm M_{encl}$, for regions that were determined to be rotating.
\begin{equation}
    M_{\rm encl} = \frac{v_{\rm c}^2r}{G}
\end{equation}
where $v_{\rm c}$ is the circular velocity determined from the measured velocity shift, $v$, at a distance, r, from the \hii region center corrected for the inclination, $i$:
\begin{equation}
    v_{\rm c} = v \, sin(i)
\end{equation}
Assuming a circular region and no preferred inclination relative to the direction of rotation, we take an average of $sin(i)$ between 0 and $\pi/2$, resulting in a factor of $\frac{\pi}{4}$. The median $\rm M_{encl}$ for rotating regions is $\rm \sim1.0\times10^4 \, M_{\odot}$, more than an order of magnitude less than $\rm M_{vir}$, but still significantly greater than $\rm M_{HII}$.

\begin{figure}
    \centering
    \includegraphics[width=0.48\textwidth]{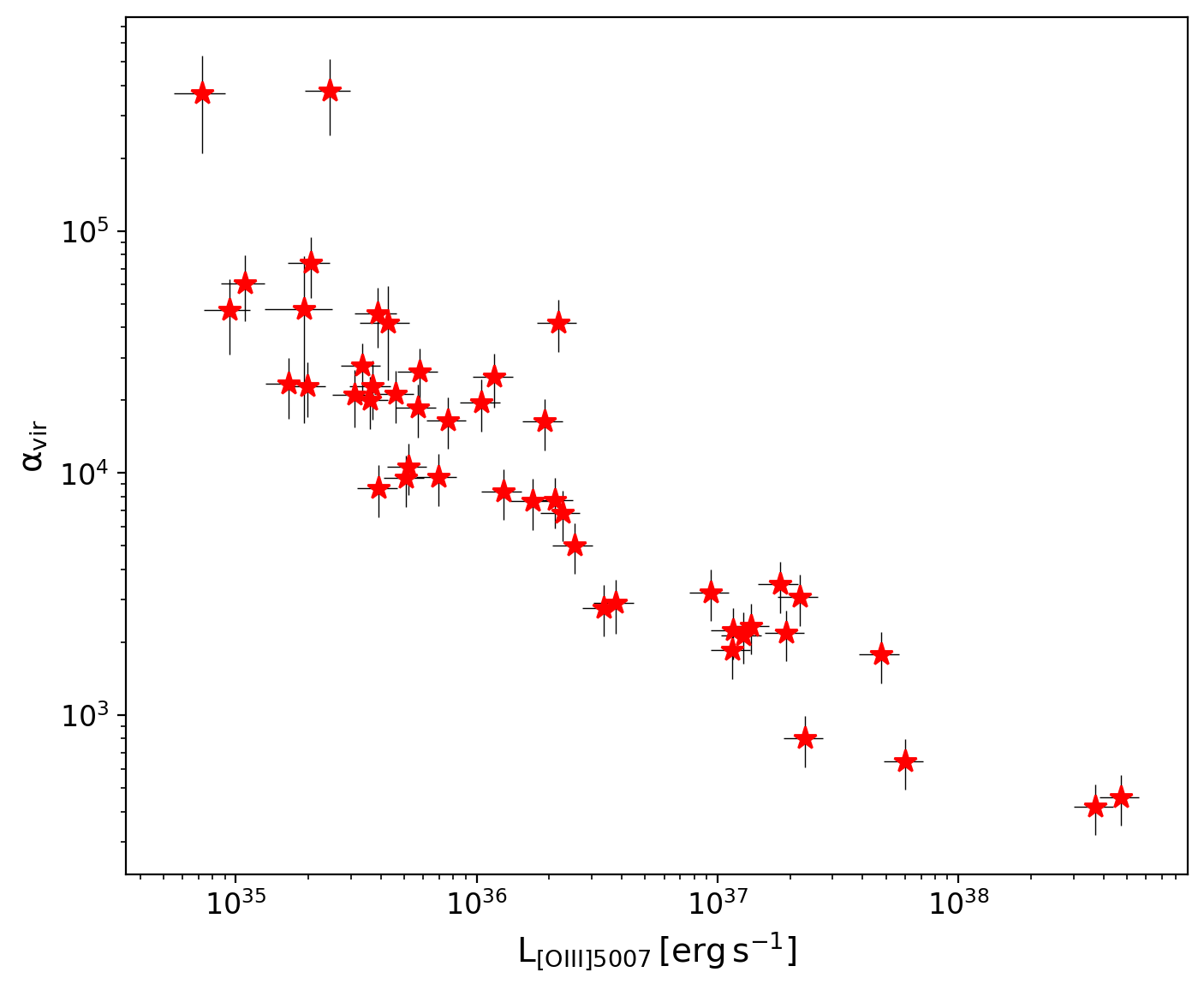}
    \caption{The virial parameter, $\rm \alpha_{vir}=\frac{5 \sigma^2 r}{G M_{gas}}$, plotted as a function of the integrated \hii region luminosity. For all regions $\rm \alpha_{vir} >>1$, indicating regions which are not virialized.}
    \label{fig:virial_param}
\end{figure}

The more than 3 orders of magnitude discrepancy between $\rm M_{HII}$ or $\rm M_{encl}$ and $\rm M_{vir}$ implies that, on average, the \hii regions are not in fact virialized and the use of the velocity dispersion is overestimating the gravitational potential. Furthermore, the measured rotational velocity is also overestimating the gravitational potential, indicating that rotation is not the dominant cause of the velocity gradient observed accross the \hii regions. To explore this further we first investigate the virial parameter, $\rm \alpha_{vir}=\frac{5 \sigma^2 r}{G M_{gas}}$, of the \hii regions. We determine $\rm \alpha_{vir}$ using the \hii region radius following the $r_{1/2}^*$ definition and the velocity dispersion, $\sigma$ measured from the Gaussian fit to the integrated spectrum. For $\rm M_{gas}$, we use $\rm M_{\hii}$, assuming that the gas in the vicinity of the identified \hii region is fully ionized. The resulting values for $\rm \alpha_{vir}$ are shown plotted as a function of $\rm L_{\oiii5007}$ in Figure \ref{fig:virial_param} with the resulting $\rm \alpha_{vir} >> 1$, falling in the regime of \hii regions which are not virialized.

\begin{figure}
    \centering
    \includegraphics[width=.48\textwidth]{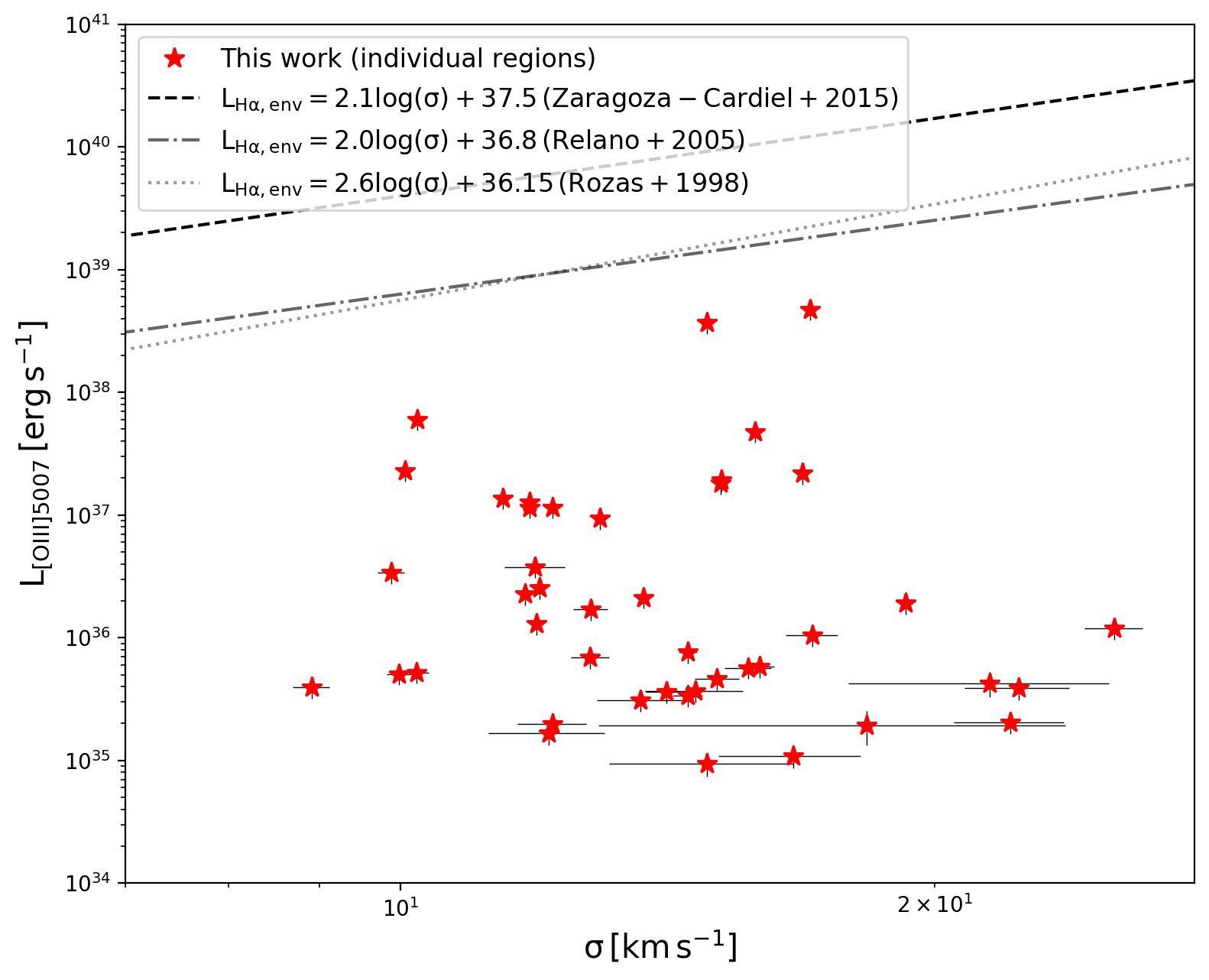}
    \caption{Measured $\rm L_{\oiii5007}$ plotted against $\sigma$ from Gaussian fits to integrated \hii region spectra. The lines show the fits to the upper envelope of this $\rm L-\sigma$ relationship for other samples of \hii regions. Regions that fall near the envelope are thought to be virialized while those below the curve are under-luminous for their velocity dispersion. This is the regime where the IC\,10 regions fall, further supporting the interpretation that they are not virialized and show significant non-gravitational motion.}
    \label{fig:L-sigma}
\end{figure}

Studies of \hii regions have often investigated the relationship between the luminosity and velocity dispersion, the $\rm L-\sigma$ relation, to study the the region dynamics. Rather than finding a correlation between these two properties for all \hii regions, studies typically fit the upper envelope of the relationship defining the area of the correlation where \hii regions in virial equilibrium would lie \citep[e.g.,][]{Arsenault1990}. We compare the identified \hii regions in IC\,10 to the relationships measured for the upper $\rm L-\sigma$ envelope in three such studies \citep{Zaragoza-Cardiel2015, Relano2005, Rozas1998} as another test of virialization. As can be seen from this comparison in Figure \ref{fig:L-sigma}, IC\,10's \hii regions fall below the envelope fits further supporting the conclusion that the \hii regions are not virialized and the dynamics are dominated by sources of energy besides gravity.

The free-fall time estimate is a useful quantity for regions which are forming stars under simple gravitational collapse. We have shown that the \hii regions of IC\,10 are not virialized, however, and thus the free-fall time may not be the best characterization in this case. The crossing time, based on the measured velocity dispersion may provide a more useful characterization of the timescale relevant for an expanding \hii region. The crossing time is defined as:
\begin{equation}
    \tau_{\rm cr} = \frac{r}{\sigma} \label{eqn:t_cr}
\end{equation}
The average $\rm \tau_{cr} \sim 2\times10^5 \, yrs$ for IC\,10's \hii regions.

\subsubsection{Energetics}\label{sec:feedback}
\begin{figure}[h]
\centering
\gridline{\fig{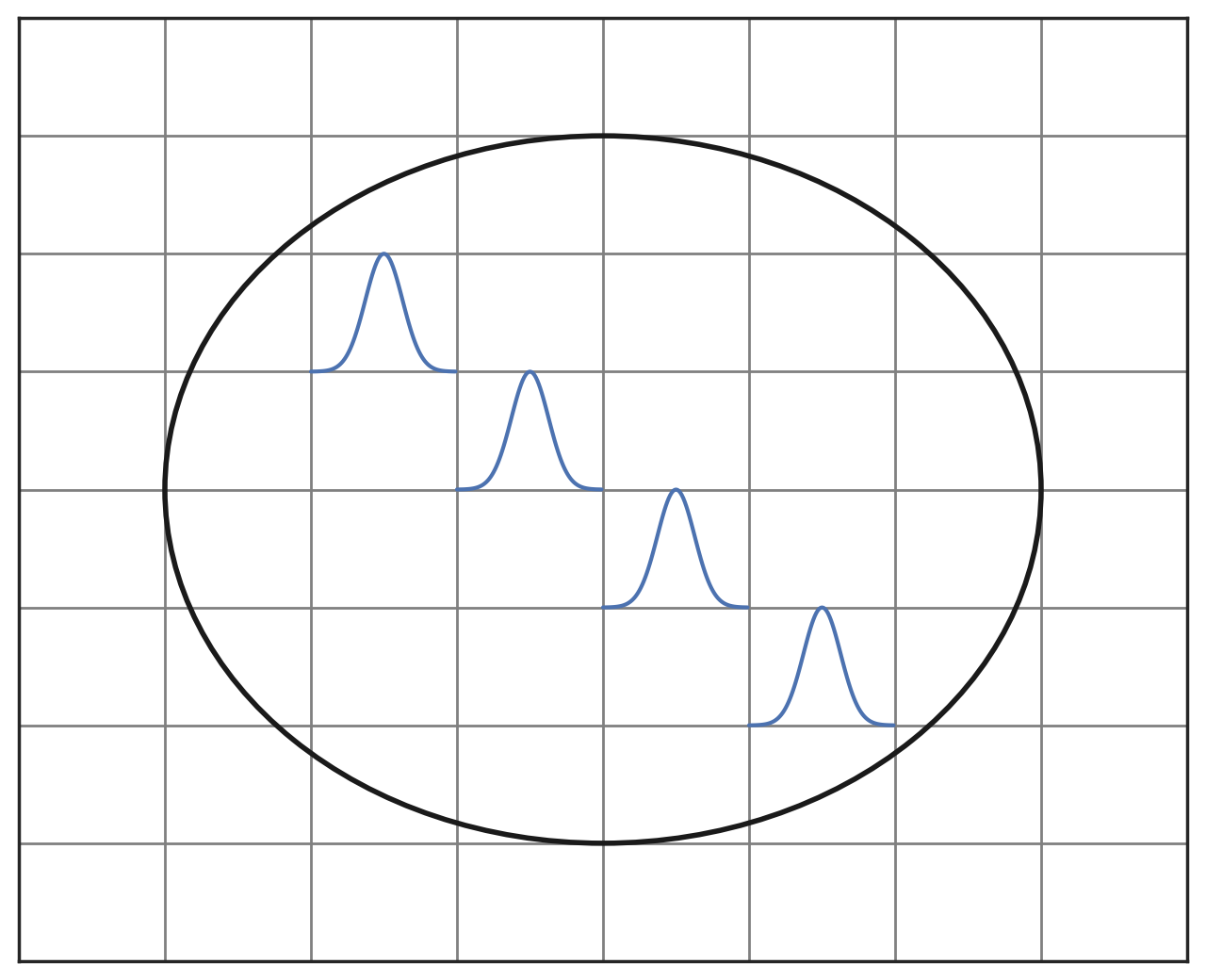}{0.45\textwidth}{(a)}}
\gridline{\fig{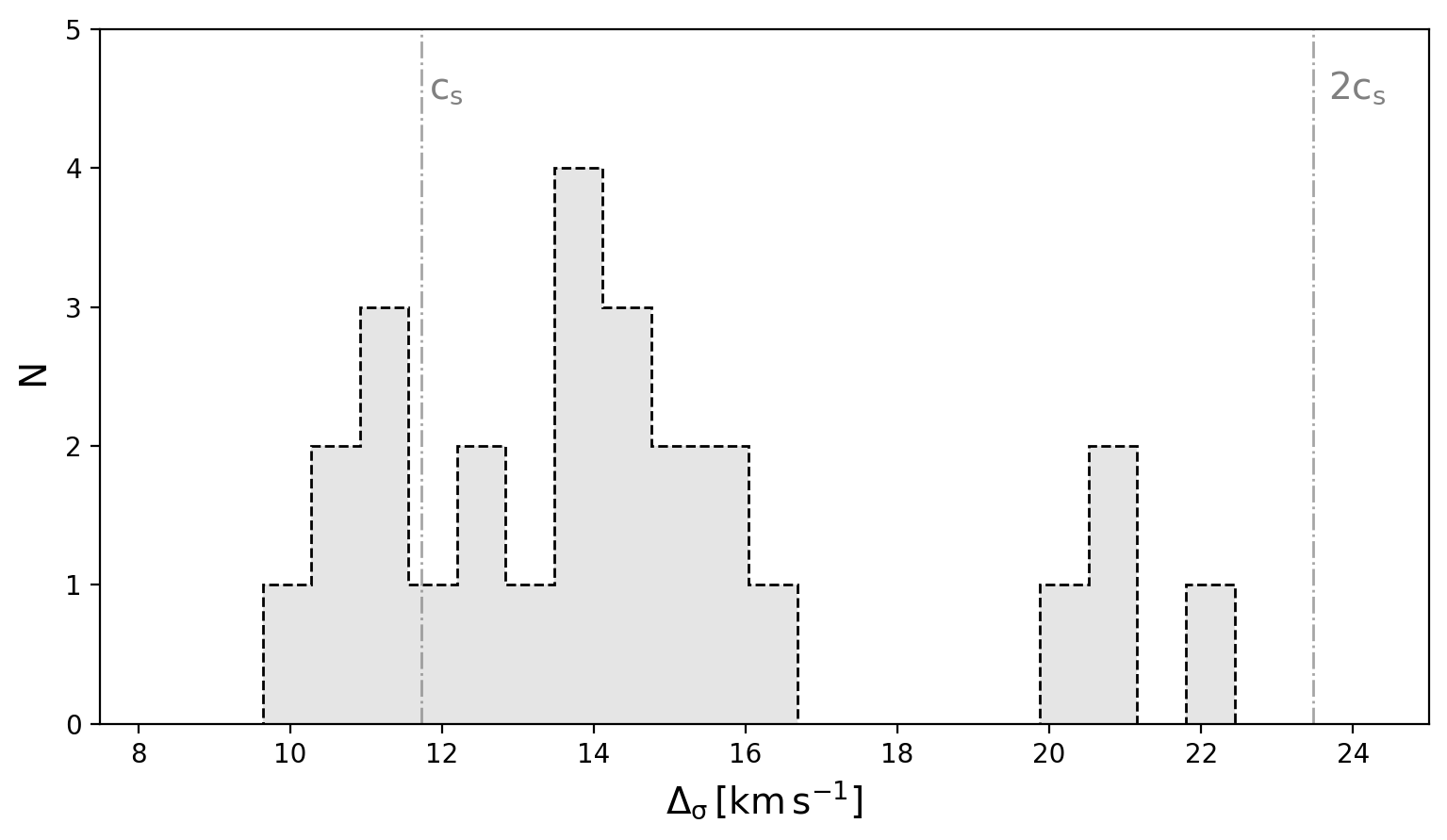}{0.49\textwidth}{(b)}}
\caption{(a): Illustration of the Gaussian components summed in the simple model of a rotating \hii region. Components are added at each spaxel along the axis of rotation with the flux of the final profile normalized to the measured spectrum. The width of each spaxel Gaussian is set by the instrumental width with the centers set by the measured rotational velocity. (b): Histogram of the non-rotational component of velocity dispersion measured in each of the rotating \hii regions where $\rm \Delta \sigma = \sqrt{\sigma_{meas}^2 - \sigma_{mod}^2}$, the difference between the measured and model velocity dispersion. The vertical dashed line shows the estimated sound speed in a typical \hii region, $\rm c_s$, treating the region as an ideal gas with temperature $\rm T=10^4 K$. \label{fig:model_spec}}
\end{figure}

As outlined in the previous section, the \hii regions identified in IC\,10 are not virialized, and the velocity dispersions are therefore not a good estimate of the gravitational potential. To quantify the amount of dispersion due to sources other than gravitational motion we generate model spectra with only rotational motion included for those regions classified as rotating. For this simple model we sum individual Gaussian profiles at each pixel along the direction of rotation with the center set by the measured $\Delta V$, and the width set only by the instrumental width, $\sim$7.5\kms. The peak flux of each component of the sum is based on a simple Gaussian flux profile for the \hii region. After each model pixel is summed the total flux is normalized by the measured \oiii5007\ang \, flux of the region for accurate comparison. These model spectra are then fit with a single Gaussian profile and the velocity dispersion compared to the measured value for the \hii region. The non-rotational motion in the measured profile is defined as $\rm \Delta \sigma = \sqrt{\sigma_{meas}^2 - \sigma_{mod}^2}$, the difference between measured and modeled. The distribution of $\rm \Delta \sigma$ is shown in Figure \ref{fig:model_spec} along with a cartoon illustrating the Gaussian components of the model spectrum. The average value for $\rm \Delta \sigma$ is $\sim$14\kms, similar to the sound speed expected in a typical \hii region.

We checked the accuracy of this simple rotating \hii region model using the software SHAPE \citep{shape}. SHAPE allows the user to generate a model with potentially complex geometry and kinematic structure. We use a spherical geometry with only rotational motion at the spatial and spectral resolution for each observation. We model this for 3 of our identified rotating \hii regions (H18d, H18e, G16) and compare the velocity dispersion from the SHAPE model and simplified Gaussian sums. For region G16, we find that the SHAPE model produces a velocity dispersion 0.3 \kms \, greater than our model, in the other two the difference is less than 0.03 \kms. With measured spectra producing velocity dispersions $\sim$14\kms \, greater than either model, this discrepancy is negligible. We therefore use the simple model summing Gaussian components for the full sample of \hii regions due to the ease of extending this to a larger sample.

The excess in velocity dispersion that is not attributable to rotational motion can not be explained by virialization and therefore may lead to expansion of the \hii regions. We estimate the amount of inward, $\rm P_{in}$, and outward pressure, $\rm P_{out}$, in the \hii region, with an imbalance indicating that the region is not in equilibrium with the ISM. As a first order approximation, we estimate the outward sources of pressure to come predominantly from thermal gas pressure, 
\begin{equation}
    P_{\rm gas}=2nkT \label{eqg:Pgas},
\end{equation}
internal turbulence in the region, 
\begin{equation}
    P_{\rm turb,int}=\frac{1}{2}\rho \sigma_{\rm t}^2 \label{eqn:Pturbint},
\end{equation}
and direct radiation pressure, 
\begin{equation}
    P_{\rm dir} = \frac{Q h \nu}{4 \pi r^2 c}. \label{eqn:Pdir1}
\end{equation}
where $Q$ is the ionizing photon production rate which we can estimate from the integrated \oiii \, luminosity of the \hii region:
\begin{equation}
    Q = \frac{L_{\rm \oiii} \lambda}{h c} = \frac{L_{\rm \oiii}}{h \nu}. \label{eqn:Q}
\end{equation}
Combining Equations \ref{eqn:Pdir1} \& \ref{eqn:Q} results in the formula for direct radiation pressure based on the measured luminosity:
\begin{equation}
    P_{\rm dir} = \frac{L_{\rm \oiii}}{4\pi r^2c} \label{eqn:Pdir}.
\end{equation}
We use the definition of $r_{1/2}^*$ for the radius, $r$, $k$ is the Boltzmann constant, and $c$ is the speed of light. We are unable to directly measure the gas temperature, $T$, and number density, $n$, from our spectra so we use a constant temperature of $\rm T=10^4 \, K$ and estimate $n$ for each region from the \stromgren sphere approximation resulting in a median $\rm n \sim 20 \, cm^{-3}$. This is significantly lower than the $\rm \sim10^2\, cm^{-3}$ density determined by \citet{Polles2019} from models of infrared cooling lines in five of the brightest \hii region complexes which are also included in this study, suggesting that we may be underestimating the actual density in the assumption of a \stromgren sphere representation. To maintain internal consistency with other measured and estimated quantities we will proceed with density determined from our KCWI spectra with the caveat that this may result in an underestimate of $\rm P_{gas}$ and $\rm P_{turb}$. This estimated value of $\rm n$ is combined with the mass of the hydrogen atom to determine the value of $\rm \rho$ used in the calculation of $\rm P_{turb,int}$. The turbulent linewidth, $\rm \sigma_t$ is evaluated by removing the thermal sound speed from the measured velocity dispersion, $\rm \sigma_t = \sqrt{\sigma^2-c_s^2}$, with $\rm c_s$ is defined as:
\begin{equation}
    c_{\rm s} = \sqrt{\frac{\gamma k T}{m_{\rm h}}}
\end{equation}
where $\rm \gamma = 5/3$ for an ideal gas.

The average values estimated for these pressure components are $\rm P_{dir}\sim4\times10^{-14} \, dyne \, cm^{-2}$, $\rm P_{turb,int}\sim1\times10^{-11} \, dyne \, cm^{-2}$, and $\rm P_{gas}\sim5\times10^{-11} \, dyne \, cm^{-2}$, making $P_{\rm gas}$ and $P_{\rm turb,int}$ the dominant factors in the outward pressure. One caveat is that the form of direct radiation pressure used here based on Q is specifically at the ionization front \citep{McLeod2019}. Another often used method to use the bolometric luminosity of all the stars in the region, which is estimated as $L_{\rm bol}\approx138L_{\rm H\alpha}$ (or $L_{\rm bol}\approx138L_{\rm \oiii}$) \citep{Lopez2014} in place of $L_{\rm \oiii}$ in Equation \ref{eqn:Pdir}. This would increase the impact of $\rm P_{dir}$, while still leaving it an order of magnitude less than $\rm P_{gas}$. However, there is some uncertainty in the correlation of $L_{\rm bol}\approx138L_{\rm H\alpha}$ based on the age and star formation history of a region. This could overestimate the bolometric luminosity for a young stellar population like that of an \hii region \citep{Lopez2014}. Further, there is some disagreement on whether this definition of radiation pressure traces the force that is actually exerted on the gas, as this may be lower than the pressure in an optically thin medium like the interior of an \hii region \citep{Pellegrini2011, Krumholz2014}. Since both definitions produce a radiation pressure here that is sub-dominant compared to $\rm P_{gas}$ we will proceed with the definition based on $L_{\rm \oiii}$, but note that the uncertainty in the definition of $\rm P_{dir}$ may result in a less drastic difference in the sources of pressure.

An additional source of outward pressure that is not included here is the hot gas pressure observed in X-rays. A diffuse X-ray component is observed in this same region of IC\,10 with an average temperature of $\rm \sim4\times10^6 \, K$ \citep{Wang2005}. Determining the temperature and number density of the X-ray emitting gas at the resolution of individual \hii regions needed to include in $\rm P_{out}$ at this scale is beyond the scope of this study, but it would likely provide a smaller contribution than $\rm P_{gas}$ as found in a sample of 32 LMC and SMC \hii regions which show evidence of leakage of this hot gas \citep{Lopez2014}. 

\begin{figure}[h]
    \centering
    \includegraphics[width=0.48\textwidth]{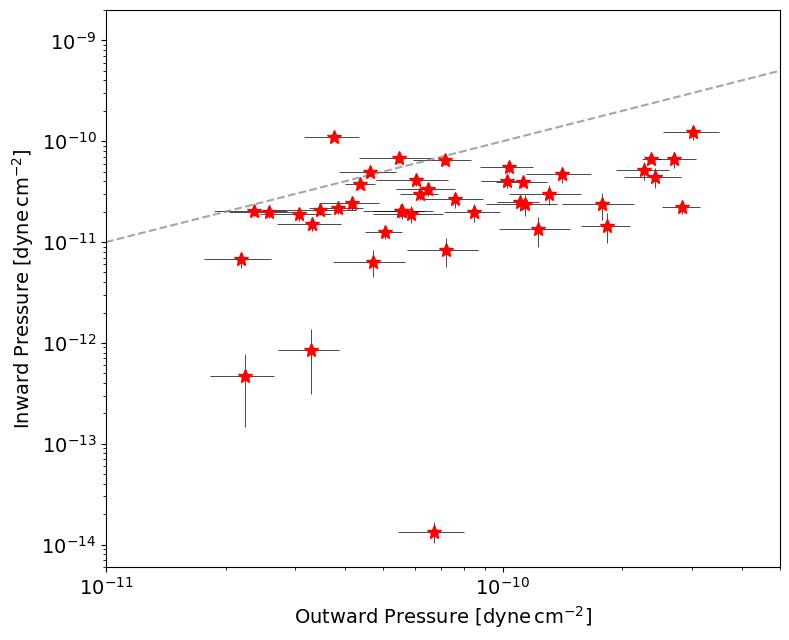}
    \caption{Comparison of inward and outward sources of pressure in the identified \hii regions with the 1:1 line shown in gray. The inward pressure estimate is a sum of $P_{\rm grav}+P_{\rm turb,ext}$, while the outward estimate is a sum of $P_{\rm gas}+P_{\rm dir}+P_{\rm turb,int}$. On average $\rm P_{out} \sim  3 P_{in}$, with 89\% of the \hii regions showing $\rm P_{out} > P_{in}$.}
    \label{fig:pressure_comp}
\end{figure}

For the inward pressure we combine the contributions of pressure due to self-gravity, 
\begin{equation}
    P_{\rm grav}= GM^2/4\pi r^4 \label{eqn:Pgrav}
\end{equation}
where $M=M_{\rm \hii}$, with external turbulent pressure, $\rm P_{turb,ext}$, from the surrounding gas evaluated following Equation \ref{eqn:Pturbint}. We estimate this latter pressure source with $\sigma_{\rm t,ext}$ evaluated in a 3.5 pc (3 pixel) border around each \hii region with the sound speed removed in the same way as the internal turbulent linewidth. The density, $\rm \rho$ used in the external turbulent pressure is the same as its internal counterpart as we do not have a direct method of measuring the gas density. This should still provide a reasonable first order estimate that may even be conservative as \citet{Polles2019} find lower typical densities in modelled zones containing diffuse gas than the bright \hii regions in IC\,10. On average the difference ($P_{\rm turb,ext}$-$P_{\rm turb,int}$), or the resulting turbulent pressure, provides an inward pressure which is $\rm \sim500\times P_{grav}$ due to the compact, low-mass nature of the identified \hii regions. As shown in Figure \ref{fig:pressure_comp}, 89\% of the \hii regions show $\rm P_{out} > P_{in}$ with on average $\rm P_{out} \sim 3 P_{in}$. As these are all approximations it does not necessarily indicate that a given \hii region exhibiting greater $\rm P_{out}$ will be expanding (and vice versa), but rather that it is likely that the majority of the \hii regions in our sample are expanding into the ISM, especially since $\rm P_{out}$ can be considered a lower estimate with the exclusion of the hot gas pressure. This additional component would increase the inbalance towards greater $\rm P_{out}$ and increase the likelihood and/or strength of the \hii region expansion.

\begin{figure*}[ht!]
    \centering
    \includegraphics[width=0.95\textwidth]{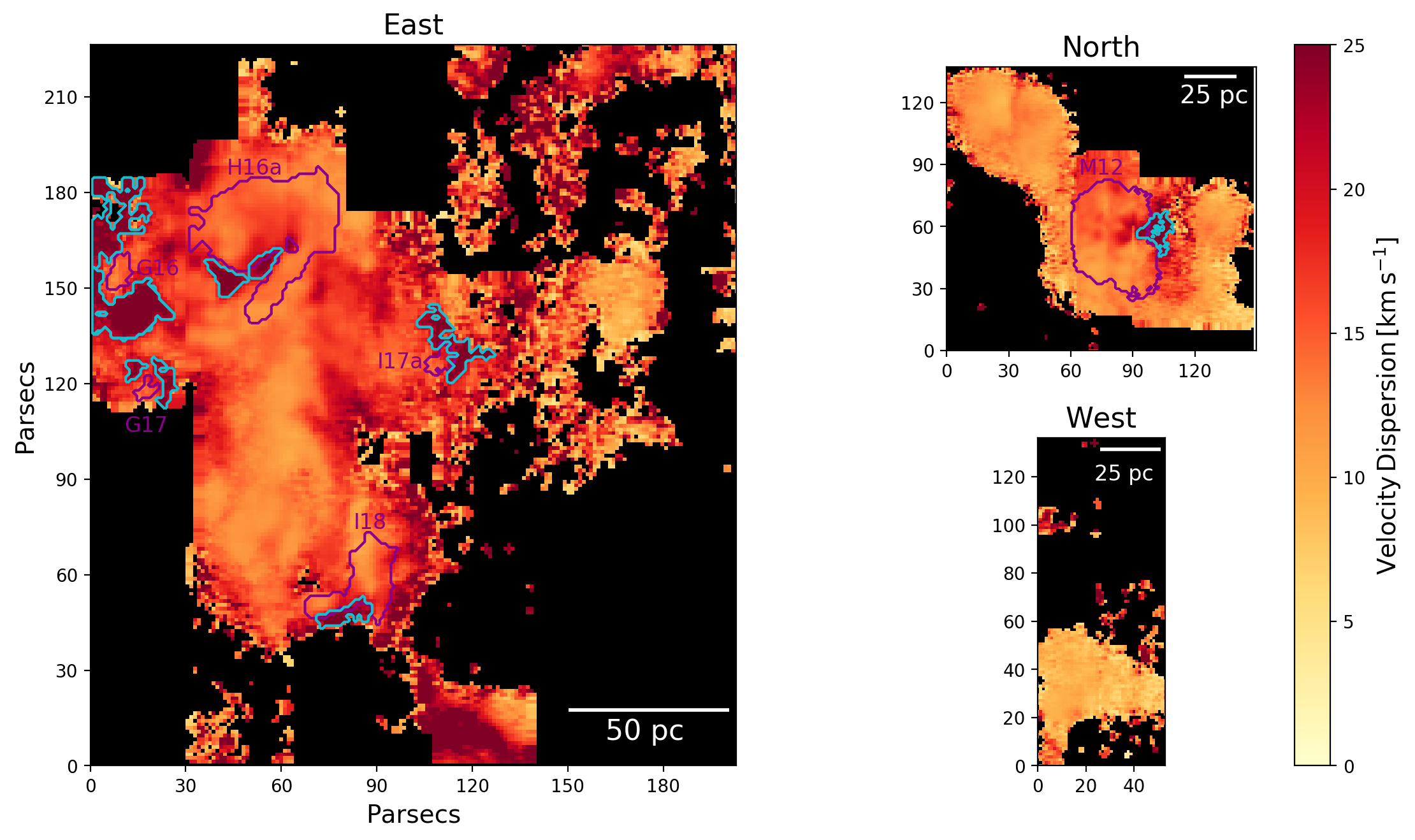}
    \caption{Maps of velocity dispersion of the ionized gas in IC\,10 measured from Gaussian fits to the \oiii5007\ang \, line at each spaxel for the high resolution ``small slicer, R$\sim$18,000" observing mode. Each map is divided into the ``North" portion of the FoV, the middle row shows the ``West" field, and the bottom row shows the ``East" region of the observed field. The purple contours mark the edges of the \hii regions which are bordered by areas of elevated velocity dispersion which may be due to possible outflows, (outlined in cyan). These potential outflows must have a maximum velocity dispersion $\ge \rm 25 km \, s^{-1}$ and a minimum of $\rm 21 km \, s^{-1}$. SNR $>$2 is required for these maps} \label{fig:outflows}
\end{figure*}

Additionally, there are 6 \hii regions where we find significantly elevated velocity dispersions at the region boundaries indicating the presence of outflows. These areas of elevated velocity dispersion are defined and identified using a similar method as identifying the \hii regions described in Section \ref{sec:astrodendro}. We use the \texttt{astrodendro} package along with the velocity dispersion map in Figure \ref{fig:disp_maps} to identify areas with a peak velocity dispersion $\rm >25 km \, s^{-1}$ and a minimum of 21$\rm km \, s^{-1}$. These regions must also be resolved with a diameter greater than the FWHM measured from standard star observations. Of the regions identified with elevated velocity dispersion only those located at the border of an \hii region are considered as potential outflows. These outflows and the host \hii regions are shown in Figure \ref{fig:outflows}.

We interpret the regions with elevated velocity dispersion as turbulent volumes. Turbulence is believed to decay on an eddy turnover time (roughly the crossing time of the turbulent region, $l/\sigma$), where $l$ is the linear size of the turbulent region. Using the average properties measured from these turbulent regions of $l\approx 5.4\,{\rm pc}$ and $\sigma\approx 25.5\,{\rm km\, s^{-1}}$  the turbulence will decay over $2\times10^5\,{\rm yrs}$, much shorter than the lifetimes or ages of the \hii regions. This raises the question of what powers the turbulence. We consider three possibilities: expansion of the \hii regions causing outflows through lower density channels; photoionization heating of neutral gas; or stellar winds mixing with dense gas leading to turbulence.

In the simplest model we would expect all three of these mechanisms to act outward in a spherically symmetric zone around the central star cluster. However, \hii regions do not typically exist in a medium of uniform density \citep[e.g.,][]{Harper-Clark2009, RogersPittard2013}. There are holes and channels for outflowing gas to escape, which would produce isolated areas of increased turbulence rather than covering the entire perimeter, just as we see in these \hii regions. This non-uniform density can clearly be seen from the irregular morphology of these 6 regions as well as the rest of IC\,10's \hii regions.

To first determine whether these regions of elevated velocity dispersion can be maintained by outflows generated from expansion of the observed \hii regions, we compare the kinetic luminosity, $L_{\rm kin}$, inside the star-forming region with the turbulent luminosity, $L_{\rm turb}$ in the potential outflow. 

The kinetic luminosity crossing the boundary of the \hii region is 
\begin{equation}
    L_{\rm kin} \equiv \dot{E}_{\rm kin} = \frac{1}{2}M_{\rm \hii}v_{\rm exp}^2 \frac{v_{\rm exp}}{r_{1/2}^*}\label{eqn:L_kin},
\end{equation}
where $v_{exp}$ is the expansion velocity of the region defined as the half-width at zero intensity (HWZI)\footnote{Note that the HWZI is proportional to the velocity dispersion, $\rm \sigma$, but is a more physically intuitive way of denoting the expansion velocity as it captures the full range of velocities contributing to the Gaussian line profile.} of the emission line. 

The turbulent luminosity in each elevated dispersion region is determined similarly from the rate of change of the kinetic energy attributable to turbulence, which we will refer to as $\rm \dot{E}_{turb}$ to differentiate it from the internal \hii region kinetic energy above. This is defined as follows:
\begin{equation}
    L_{\rm turb} \equiv \dot{E}_{\rm turb} = \frac{1}{2}M_{\rm t}\sigma^2\frac{\sigma}{l} \label{eqn:E_turb}
\end{equation}
where $\rm M_{t}$ and $\rm \sigma$ are the ionized gas mass and velocity dispersion in the potential turbulent region, measured in the same way as their counterparts inside the \hii regions and $l$ is the average of the turbulent region major and minor axes. $\rm \sigma$ is measured from the Gaussian fit to \oiii5007\ang\, and $\rm M_t$ is calculated from the density determined via the \stromgren sphere approximation. This relies on the assumption that most of the gas in the areas of elevated velocity dispersion is ionized, which may not be reliable outside of an \hii region and particularly for the more extended areas of elevated dispersion. We may then be underestimating $\rm M_t$ and thus $\rm L_{turb}$.

The second possible scenario for these turbulent regions at the edges of the identified \hii regions are ``champagne" or ``blister" flows \citep[e.g.,][]{Israel1978, Tenorio-Tagle1979} in which radiation from the central star cluster heats and ionizes neutral gas. This causes an area of overpressure, resulting in rapid expansion of the gas and could explain the observed elevated velocity dispersions. We estimate the amount of energy available for this mechanism from the type and number of ionizing stars in each \hii region determined in Section \ref{sec:SFR_limit}. We take the effective temperature, $T_{\rm eff}$ for the stellar type from the models of \citet{Martins2005} for O stars and \citet{Smith2002} for B stars, combined with Wien's Law to estimate the wavelength at the peak of the black-body curve. We assume this represents the average energy of the emitted photons. We take the amount of energy above 13.6 eV as the energy imparted to the electron after ionizing a hydrogen atom. This is combined with the ionizing photon production rate for the determined stellar type and the number of stars required, $N_{\rm stars}$. This is 1 except for the case of H16a and M12 which require multiple O3 stars to produce the measured luminosity:
\begin{equation}
    L_{\rm champagne} = \left (E_{\rm peak} - 13.6eV\right) Q N_{\rm stars} \label{eqn:L_champ}
\end{equation}
where $E_{\rm peak}$ is the energy of a photon at the peak wavelength.

The third possible explanation for the observed velocity dispersions are winds from massive stars. This can result in turbulence at the interface of the hot wind and cold dense gas where mixing occurs and thermal energy is dissipated \citep{Krumholz2019}. The wind luminosity is described by the following equation
\begin{equation}
    L_{\rm wind} = \frac{1}{2}\dot{M}V_{\rm wind}^2N_{\rm stars} \label{eqn:L_wind}
\end{equation}
where $\dot{M}$ is the mass loss rate in the stellar wind and $V_{\rm wind}$ is the wind speed. For the B0 star in \hii region G17, $\dot{M}$ and $V_{\rm wind}$ are taken directly from the \citet{Smith2002} models. For the O stars in the other 5 \hii regions, these quantities are estimated from the stellar mass, $M$, luminosity, $L$, and radius, $R$, determined in the \citet{Martins2005} models. The wind velocity is then estimated as 
\begin{equation}
    V_{\rm wind}=3\sqrt{2GM/R},
\end{equation}
with a maximum mass loss rate of
\begin{equation}
    \dot{M}=\frac{L}{V_{\rm wind}c}  
\end{equation}
following \citet{Lamers2017}.

The required $L_{\rm turb}$ as well as the luminosities available from each scenario to power the observed turbulence are reported in Table \ref{tab:outflow}. It should be noted that each of these scenarios assumes a spherically symmetric deposition of energy and thus should be multiplied by the factor, $\Omega$, the fraction of a sphere covered by the outflow as seen from the stars, determined from the ratio of the projected area of the outflow and \hii region. If the luminosity available in the outflow mechanism multiplied by $\Omega$ is greater than $L_{\rm turb}$, it indicates that the mechanism could provide sufficient energy to support the energy dissipated in the turbulent region. 

For the three largest regions with potential outflows, H16a, I18, and M12 $L_{\rm kin}\Omega > L_{\rm turb}$ with H16a $L_{\rm kin}\Omega \sim1.5 L_{\rm turb}$, M12 with $L_{\rm kin}\Omega \sim3 L_{\rm turb}$, and $L_{\rm kin}\Omega \sim L_{\rm turb}$ for I18. However, for these same three \hii regions the early O stars needed to produce the measured $\rm L_{\oiii}$ produce estimated $L_{\rm wind}$ and $L_{\rm champagne}$ that are greater than $L_{\rm kin}$. For region I18 which requires a single O5 star, $L_{\rm wind}\sim2\times L_{\rm kin}$ while $L_{\rm wind}\sim10\times L_{\rm kin}$ for H16a and M12 which require multiple O3 stars. In all three of these \hii regions, $L_{\rm champagne}$ is estimated to be $\sim$2 orders of magnitude greater than $L_{\rm kin}$. This scenario of observing champagne flows in the turbulent regions around these 3 \hii regions is therefore the most likely.

For the three smaller regions, G16, G17, and I17a, $L_{\rm kin}$ is not high enough to sustain the turbulence we see. This is unsurprising as the turbulent regions are comparable in size or larger than the \hii region they border (see Figure \ref{fig:outflows}). G16 and I17a both require a single O9 star to produce the required ionization, and the estimated winds from this type of star produce $L_{\rm wind}\Omega \sim 100-1000\times L_{\rm turb}$ indicating that stellar winds are a possibly sufficient source of energy to sustain the turbulent regions around G16 and I17a. For G17, none of three scenarios considered here produce a sufficient amount of energy to support the measured turbulence in the surrounding region. While these estimates are approximate, this along with the extended nature of the turbulence around G17 indicate an external source of energy.

\begin{deluxetable*}{ccc|CCCC|CCCCC} 
\tablecaption{Outflow Properties\label{tab:outflow}} 
\tablehead{ & & & \multicolumn{4}{c|}{Turbulent Region} & \multicolumn{5}{c}{\hii Region} \\ 
\colhead{ID} & \colhead{radius} & \colhead{stellar type} & \colhead{$\rm \sigma$} & \colhead{$\rm \tau_{eddy}$}  & \colhead{$\rm L_{turb}$} & \colhead{$\Omega$} & \colhead{$\rm v_{exp}$} & \colhead{$\rm \tau_{dyn}$} & \colhead{$\rm L_{kin}$} & \colhead{$\rm L_{champagne}$} & \colhead{$\rm L_{wind}$} \\ 
 & \colhead{(pc)} & & \colhead{$\rm (km \, s^{-1})$} & \colhead{$\rm (10^6\, yrs)$} & \colhead{$\rm (erg \, s^{-1})$} & & \colhead{$\rm (km \, s^{-1})$} & \colhead{$\rm (10^6\, yrs)$} & \colhead{$\rm (erg \, s^{-1})$} & \colhead{$\rm (erg \, s^{-1})$} & \colhead{$\rm (erg \, s^{-1})$}}  
\startdata 
G16 & 2.60 & O9 & 27.12 & 0.35 & 1.94\times10^{36} & 0.30 & 44.06 & 0.07 & 4.08\times10^{35} & 5.99\times10^{35} & 6.31\times10^{38} \\ 
G17 & 2.11 & B0 & 22.92 & 0.11 & 1.59\times10^{35} & 0.19 & 44.05 & 0.05 & 2.44\times10^{35} & -2.73\times10^{35}\tablenotemark{a} & 5.46\times10^{34} \\
H16a & 10.32 & O3 & 25.98 & 0.11 & 7.02\times10^{35} & 0.03 & 51.65 & 0.22 & 3.27\times10^{37} & 3.06\times10^{39} & 1.40\times10^{38} \\ 
I17a & 2.56 & O9 & 25.31 & 0.24 & 3.27\times10^{35} & 0.20 & 51.76 & 0.06 & 7.70\times10^{35} & 5.99\times10^{35} & 6.31\times10^{38} \\ 
I18 & 7.02 & O5 & 25.19 & 0.15 & 2.12\times10^{35} & 0.05 & 46.02 & 0.17 & 3.84\times10^{36} & 1.13\times10^{38} & 7.04\times10^{36} \\ 
M12 & 9.24 & O3 & 27.01 & 0.19 & 4.79\times10^{35} & 0.08 & 45.17 & 0.23 & 1.82\times10^{37} & 2.29\times10^{39} & 1.05\times10^{38} \\ 
\enddata 
\tablecomments{Measured properties and estimated energies in the turbulent regions and associated \hii regions. \tablenotetext{a}{The negative value of $\rm L_{champagne}$ for region G17 is due to the method of estimation (Equation \ref{eqn:L_champ}). The peak photon energy for a B0 star used to estimate the average is $<13.6$eV, giving a negative estimate of $\rm L_{champagne}$. This merely indicates that the turbulent region observed is unlikely to be due to a champagne flow.}}
\end{deluxetable*}
\newpage
\subsection{Metallicity}\label{sec:metallicity}
The KCWI small slicer, R$\sim$18,000 mode observations used throughout this analysis provide a detailed look at the structure and kinematics of the ionized gas, but the wavelength coverage is extremely limited. Our supplementary observations in the large slicer, R$\sim$900 mode rectify this shortcoming with coverage from 3500-5500\ang, at the expense of more limited spatial sampling and spectral resolution. This wavelength range allows us to estimate the gas-phase metallicity throughout IC\,10. Ideally, the auroral \oiii4363\ang \, line would be used with \oiii5007\ang \, to infer the electron temperature and metallicity \citep[e.g.,][]{Kewley2019}, but this is a very weak emission line and is unfortunately not detected in our stacked or individual spaxel spectra. Instead, we use the empirical $R_{23}$ strong line calibration. The commonly used $R_{23}$ was proposed by \citet{Pagel1979} as a calibration with the oxygen abundance as it is less sensitive to geometric factors than the \oiii/\hb \, ratio alone. It is defined as:
\begin{equation}
    \small
    \rm R_{23} = \frac{F(\oii3727,3729\AA)+F(\oiii4959\AA)+F(\oiii5007\AA)}{F(H\beta)}
\end{equation}
One caveat with this diagnostic is that it is degenerate, providing two possible values of the metallicity for a given $R_{23}$. Another diagnostic ratio is therefore required to determine the correct solution. These nebular ratios often make use of the \nii6584\ang \, line \citep{Nagao2006}, but since this is not in the observed wavelength range we will instead use the line flux ratios utilizing \oiii5007\ang, \oiii4959\ang, and the \oii3727,3729\ang\, doublet; an indicator of the ionization parameter. 

\begin{figure*}[ht]
    \centering
    \gridline{\fig{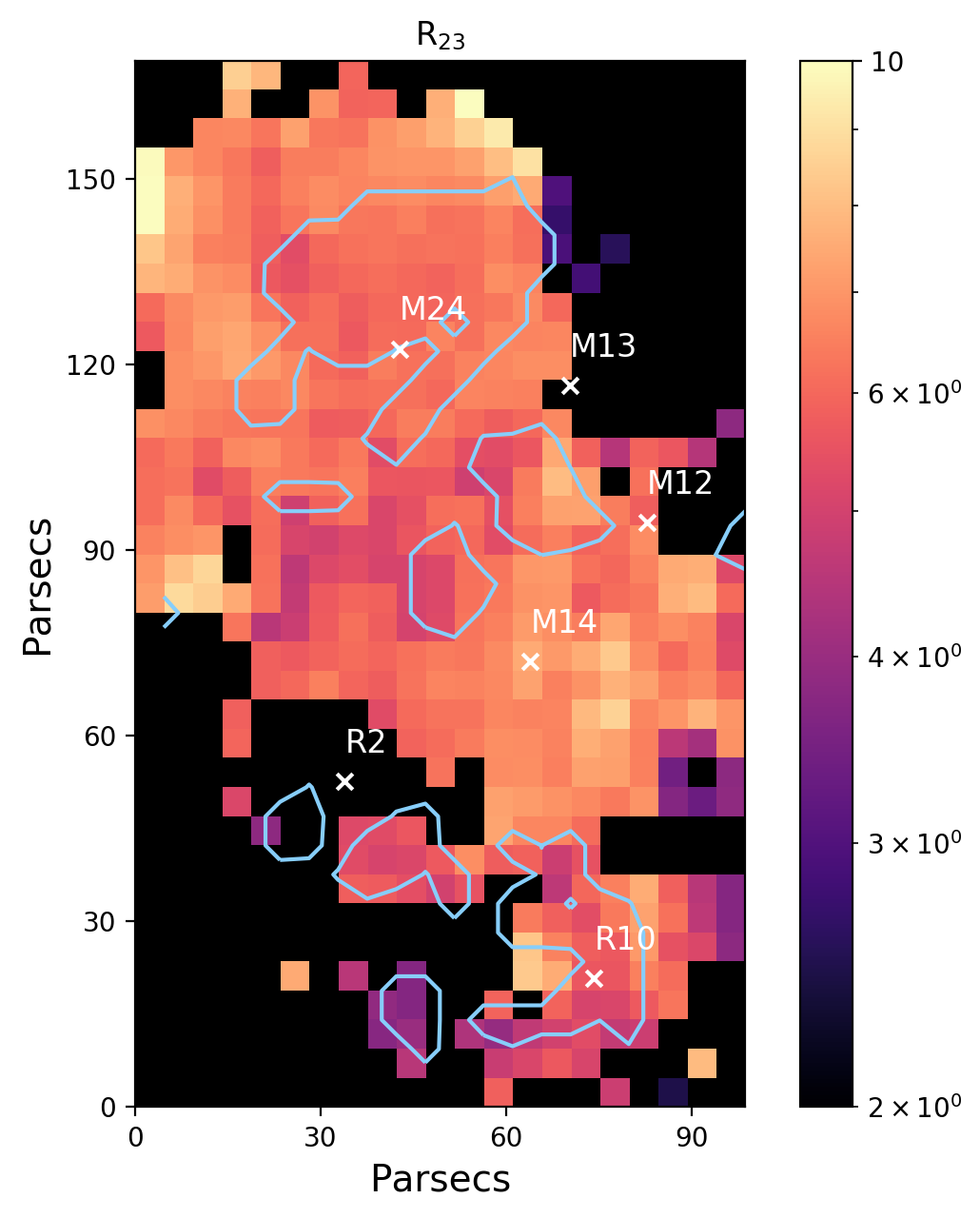}{0.32\textwidth}{(a)}
            \fig{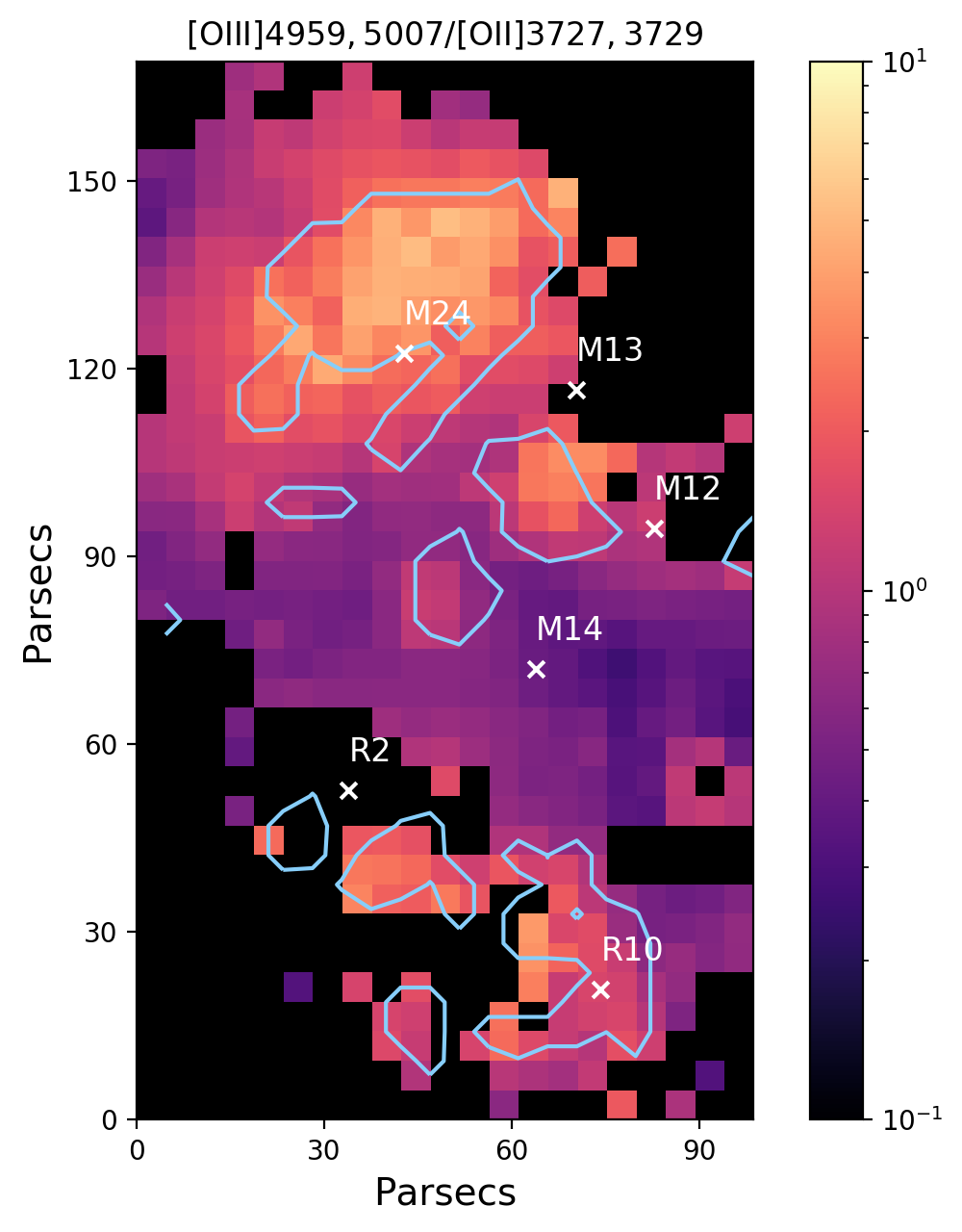}{0.32\textwidth}{(b)}
            \fig{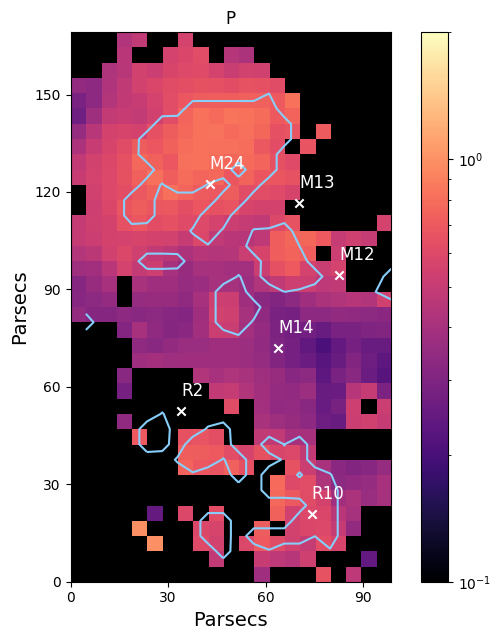}{0.32\textwidth}{(c)}}
    \gridline{\fig{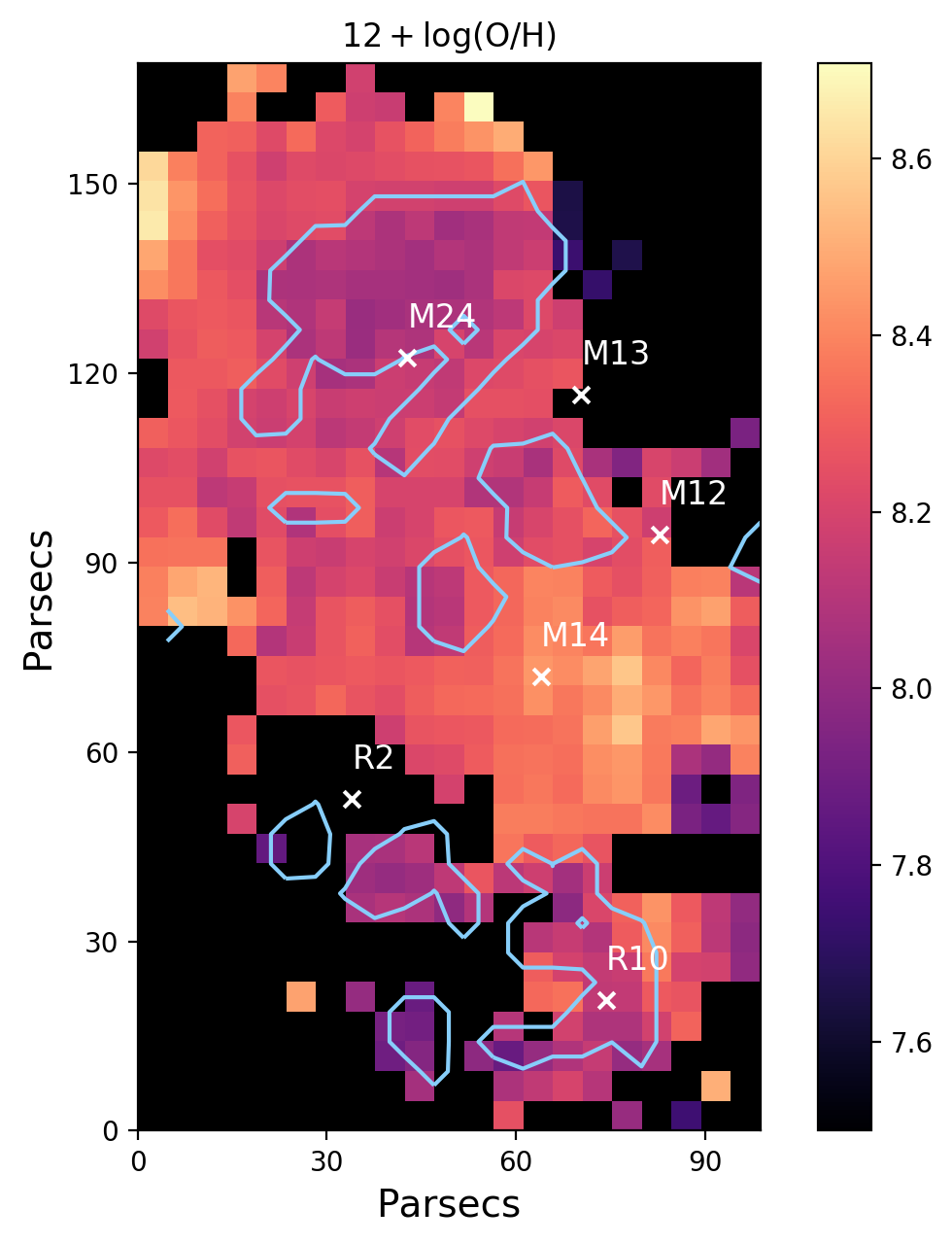}{0.42\textwidth}{(d) KK04}
            \fig{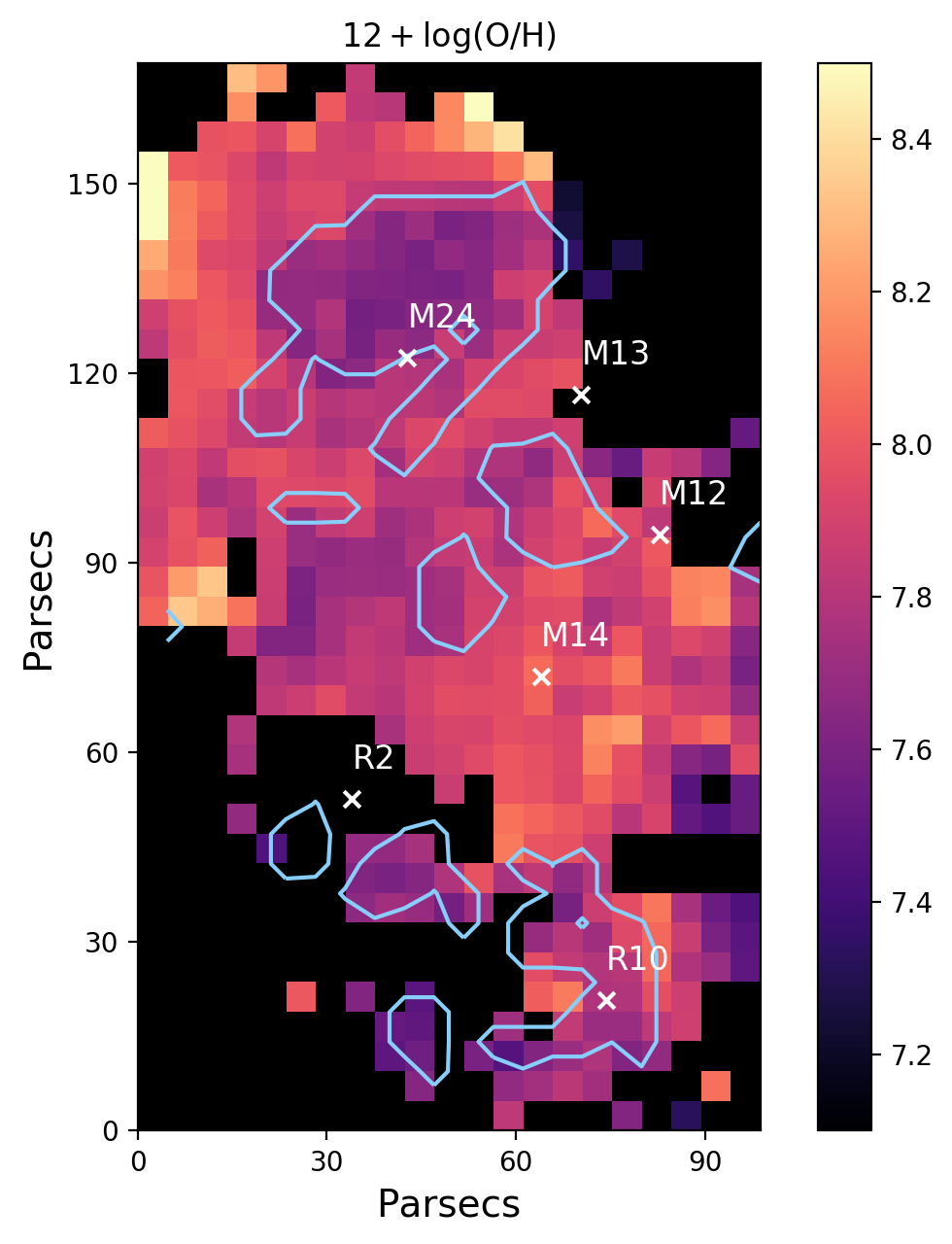}{0.42\textwidth}{(e) PT05}}
    \caption{Diagnostic line ratio maps from the ``large slicer, R$\sim$900" mode observations. Due to the lower SNR of the \oii3727\ang \, line only the lower left portion of the field is shown, coinciding with the ``East" field of the high resolution observations discussed in the majority of this study. (a): $R_{23}$ metallicity diagnostic. (b): $O_{32}$ line ratio (Equation \ref{eqn:KK04_O32}) used in the KK04 metallicity calibration. (c) P line ratio (Equation \ref{eqn:PT05_p}) used in the PT05 calibration. (d) Metallicity as determined by the KK04 diagnostic. (e) Metallicity as determined by the PT05 diagnostic. The metallicities are systematically lower than in the KK04 map, but the locations of lower \textit{relative} metallicity are consistent. The \hii regions identified in the high resolution mode are shown with the light blue contours (binned to the resolution here) and the locations of known WR stars are marked with white X's.}
    \label{fig:metallicity}
\end{figure*}

After applying the dereddening correction and DIG subtraction described in Section \ref{sec:spectra}, we employ a similar fitting method as for the \oiii5007\ang/\hb \, line ratio at each spaxel: fitting a Gaussian profile to the \hb \, line and using the center to define the systemic velocity shift at that spaxel and the width to define the number of wavelength channels over which to integrate the emission line fluxes. These fluxes are determined by a direct sum of the flux at that wavelength channel in the spectrum rather than integrating over the fitted Gaussian so as not to skew the resulting flux of the \oii3727,3729\ang \, doublet by the fitting of a single Gaussian. Defining the same central velocity and line width ensures that each line flux is evaluated over the same gas column. Any spaxel with a SNR of the \oii3727,3729\ang \, doublet ($\rm SNR_{\oii}$) $<3$ is removed from the analysis. Due to this cut only the ``East" (lower left) portion of the field covering the HL111 and HL106 complexes of \citet{HL1990} is included here and shown in Figure \ref{fig:metallicity}, with the vast majority of this field showing $\rm SNR_{\oii}>4$.

There are a number of calibrations in the literature utilizing the $R_{23}$ parameter, but we will limit our discussion to just three: the theoretical calibration of \citet{KK04} (hereafter KK04), and empirical calibrations from \citet{PT05} (PT05) and \citet{Nagao2006} (N06). Both the KK04 and PT05 calibrations rely on separate equations for what are referred to as the ``upper" and ``lower" branches of the diagnostic, while N06 uses a single continuous calibration. All three employ a diagnostic ratio involving one or more \oiii \, lines and \oii3727,3729\ang \, in addition to $R_{23}$.

The KK04 equations for oxygen abundance on each branch are dependent on $R_{23}$ as well as the ionization parameter, q, which is in turn dependent on the oxygen abundance and $O_{32}$, defined as:
\begin{equation}
    O_{32}=\frac{F(\oiii4959\AA)+F(\oiii5007\AA)}{F(\oii3727,3729\AA)} \label{eqn:KK04_O32}
\end{equation}
Since the equations for oxygen abundance and ionization parameter are dependent on each other this method requires an iterative solution, but the result converges after a few iterations. The transition between the two branches is noted to be at $\rm 12+log(O/H)=8.4$. This is slightly higher than the global metallicity typically measured for IC\,10 at $\rm 12+log(O/H)\sim8.2$ \citep[e.g.,][]{Skillman1989, Lebouteiller2012}, making the lower branch likely a better match.

The PT05 calibration does not require iteration, with each branch simply being dependent on the value of $R_{23}$ and the line ratio P:
\begin{equation}
    \footnotesize
    P= \frac{F(\oiii4959\AA)+F(\oiii5007\AA)}{F(\oiii4959\AA)+F(\oiii5007\AA)+F(\oii3727,3729\AA)} \label{eqn:PT05_p}
\end{equation}
For this calibration there is a ``transition zone" between $\rm 8.0<12+log(O/H)<8.5$ where the lower branch applies for values $\rm <8.0$ and the upper branch applies for values $\rm >8.5$. This transition zone does fall at the expected metallicity for IC\,10, so the results from this method should be taken with caution.

The N06 calibration, by contrast does not have a separate equation for the upper and lower branches. The equation for the oxygen abundance is also only dependent on the value of $R_{23}$, however the solution is double valued, relying on the  \oiii5007\ang/\oii3727,3729\ang \, (or other) line ratio to break this degeneracy. We evaluate the oxygen abundance in our IC\,10 field using this method to determine the most likely solution for each spaxel independently and also by constraining the solution for the entire field to either the upper or lower portion of the curve, mimicking the two branches of the KK04 and PT05 methods. 

The results for all three calibrations and the branches of each are summarized in Table \ref{tab:metallicity} which includes the mean metallicity throughout the field, for spaxels inside detected \hii regions, and for spaxels outside \hii regions. The upper branch solutions seem to systematically overestimate the oxygen abundance, a not unexpected result given the low global metallicity of IC\,10. The continuous N06 calibration gives an unrealistic result when the degeneracy is broken at each spaxel based on the \oiii/\oii\, ratio, resulting in an abrupt jump from low to high solutions at the edges of \hii regions rather than a smooth transition.  Even when constrained to the lower solutions, the N06 calibration appears to give less realistic estimates with a $\sim0.9$dex lower average metallicity than the previous estimates for IC\,10.

\begin{deluxetable*}{lccc}
     \tablecaption{Metallicity Calibration Results\label{tab:metallicity}}
     \tablehead{\colhead{Method} & \colhead{Mean Inside Regions} & \colhead{Mean Outside Regions} & \colhead{Total Mean} \\
     \colhead{} & \colhead{$\rm 12+log(O/H)$} & \colhead{$\rm 12+log(O/H)$} & \colhead{$\rm 12+log(O/H)$}}
\startdata 
\multicolumn{4}{c}{Lower Branch} \\ 
\tableline 
KK04 & 8.14 $\pm$ 0.01 & 8.26 $\pm$ 0.02 & 8.22 $\pm$ 0.02 \\ 
PT05 & 7.76 $\pm$ 0.01 & 7.89 $\pm$ 0.03 & 7.85 $\pm$ 0.03 \\ 
N06\tablenotemark{a} & 7.50 $\pm$ 0.01 & 7.53 $\pm$ 0.01 & 7.52 $\pm$ 0.01 \\ 
\tableline 
\multicolumn{4}{c}{Upper Branch} \\ 
\tableline 
KK04 & 8.85 $\pm$ 0.01 & 8.79 $\pm$ 0.01 & 8.81 $\pm$ 0.01 \\ 
PT05 & 8.40 $\pm$ 0.01 & 8.22 $\pm$ 0.02 & 8.27 $\pm$ 0.02 \\ 
N06\tablenotemark{a} & 8.61 $\pm$ 0.01 & 8.58 $\pm$ 0.01 & 8.59 $\pm$ 0.01 \\ 
\tableline 
\multicolumn{4}{c}{Continuous Calibration} \\ 
\tableline 
N06 & 8.26 $\pm$ 0.03 & 8.10 $\pm$ 0.04 & 8.13 $\pm$ 0.04 \\ 
\enddata 
\tablecomments{Average oxygen abundance derived from the three different calibrations for the upper and lower branches. The means are derived inside and outside the contours of the identified \hii regions as well as over all spaxels. The lower branches of the PT05 and KK04 branches provide the most reasonable solutions for the metallicity given the existing global measurements of IC\,10 which are below the branch transitions. The results from these two methods likely bracket the true metallicity. \tablenotetext{a}{The ``upper" and ``lower" branches of the N06 calibration are evaluated from the same equation with the root in the desired range taken as the solution rather than using the \oiii5007\ang/\oii3727,3729\ang \, line ratio to break the degeneracy.}}
\end{deluxetable*}

The lower branches of the KK04 and PT05 calibrations give the most reasonable results for the metallicity throughout this field of view in IC\,10 with average metallicities $\rm 12+log(O/H)_{PT05,lower}\approx7.85\pm0.03$ and $\rm 12+log(O/H)_{KK04,lower}\approx8.22\pm0.02$. There is a well studied offset between these two calibrations and it is thought that they span the range of potential ``true" values \citep[e.g.,][]{Moustakas2010, Kewley2008}, making this 0.4dex range a good indicator of the likely metallicity in this region of IC\,10.  While the values differ, the variation in metallicity across the field is consistent between both the KK04 and PT05 results. With both calibrations there is an average $\sim$0.1dex lower metallicity inside the \hii regions than in the surrounding gas. On average this difference is within the uncertainties, with some areas of higher metallicity in the diffuse gas being more apparent in the maps of Figure \ref{fig:metallicity}. This trend is consistent with the study by \citet{McLeod2019} of two \hii region complexes in the LMC in which they find lower oxygen abundance within the compact \hii regions than elsewhere in the complex. The difference in metallicity in this study is larger than in IC\,10, although the authors note that there is a dependence in their calibrations on ionization parameter (as the MUSE spectra do not cover \oii3727\ang \, needed for $R_{23}$) and the abundances may therefore be underestimated in the \hii regions. An earlier study by \citet{Russell1990} also found slightly lower metallicity in individual \hii regions in the SMC and LMC (0.1dex and 0.22dex respectively) than the global measurements. Deeper and wider field observations in IC\,10 and other local galaxies are needed in order to form a clearer picture of the metallicities of \hii regions relative to the surrounding gas.

\subsection{Diffuse Ionized Gas}\label{sec:DIG}
A significant fraction of the ionized gas emission in star forming galaxies has been observed outside of the \hii regions in the DIG. Often studies will differentiate the DIG from the star formation based on the \ha \, surface brightness. For example, in a sample of 109 galaxies, \citet{Oey2007} attribute $\sim$60\% of the \ha \, flux to the DIG with no correlation based on the galaxy Hubble type. \citet{Lacerda2018} on the other hand, propose a system of differentiating the DIG based on the equivalent width instead. They do find a correlation in the DIG fraction with Hubble type, with the highest contribution in ellipticals and lowest in late type galaxies, resulting in a similar average DIG fraction but with a wide distribution, $56\%\pm38\%$. Part of the cause for the differing conclusions from these two large studies is likely due to the two methods of differentiating DIG from star forming regions.

Before the identification and spectral analysis of IC\,10's \hii  regions we performed a conservative subtraction of the DIG contribution based on the choice of a low surface brightness region free of known \hii regions. This resulted in an average DIG contribution of $\sim$21\% of the \oiii5007\ang \, flux per spaxel across the FoV, with only $\sim$1\% at the \hii regions. Of the total \oiii5007\ang \, flux observed in our FoV, 78\% is contained in the \hii regions, 20\% in the connecting complexes, and the remaining 2\% from the DIG. This is significantly lower than what would be expected for the overall DIG contribution in IC\,10, but our field of view intentionally selected an area dense with \hii regions and complexes. To estimate the flux contribution of these three components throughout IC\,10 as a whole the measured flux is scaled based on the ratio of the number of \hii regions observed to total identified previously in IC\,10 \citep{HL1990} and the ratio of area observed to total area. This reduces the total estimated flux contribution from \hii regions to only 26\%, complexes to 7\%, and increases the estimated DIG contribution to 57\% throughout IC\,10. This estimate for the galaxy as a whole is more in line with values seen in previous studies for the DIG contribution to galaxy flux \citep[e.g.,][]{Oey2007}, but highlights the irregular distribution of \hii regions and DIG in irregular galaxies such as IC\,10.

We differentiate the \hii regions and DIG based on the ionized gas surface brightness, but compare the equivalent width between these areas of emission. Interestingly we find no significant difference between the distribution of \hb \, equivalent width between spaxels identified as belonging to an \hii region and that belonging to DIG in IC\,10. This may be a selection effect of our study. The high density of \hii regions not only gives a small sample of diffuse gas, but also makes it likely to be more closely associated with the inter-dispersed \hii regions than is typical of DIG studies. However, the DIG in IC\,10 may be ionized by different sources than typically observed. \citep{Hidalgo-Gamez2005} finds higher excitation in the IC\,10 DIG than for spiral galaxies which they find can be produced by leakage from \hii regions and the large number of WR stars. The gas throughout IC\,10, and particularly in the KCWI FoV, may be more similar to what \citet{Lacerda2018} refers to as mDIG, or 'mixed' DIG in which the ionization source is due to a combination of processes such as emission from an older stellar population in addition to photon leakage from \hii regions.  This is consistent with the small difference in metallicity we see between the \hii regions and surrounding gas in comparison with other studies, as well as \citet{Polles2019} Cloudy simulations showing matter-bounded regions in IC\,10 which would result in escaping photons ionizing the DIG.

\subsection{Scaling Relations}\label{sec:scaling}
In \citet{Cosens2018} we developed a framework to use Bayesian inference via PyStan to fit the scaling relationships between the properties of local and high-redshift star-forming regions in the literature. In this study we focused primarily on the relationship between star forming region size and \ha \, luminosity (\lha), which takes on the form of a power law:
\begin{equation}
    L_{\rm H\alpha}=exp(\beta)r^{\alpha}_{\rm clump}
\end{equation}
with $r_{\rm clump}$ giving the radius of the star forming region and $\beta$ giving the intercept of the fit. The key model parameter of interest is the slope, $\alpha$. The value of this slope holds information about the driving formation mechanism of the star-forming regions. A slope of $r^3$ is often explained by a region which forms under Jeans collapse and is then well-represented by a \stromgren \, sphere. On the other hand, a slope of $r^2$ is often explained by a region which forms under Toomre instability and undergoes the fastest mode of Jeans collapse resulting in a different form for the characteristic mass and size \citep[e.g.,][]{Genzel2011}. In \citet{Cosens2018} we found that this $r^2$ slope could also be explained by a \stromgren \, sphere argument where the ionizing photon production rate is large enough that the radius of the region is larger than the scale height of the galaxy disk. This would lead to a non-spherical geometry and an observed relationship of $L\sim r^2$.

Using Bayesian inference provides a number of advantages over standard least-squares fitting. First, uncertainties in every dimension can be incorporated in the fitting; there is no need to estimate a single overall uncertainty assigned to one dimension. Second, we can use our existing knowledge of the scaling relationships between parameters to inform our model through the use of Bayesian priors. Third, this method reproduces a distribution for each model parameter, allowing the determination of not only the best fit model, but robust determination of uncertainties for each model component as well. Using this framework allowed us to perform robust fits to the overall scaling relationships as well as investigate potential differences in smaller subsamples such as redshift bins and lensed versus field galaxies. Interestingly, we were able to identify a possible break in the size-luminosity scaling relationship based on the \sfrd\ of the star-forming region.

However, a key missing area of the parameter space in our previous investigation was small, low $\rm L_{H\alpha}$ star-forming regions ($<50$ pc, $<10^{34-36}$ \ergs). This sets the limit on the low-mass end of the relationship and helps to constrain the intercept when performing fits. One challenge in interpreting fitting results for sub samples of the local and high-redshift clumps was that the best fit slope and intercept are not entirely independent parameters. Therefore, missing constraints on the low-mass intercept of the size-luminosity relationship makes it difficult to be certain whether a change in slope is really a change in that parameter or just in the lever arm of the fit. With the proximity of IC\,10 and the sensitivity of KCWI we are able to target a large sample of \hii regions at this crucial scale.
\subsubsection{Size-Luminosity Relationship}
\begin{figure*}[h]
\gridline{\fig{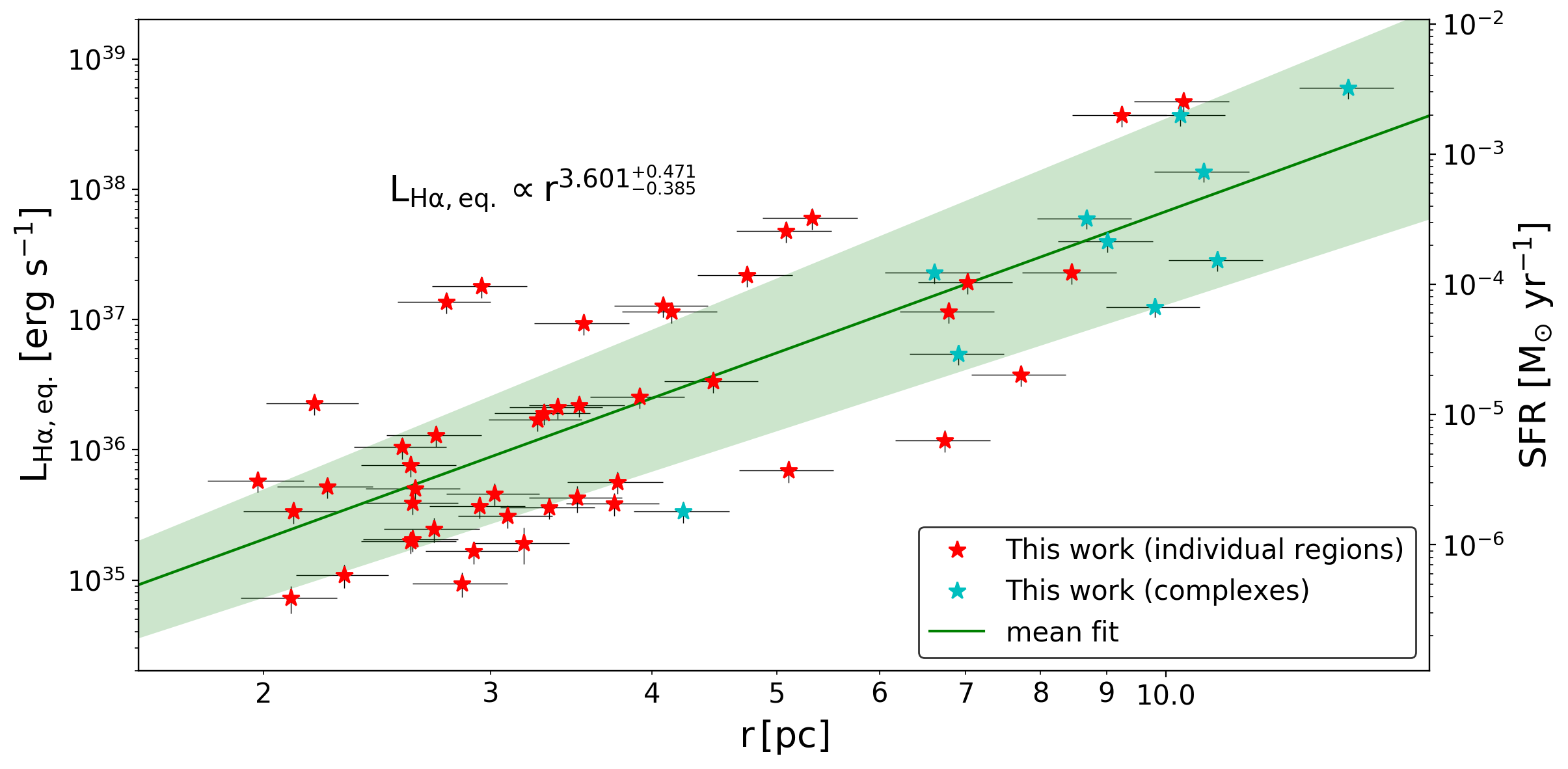}{0.98\textwidth}{(a)}}
\gridline{\fig{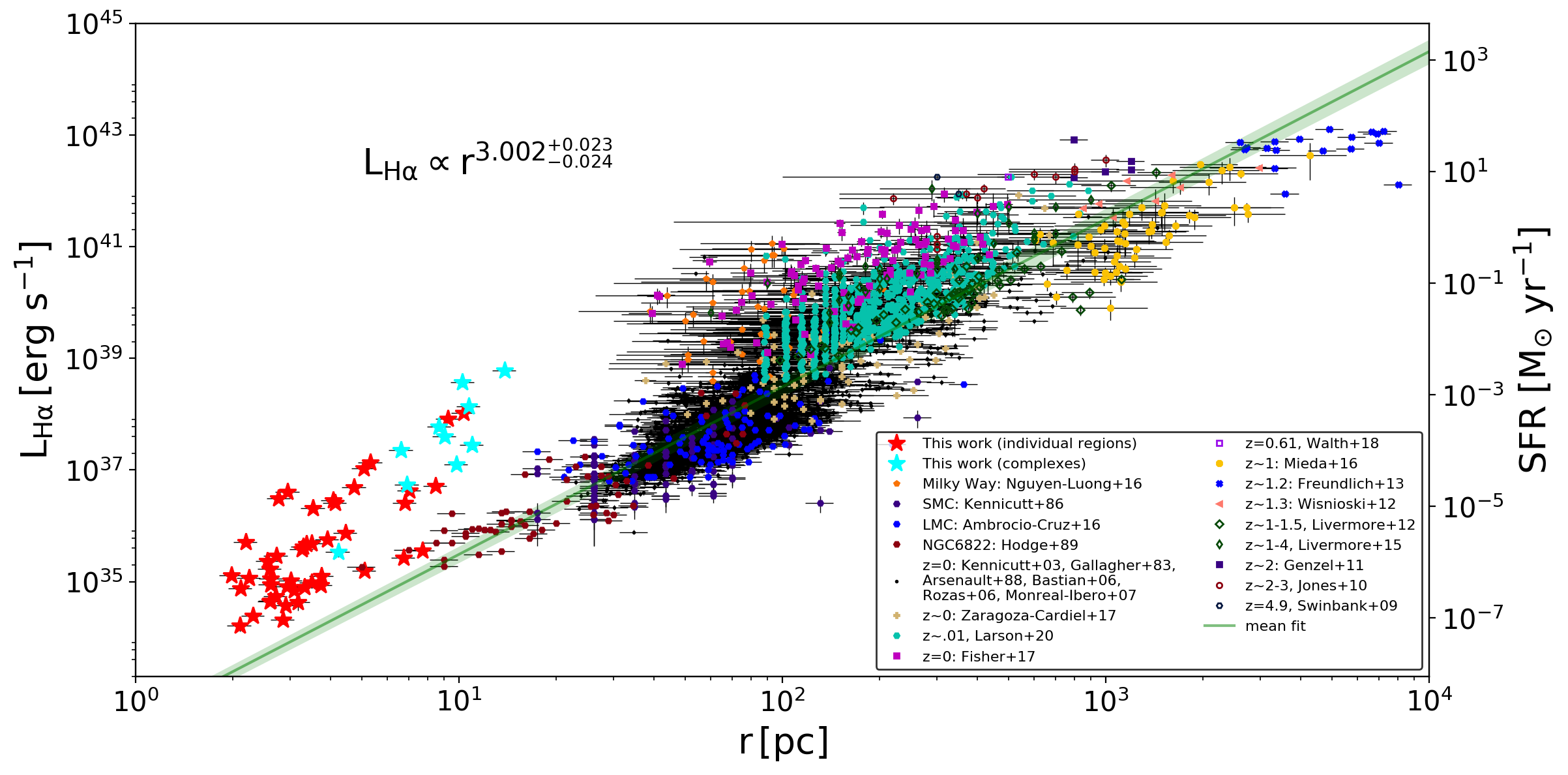}{0.98\textwidth}{(b)}}
\caption{(a): Results of fitting the size-luminosity scaling relationship for IC\,10 \hii regions identified in our KCWI observations. (b): Results of fitting the size-luminosity scaling relationship for IC\,10 \hii regions (red stars) as well as the rest of the local and high-redshift sample. The IC\,10 region complexes are shown for comparison (cyan stars) but are not included in the determination of the fit. \label{fig:ic10_mcmc}}
\end{figure*}

While these small, low-mass star-forming regions are critical for constraining the intercept of the size-luminosity relationship, one must first check that the possibility of stochastic sampling is not biasing the measured properties. The lower mass limit to avoid stochastic effects is typically determined to be \mstar$\sim 10^3$\msun \, \citep[e.g.,][]{Hollyhead2015, Krumholz2015}. Below these masses random sampling of the IMF can lead to deviations between the actual physical properties and those determined from photometric measurements. The average stellar mass distribution determined for the IC\,10 \hii regions from the measured \oiii5007\ang \, flux is $\rm M_* \sim (4\pm14)\times10^2 \, M_{\odot}$, with 96\% falling below the $10^3$\msun \, limit. Further, IC\,10's \hii regions are largely consistent with ionization predominantly from a single O or B type star which will be stochastic by nature. \citet{Hannon2019} found themselves faced with a similar dilemma studying a sample of $\sim$700 young star clusters with over 90\% below the stochastic limit. They investigate a method of mitigating the impact of stochastic sampling by stacking the fluxes of individual clusters with similar properties. These composite clusters no longer fall into the regime of stochastic sampling, but \citet{Hannon2019} still find results consistent with the individual clusters. We apply a similar check to our sample, but rather than stacking the \hii region spectra, we instead make use of the hierarchical structure determined from \texttt{astrodendro} which identifies the \hii region complexes. Using the complexes results in a smaller sample than with individual regions, but one that lies above the stochastic limit with a mean \mstar $\sim 2\times10^3$\msun. We do not find any significant deviation in the trends determined for the complexes vs. the individual \hii regions, finding proportional increases in the mass and luminosity to the increase in radius (e.g., Figure \ref{fig:ic10_mcmc}). The measured properties of individual regions may show an increase in scatter due to stochastic sampling, but the average sample properties and trends do not appear to be affected. Therefore we will proceed with the determination of the scaling relationships using the individual regions in the fitting, but we will include the complexes in all figures for comparison.

Fitting the size-luminosity relationship using our MCMC framework with just IC\,10 \hii regions yields a slope of $\rm L_{H\alpha} \propto r_{clump}^{3.6}$. The complexes do shift to slightly larger radii than the individual \hii regions but with a proportional increase in luminosity, placing them along the same relationship. We also combine these regions with the full sample of \hii regions and high-redshift clumps outlined in \citet{Cosens2018}, now including additional published samples of star-forming regions in the SMC \citep{Kennicutt1986}, LMC \citep{Ambrocio-Cruz2016}, NGC6822 \citep{Hodge1989}, and local LIRGs \citep{Zaragoza-Cardiel2017, Larson2020}. These LIRG studies also make use of \texttt{astrodendro} to identify star-forming regions and their properties\footnote{The effective radii, $\rm r_{eff}$, of \citet{Zaragoza-Cardiel2017} are adjusted to match our definition of $r_{1/2}^*$ for a consistent comparison.}. Collating all low and high-redshift samples results in a significantly shallower slope of approximately \lha $\propto r^3$; matching the result of fitting the full sample in \citet{Cosens2018} and indicating the IC\,10 \hii regions may be an outlier. The IC\,10 fit and entire sample fit are shown in Figure \ref{fig:ic10_mcmc}, where it can be can see that the IC\,10 \hii regions lie above the size-luminosity relation found for the full sample. This is also true of some other local samples, particularly in more extreme environments such as the LIRGs \citep[e.g.][]{Larson2020} and turbulent galaxies \citep{Fisher2017}. Whether the offset of the IC\,10 \hii regions is then due to improved resolution breaking the IC\,10 regions down into more compact components, or due to a fundamental difference in the environments of star-forming regions driving scatter in the scaling relationships is not fully clear. It may be that active expansion of these regions as discussed in Section \ref{sec:feedback} leads to the offset we see here with the \hii regions currently being under-sized for their luminosity.

The full results of these fits and those described in Section \ref{sec:scaling_sfrd} are shown in Table \ref{tbl:fit_params2D} for the intercept, slope, and intrinsic scatter along with their uncertainties. Results of fitting additional data subsets (as detailed in column 1) are also included for completeness.

\begin{deluxetable*}{lccrrrrr} 
	 \tablecaption{Size - Luminosity Relation Fit Parameters: ($L_{H\alpha} = e^{\beta} r_{clump}^{\alpha}$)} \label{tbl:fit_params2D} 
	 \tablehead{\colhead{Sample} & \colhead{Subset Criteria} & \colhead{Figure} & \colhead{$\alpha$} & \colhead{$\beta$} & \colhead{Scatter (r)} & \colhead{Scatter (L)} & \colhead{\# of Clumps}} 
\startdata 
IC\,10 & \nodata & \ref{fig:ic10_mcmc} & $3.601^{+0.471}_{-0.385}$ & $78.812^{+0.455}_{-0.652}$ & $0.147^{+0.154}_{-0.103}$ & $0.156^{+0.144}_{-0.103}$ & 46 \\
\tableline 
\multirow{3}{12em}{all comparisons} & \nodata & \nodata & $3.215^{+0.023}_{-0.030}$ & $73.782^{+0.129}_{-0.119}$ & $0.184^{+0.118}_{-0.112}$ & $0.171^{+0.115}_{-0.114}$ & 3858 \\ 
 & high \sfrd & \nodata & $1.788^{+0.063}_{-0.059}$ & $84.617^{+0.314}_{-0.357}$ & $0.461^{+0.317}_{-0.316}$ & $0.470^{+0.320}_{-0.314}$ & 264 \\ 
 & low \sfrd & \nodata & $2.967^{+0.024}_{-0.027}$ & $74.670^{+0.123}_{-0.111}$ & $0.142^{+0.092}_{-0.095}$ & $0.136^{+0.092}_{-0.093}$ & 3573 \\ 
\tableline 
\multirow{3}{10em}{IC\,10 \& all comparisons} & \nodata & \ref{fig:ic10_mcmc} & $3.002^{+0.023}_{-0.024}$ & $74.800^{+0.100}_{-0.115}$ & $0.240^{+0.147}_{-0.154}$ & $0.249^{+0.154}_{-0.151}$ & 3904 \\ 
 & high \sfrd & \ref{fig:sfrd_mcmc} & $1.841^{+0.070}_{-0.059}$ & $84.310^{+0.335}_{-0.384}$ & $0.426^{+0.279}_{-0.285}$ & $0.360^{+0.323}_{-0.268}$ & 266 \\ 
 & low \sfrd & \ref{fig:sfrd_mcmc} & $2.775^{+0.021}_{-0.018}$ & $75.577^{+0.086}_{-0.090}$ & $0.194^{+0.136}_{-0.132}$ & $0.194^{+0.132}_{-0.138}$ & 3617 \\    
 \enddata
\tablecomments{Results of model parameters determined from MCMC fitting of IC\,10 \hii regions and the local and high-redshift comparison sample for all fits discussed in Section \ref{sec:scaling} as well as additional fits to only the comparison sample (rows 2-4).}
\end{deluxetable*}

\subsubsection{\sfrd \, break} \label{sec:scaling_sfrd}
In \citet{Cosens2018}, we found that there was a potential break in the size-luminosity relationship that divides star-forming regions into two samples: one with high \sfrd\ and one with low \sfrd \, with the break nominally located at a value of $\rm\Sigma_{SFR}= 1 \, $\myrkpc. Both locally and at high redshift we found that lower \sfrd \, star-forming regions followed a size-luminosity relationship of  $L\sim r^3$, while the high \sfrd \, sub-sample followed a relationship closer to $L\sim r^2$.

In our observations of IC\,10's \hii regions, the average size of identified \hii regions is $\rm r_{avg}=4.0 \, pc$ with a $\rm \Sigma_{SFR,avg}=0.20 \,$\myrkpc \, with only 2 identified \hii regions falling above the $\rm\Sigma_{SFR}> 1 \,$\myrkpc \, limit. When combined with the full comparison sample this does still provide some improvements on constraining the size-luminosity relationship at low masses by reducing the uncertainties on the fitted parameters. The resulting slopes for the two populations remain consistent with the previous results, with nominal values differing by $\rm <2\sigma$.

\begin{figure*}[h]
\gridline{\fig{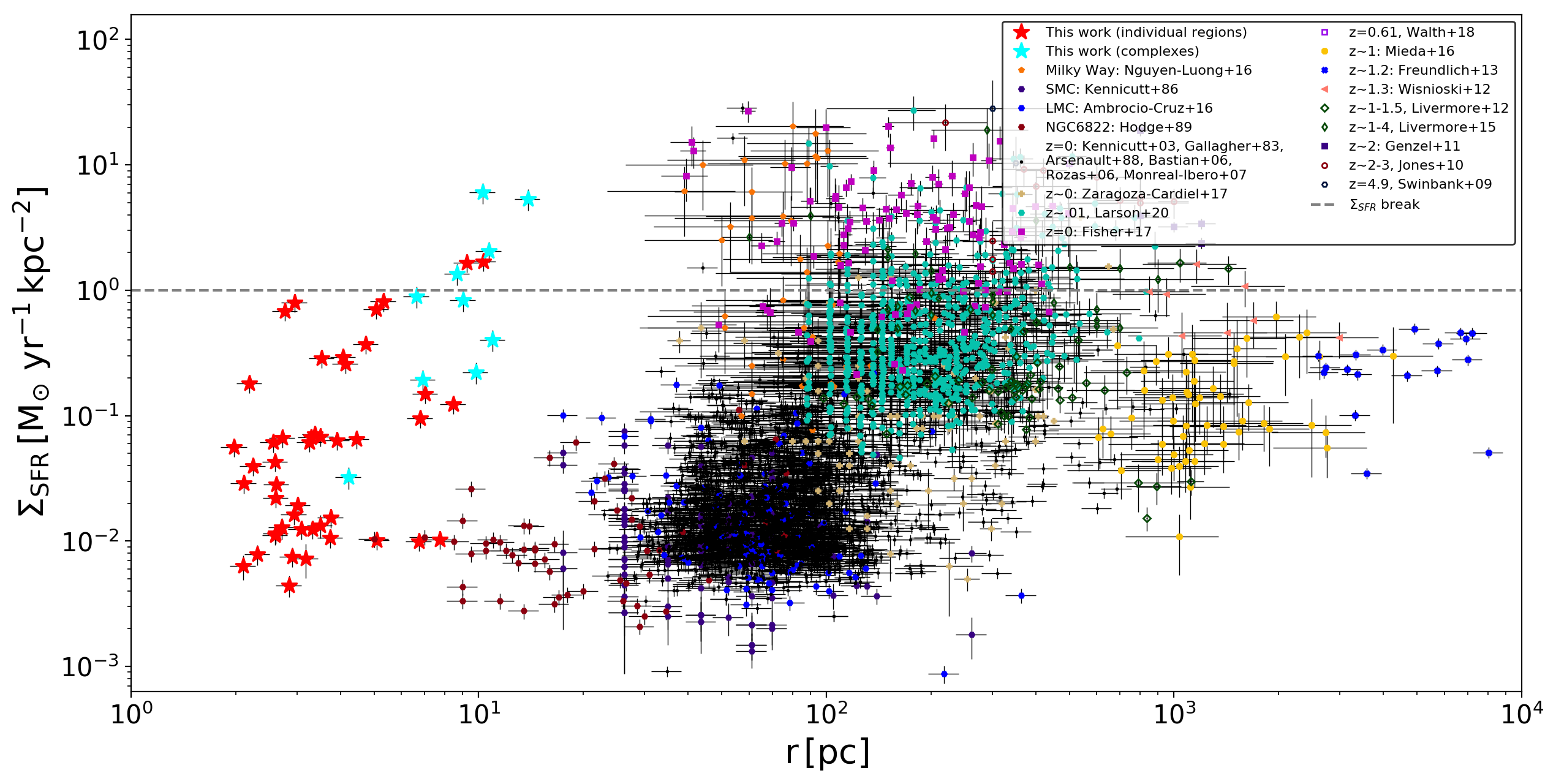}{0.98\textwidth}{(a)}}
\gridline{\fig{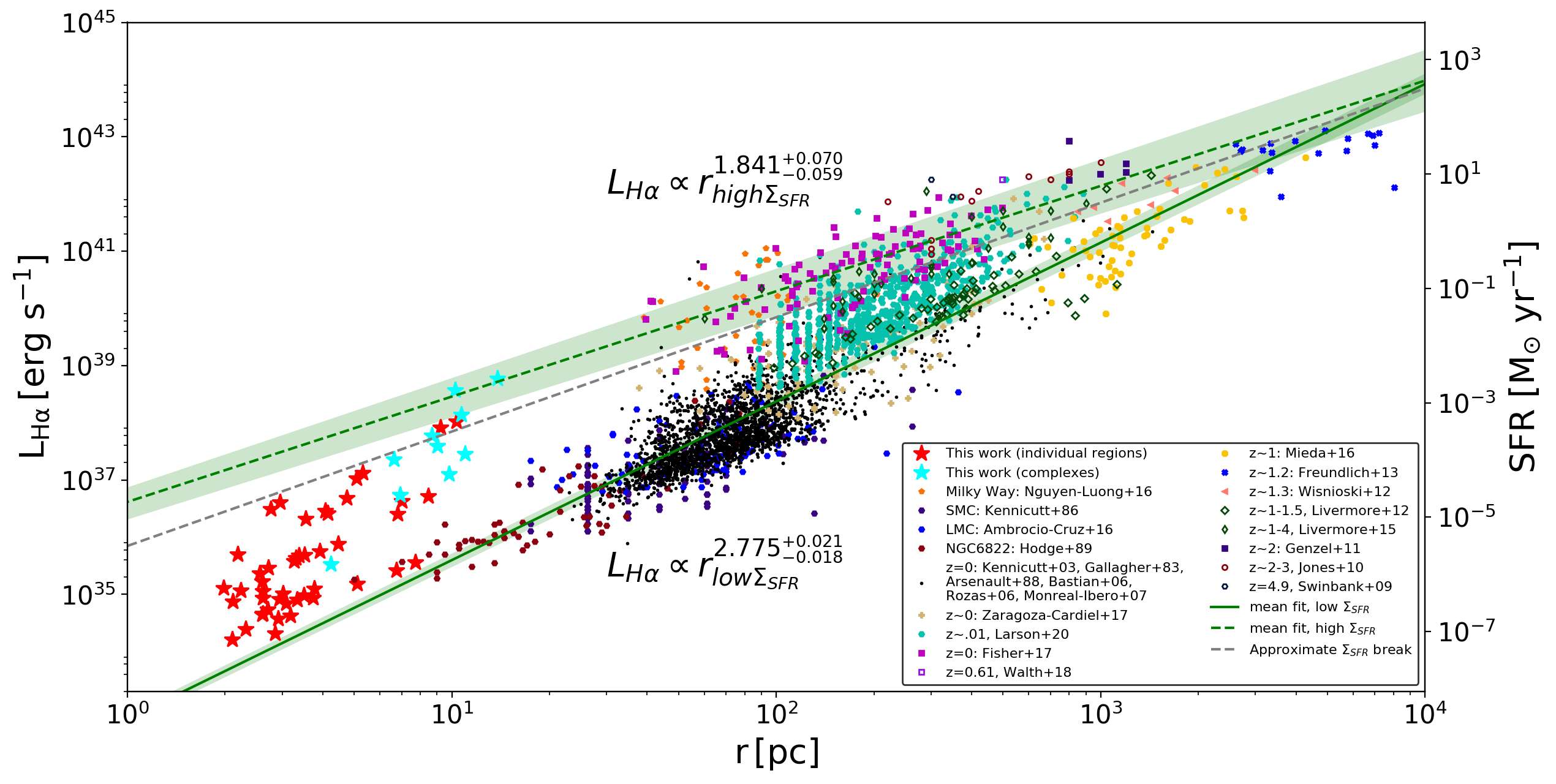}{0.98\textwidth}{(b)}}
\caption{(a): Size vs. \sfrd \, for \hii regions identified in our observations of IC\,10 as well as each comparison sample. These data are divided into high and low \sfrd \, based on a break at $\rm \Sigma_{SFR} = 1 \, $\myrkpc (dashed line). (b): Results of fitting the size-luminosity scaling relationship for these two groups. The low \sfrd \, subset tends towards a slop of $L\sim r^3$, while the high \sfrd \, subset produces a slope closer to $L\sim r^2$. The IC\,10 region complexes are shown for comparison (cyan stars) but are not included in the determination of the fit. \label{fig:sfrd_mcmc}}
\end{figure*}

\subsubsection{Size - Velocity Dispersion}\label{sec:scaling_sigma}
\citet{Larson1981} found an empirical relationship between the size of molecular clouds and their velocity dispersion that is tied to turbulence in the clouds and ISM. For resolved Milky Way GMCs in virial equilibrium, \citet{Larson1981} found a relationship:
\begin{equation}
    \sigma = 1.1r^{0.38}
\end{equation}
where $r$ is the radius of the GMC in pc and $\sigma$ is in \kms. This has been refined and updated with expanded galactic \citep{Solomon1987} and extra-galactic \citep{Bolatto2008} data sets of molecular clouds, arriving at slight changes to the scaling but the same basic conclusion that the line width increases with cloud size approximately $\rm \propto r^{1/2}$. This scaling is typically interpreted as being due to turbulence in the molecular clouds \citep[e.g.,][]{Larson1981, Bolatto2008}.

\begin{figure*}[h]
\centering
\gridline{\fig{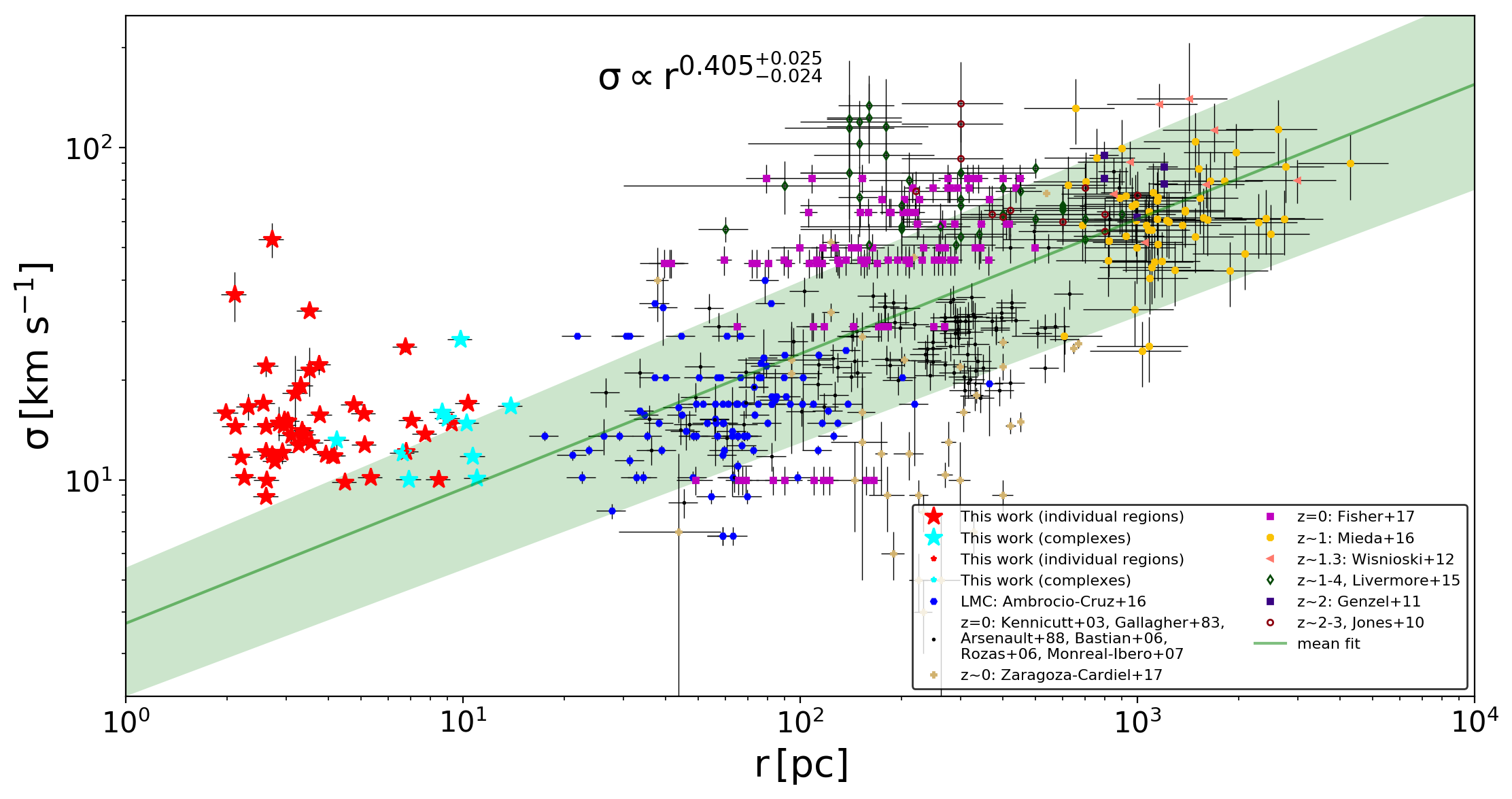}{0.9\textwidth}{(a)}}
\gridline{\fig{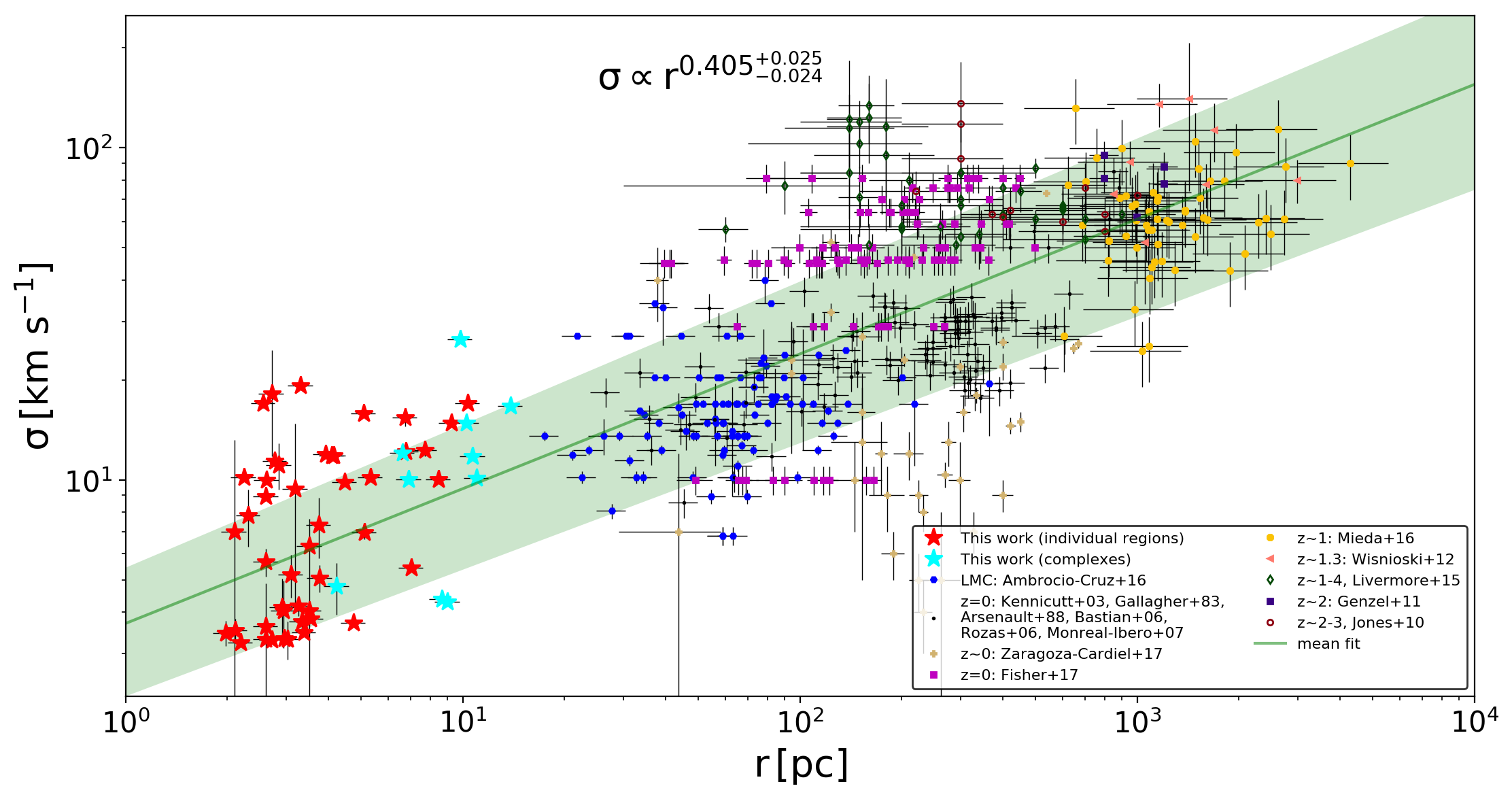}{0.9\textwidth}{(b)}}
\caption{Size vs. $\sigma$ \, relationship for the star-forming regions in each comparison sample containing spectral information. The fit is computed using our MCMC framework excluding the IC\,10 \hii regions (red stars) and complexes (cyan stars) from the analysis. The mean fit is shown with the green line while the uncertainty in the fit is denoted by the shaded green region. (a): Both the IC\,10 \hii regions and complexes fall above this mean fit, and in most cases above the uncertainty as well, indicating that these \hii regions are not virialized. The complexes lie closer to the mean relationship while still tending towards higher $\sigma$ for a given size, further supporting the conclusion that the individual regions are not virialized as the increase in size does not result in a proportional increase in measured line width. (b): The IC\,10 \hii regions and complexes lie along the mean size - $\sigma$ \textit{after} the excess velocity dispersion determined in Section \ref{sec:feedback} is removed. The excess dispersion was found to be in addition to that produced by rotational motion of the \hii region gas and is therefore likely due to outflows or expansion of the ionized gas. \label{fig:size_sigma_mcmc}}
\end{figure*}

While Larson's Law is usually discussed in relation to GMCs it may also apply to ionized gas in star-forming regions if they are in virial equilibrium and exhibit turbulent motions. \citet{Wisnioski2012} fit this relationship with a portion of the local and high-redshift data sets used in the present study and find a scaling of $\sigma \propto r^{0.42}$, but determine that their $\rm z\sim1$ clumps are likely not virialized. We test this now with the addition of local and high-z star-forming regions measured with IFS observations since the study by \citet{Wisnioski2012}. Unfortunately, not all data sets used to investigate the size-luminosity relationship have spectral information, but it is still a large enough sample (515 regions) to investigate the presence of a similar size-velocity dispersion relationship. In fact, we do see a clear trend where the velocity dispersion increases with larger size regions, except with IC\,10, which is still  a notable outlier. On their own, IC\,10 \hii regions do not show any evidence of this power law scaling, and instead when included significantly skew the resultant slope. Without IC\,10 included we arrive at a relationship of $\sigma \propto r^{0.4}$ using the same MCMC framework as described in Section \ref{sec:scaling}, in line with the results of \citet{Wisnioski2012} and similar to the GMC results. However, when the IC\,10 \hii regions are included, they clearly lie above the size - $\sigma$ relationship found for the other samples as shown in Figure \ref{fig:size_sigma_mcmc}. Furthermore, including the IC\,10 regions in the fitting reduces the slope to $\sigma \propto r^{0.2}$ . This further supports the conclusion in Sections \ref{sec:mass} \& \ref{sec:feedback} that the \hii regions identified in this study are not generally virialized. In fact, if we retain only the component of velocity dispersion due to rotational motion as identified in Section \ref{sec:feedback}, then the identified \hii regions in IC\,10 follow the same relationship identified for the rest of the samples (see Figure \ref{fig:size_sigma_mcmc}b). There is a significant amount of scatter in the $\rm r - \sigma$ relationship, but it suggests that rotational motion is a much more significant source of the measured line widths in the other star-forming regions than in IC\,10's \hii regions.

\begin{deluxetable*}{lcrrrrr} 
	 \tablecaption{Size - Velocity Dispersion Relation Fit Parameters: ($\sigma = \beta r_{clump}^{\alpha}$)} \label{tbl:fit_params_sigma} 
	 \tablehead{\colhead{Sample} & \colhead{Figure} & \colhead{$\alpha$} & \colhead{$\beta$} & \colhead{Scatter (r)} & \colhead{Scatter ($\sigma$)} & \colhead{\# of Clumps}} 
\startdata 
all comparisons & \ref{fig:size_sigma_mcmc}b & $0.406^{+0.025}_{-0.025}$ & $3.704^{+1.129}_{-1.154}$ & $0.444^{+0.288}_{-0.276}$ & $0.414^{+0.275}_{-0.294}$ & 469 \\ 
\tableline 
IC\,10 \& all comparisons & \ref{fig:size_sigma_mcmc}a & $0.216^{+0.013}_{-0.012}$ & $10.442^{+1.072}_{-1.060}$ & $0.262^{+0.165}_{-0.171}$ & $0.257^{+0.176}_{-0.172}$ & 515 \\ 
\enddata
\tablecomments{Results of model parameters determined from MCMC fitting of IC\,10 \hii regions and the local and high-redshift comparison sample for the size-velocity dispersion relationship discussed in Section \ref{sec:scaling_sigma}.}
\end{deluxetable*}

\section{Discussion} \label{sec:discussion}
Observational studies of the stars and \hii regions in IC\,10 suggest that the starburst occurred relatively recently, likely in the past $\lesssim10$Myr \citep[e.g.,][]{HL1990, Hunter2001}. Clustering of the young stellar population \citep{Vacca2007} and similar properties across multiple regions of the starburst \citep{Polles2019} indicate a potential common origin for the recent star formation.

In our KCWI observations of the \hii regions in IC\,10's central starbursting region, we find further evidence that these regions are young, with average crossing times $\rm<Myr$. Feedback plays an important role in the regulation of star forming environments and the dissipation of material. There are many forms this feedback can take and their contribution can be estimated from their energy or momentum input into the ISM. In our KCWI observations, we only observe the ionized gas and therefore cannot estimate the contribution of every form of potential feedback, but we do estimate the contribution of two important factors: direct radiation pressure, $\rm P_{dir}$, and warm gas pressure, $\rm P_{gas}$. Of these two we find the contribution of $\rm P_{gas}$ to be dominant over $\rm P_{dir}$ by $\sim$3 orders of magnitude. It should be noted that radiation pressure is sensitive to the effects of stochastic sampling of the initial mass function since the majority of the radiation is produced in the most massive stars. The impact of radiation pressure can therefore be somewhat uncertain in smaller clusters like those powering IC\,10's \hii regions where the stellar population of the ionizing cluster is not well represented by the assumed IMF. Though, since $\rm P_{gas}$ is $\sim$3 orders of magnitude greater than the estimated $\rm P_{dir}$ we do expect radiation pressure to be comparably negligible even if the estimate is impacted by stochasticity. The same trend with a minor contribution from $\rm P_{dir}$ is also found in the giant \hii region 30 Doradus \citep{Pellegrini2011, Lopez2011}, a sample of 32 LMC and SMC \hii regions \citep{Lopez2014}, and an additional sample 11 LMC \hii regions \citep{McLeod2019}, although to a lesser extent.

The total $\rm P_{out}$ is on average $3\times$ the inward pressure which is predominantly due to turbulence in the surrounding gas. The self-gravity of the \hii regions is comparably weak in the compact \hii regions found in IC\,10. This is in contrast to results found in the molecular ISM of nearby galaxies showing kpc and sub-kpc scale equilibrium between gravitational potential and outward pressure \citep[e.g.,][]{Sun2020}. However, our \hii regions observations are on much smaller size scales, resulting in the average $\rm P_{out} \sim 3\times P_{in}$,  and with 89\% of the \hii regions identified in IC\,10 showing greater $\rm P_{out}$ indicating expansion. This is somewhat at odds with theoretical expectations of feedback effectiveness. For both direct radiation pressure and warm gas pressure, massive stars are expected to be the dominant source of the required ionizing photons, and thus these mechanisms are not expected to limit the star formation efficiency in populations with low stellar mass clusters \citep[$\rm \lesssim400\, M_{\odot}$,][]{Krumholz2019}. This may vary by cluster due to stochasticity of the IMF, but given we find $\rm P_{gas}$ to be an effective counter to inward pressure in the majority of IC\,10's \hii regions this may require further exploration. We should note that discussions of limiting the star formation efficiency are usually explored in the context of the larger molecular cloud, while we are limited to studying these mechanisms within the \hii regions and therefore cannot directly measure how much impact these mechanisms will have on clearing gas further from the ionizing cluster, but this still implies that $\rm P_{gas}$ may effectively limit star formation efficiency in a wider range of environments than previously expected. This is compounded by another potential source of expansion in the detection of diffuse X-ray emission by \citet{Wang2005} for which they note a morphological similarity with the \ha \, gas in the region of IC\,10 that we observe with KCWI. They argue that this indicates the hot gas is still confined and may be driving expansion of the surrounding ionized gas structures. However, the effectiveness of this hot gas in driving feedback is dependent on leakage and mixing and likely sub-dominant compared to $\rm P_{gas}$ \citep{Lopez2014}.

The stacked spectrum of all the identified \hii regions shows further evidence of expansion and/or outflowing gas, with an underlying broad component to the \oiii5007\ang \, emission line with $\sigma\sim35$\kms. Even the regions with evidence of rotation do not seem to have reached an equilibrium state with the surrounding gas as seen in the $\sim$14\kms \, higher measured velocity dispersions than what would be due to just rotation as well as virial parameters, $\rm \alpha_{vir} >>1$. The  higher velocity dispersions and elevated $\rm log \left([O{\sc III}]/H\beta\right)$ at the edges of many of the \hii regions indicate that there may be shocks present which could be due to expansion of the region, champagne flows, or hot stellar winds mixing with cold gas. Over 35\% of the regions show these pockets of elevated velocity dispersion at or near the edges of the \hii region. The velocity dispersions are not high enough to be produced by shocks in the \hii regions themselves as the elevated dispersions are $\sim$2$\times$ the sound speed, $c_s$. However, at the \hii region boundaries the expanding gas may collide with cold molecular gas in the ISM which could induce a shock. With typical ISM temperatures of $\sim$100K, $c_{\rm s}$ would be on the order of 1\kms \, for an ideal gas; 20$\times$ less than the areas considered to have elevated velocity dispersion. A previous study of IC\,10 ionized gas by \citet{Thurow2005} also found systematically larger line widths outside of \hii regions, but they attribute this to superposition of different filaments or shells. These IFU observations have similar spectral resolution to our small slicer, R$\sim$18,000 observations, but with KCWI we are able to achieve 3$\times$ greater spatial resolution, limiting the potential for superposition of structures with different velocities. Given that we observe areas of higher velocity dispersion at spaxels with the highest $\rm log \left(\oiii5007\AA/H\beta\right )$ we find shocked gas to be a plausible alternative to the superposition of independent filaments, particularly with the significant improvement in spatial resolution of KCWI requiring these structures to be aligned along the line of sight on $\sim$few pc scales.

Resolved areas of significantly elevated velocity dispersion ($\sigma > 25$\kms) were identified around 6 \hii regions with average eddy turnover times $\rm \tau_{eddy}=l/\sigma \sim 0.2Myr$. The turbulent volumes around the three largest of these \hii regions are most likely caused primarily by champagne flows, but stellar winds from early O stars could also provide sufficient energy to support the rate of turbulent dissipation ($\rm L_{turb}$). Two of the smaller \hii regions with resolved turbulent volumes at the border are ionized by lower mass stars with therefore lower rates of ionizing photon production. In this case we find that stellar winds are more likely to support the observed turbulence and are again sufficient to support $\rm L_{turb}$. One of the six turbulent volumes has an estimated $\rm L_{turb}$ greater than what we estimate could be provided by the \hii region and therefore is likely due to some external source of turbulence.

$\rm \tau_{eddy}$ measured for these turbulent regions is at the low end of the range observed in galactic scale outflows driven by starbursts \citep[0.1 - 10 Myr;][]{Veilleux2005}. However, the velocity measured in these ionized gas outflows, FWHM $\sim 61$\kms, is significantly lower than that measured in galactic winds and outflows from individual star forming clumps in other studies. In a sample of 25 star-forming clumps located in LIRGs, \citet{Arribas2014} identify outflows in 83\% of clumps with typical $\rm FWHM \sim 200$\kms. Similarly, in the \citet{RodriguezdelPino2019} study of ionized gas outflows in MaNGA galaxies, finding typical $\rm FWHM \sim 350$\kms \, in outflows originating from star-forming regions. Both of these studies find significantly higher velocities to the ionized gas outflows than the IC\,10 \hii regions which could be due to larger SFR (particularly in the case of the LIRGs), but they also have $\sim100\times$ lower spatial resolution to what is achieved in IC\,10 with KCWI. These measured outflows may then be due to aggregate measurements of multiple compact \hii regions and outflows. Some of this variation may also be due to biases introduced by differences in outflow detection methods. Both \citet{Arribas2014} and \citet{RodriguezdelPino2019} rely on separating a broad outflow component in the emission line of the integrated star-forming region spectra while the resolution of this study allows direct detection of the influence of outflows in the ionized gas surrounding the \hii regions. This may then simply probe more localized, lower velocity outflows than has been possible in previous studies.

The velocity dispersions measured inside the identified \hii regions are also elevated relative to what would be expected for a rotating region in equilibrium for all cases showing evidence of rotational motion. The \hii regions we identify in IC\,10 tend to be offset from the scaling relationships found between the region size \& luminosity as well as from the Larson's Law between size \& velocity dispersion. \citet{Krieger2020} find a similar offset to larger line widths for a given size when comparing molecular clouds in the starburst NGC 253 to clouds in the Milky Way's Galactic center, which they attribute to gas that is not gravitationally bound and rather lies in transient structures. Based on the discrepancies in mass estimates and the young ages of IC\,10's \hii regions, it seems likely that these \hii regions are still young enough to be undergoing expansion. They may then evolve onto the typical size-luminosity and size-velocity dispersion sequences after reaching an equilibrium state with the surrounding ISM.

\begin{figure*}
    \centering
    \includegraphics[width=.98\textwidth]{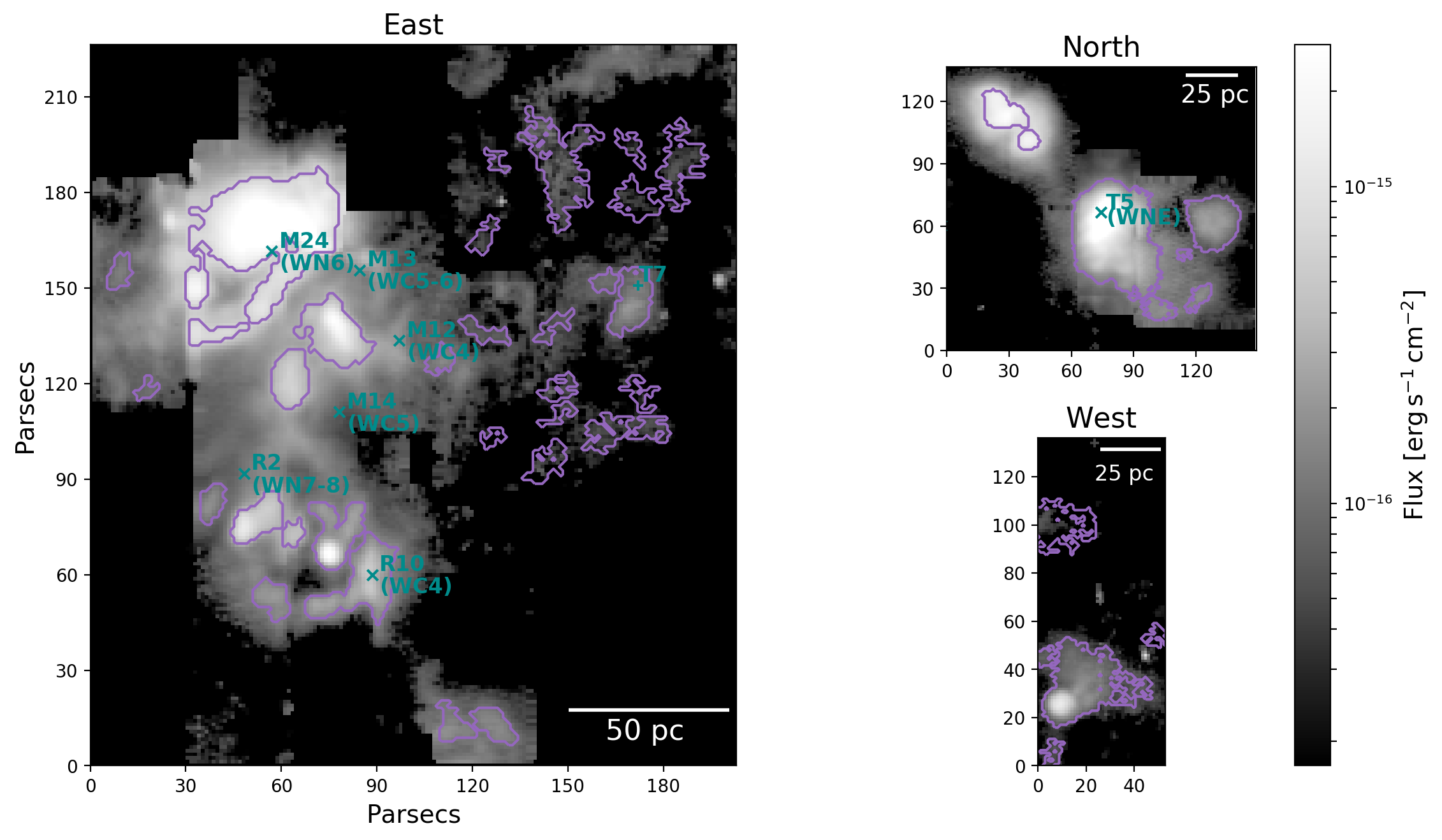}
    \caption{\oiii5007\ang \, integrated flux map with region contours reproduced from Figure \ref{fig:flux_maps} with marked locations of all known WR stars ('X') and WR candidates ('+'). The stars designation is included to the right of the location marker with the WR type in parentheses if it has been spectroscopically confirmed.}
    \label{fig:WR_locs}
\end{figure*}

The evolutionary stage of the Wolf-Rayet stars in the field of view provide another clue that the \hii regions which host them are young. Wolf-Rayet stars are a later evolutionary stage of O stars which occurs before a type I SNe. In general, high mass O stars ($\rm >40M_{\odot}$) evolve to a WN type WR, then a WC, and then a SNIc, whereas lower mass O stars are thought to not reach the WC stage and simply explode as a SNIb  \citep{Crowther2007}. Of the eight WR stars in the field covered by our KCWI observations, four of these fall in an \hii region. Two have been identified as WN spectral types (M24 and T5 in \hii regions H16a and M12 respectively), one is identified as an early WC (R10 in region I18), and the last has not yet been spectroscopically confirmed. Of the four WR stars outside of IC\,10 \hii regions, one is a late WN type and the rest are identified as WC. These stars are located in areas where ionized gas is nearby, but more diffuse and filamentary in appearance. The locations of these WR stars and their spectral type are shown in Figure \ref{fig:WR_locs} with the observed \oiii5007\ang\, flux and \hii region contours. Though this is a small sample, the trend implies that WR stars in the WN stage could be more likely to be found in current \hii regions, whereas WC types might be more likely found in areas of diffuse gas where an \hii region may have previously been present. This would imply that the star clusters the WN stars belong to are younger and the surrounding gas has not yet been disrupted. 

\section{Summary} \label{sec:conclusion}
We made use of the highest resolution mode of the Keck Cosmic Web Imager IFS $-$ 0.35$\arcsec$ spatial sampling with 1$\arcsec$ FWHM and R$\sim$18,000 $-$ to study the population of \hii regions in our nearest starburst galaxy IC\,10. These high quality IFS observations allowed us to study the spatial and kinematic properties of the \hii regions in detail. We identified 46 individual \hii regions in the central burst of the irregular galaxy with a total SFR$\sim6\times10^{-3}$ \myr. The average \hii region identified has a size of 4.0 pc, an SFR of $\rm 1.3\times10^{-4}$ \myr, an ionized gas mass $\rm M_{HII}\sim 56 \, M_{\odot}$, and a velocity dispersion of $\sim$16\kms. 

Over 95\% of the identified \hii region luminosities are consistent with the ionizing photon production rate of a single O or B star. 10 of the \hii regions ($\sim$22\%) can be powered by a single B star, with the lowest luminosity region requiring at minimum a star of spectral type B0.5.

The \hii regions appear to be blue-shifted relative to the systemic velocity of IC\,10 ($\sim$12\kms) while the diffuse surrounding gas often shows a relative red-shift. Supplementary KCWI observations with lower resolution and wider wavelength coverage are used to estimate the oxygen abundance via the KK04 and PT05 metallicity calibrations, both of which make use of the $R_{23}$ strong-line calibration. These estimates yield averages of $\rm 12+log(O/H)_{PT05,lower}\approx7.85\pm0.03$ and $\rm 12+log(O/H)_{KK04,lower}\approx8.22\pm0.02$ with the ``true" metallicity expected to lie within this range.

IC\,10's \hii regions have very short crossing times ($\rm \tau_{cr} < Myr$) and are not virialized ($\rm \alpha_{vir}>>1$ and $\rm M_{vir}>>M_{HII}$). The measured velocity dispersions in the \hii regions are also too high to be due to rotational motion alone (by $\sim$11-12 \kms). We see evidence that these regions are generally still undergoing expansion. The IC\,10 \hii regions are offset from the scaling relationships found between the region size \& luminosity as well as the size \& velocity dispersion relationship. As these regions expand they may evolve onto the scaling relationships determined for the other samples of \hii regions and star-forming clumps.

We estimated the contribution of the thermal gas pressure, $P_{gas}$, and direct radiation pressure, $\rm P_{dir}$, to the outward pressure in the \hii regions. We find $\rm P_{gas}$ to be the dominant force of expansion in IC\,10's \hii regions, being $\sim$3 orders of magnitude greater than $\rm P_{dir}$ using the definition of $\rm P_{dir}$ based on ionizing photon production rate. We also find $\rm P_{out} > P_{in}$ in 89\% of the \hii regions before accounting for additional expansion from hot gas pressure, a somewhat surprising result given the low stellar masses estimated for the ionizing stars. Five of the \hii regions show evidence of outflows that may be supported by energy in the ionizing cluster either in the form of stellar winds or champagne flows. These pressure and energy estimates add further evidence that the \hii regions in IC\,10 are young and undergoing expansion into the ISM and suggest that thermal gas pressure may be a more effective form of feedback than previously expected from low mass clusters.

These high resolution and SNR observations were possible in just 1.5 nights of Keck observing time. From just this short time we were able to obtain detailed kinematic and flux maps of a significant number of \hii regions. With additional observations the remainder of IC\,10's \hii regions can be observed in the same modes, with deeper observations of the diffuse gas and the supplementary low resolution mode. More expansive coverage of \hii regions and the DIG in IC\,10 will allow for detailed study of the kinematic and ionization state differences between these unique regions of ionized gas. KCWI observations at high spectral resolving power of the remainder of IC\,10 would also double the number of \hii regions in this unique starburst environment in which the impact of different modes of feedback and outflows can be investigated.

A relatively small investment of time with optical IFS's such as KCWI and VLT/MUSE can quickly yield a large sample of local star-forming regions in a wide range of environments. The large field of view and moderate spectral resolving power (R$\sim$2000-4000) of MUSE provides an efficient tool for mapping ionization states of \hii regions and the ISM, while the R$\sim$18,000 mode of KCWI can be leveraged for a detailed look at the gas kinematics in compact regions. Utilizing these powerful IFU's across a wide sample of star forming galaxies will allow a detailed and statistically significant study of how environmental conditions impact the effectiveness of feedback mechanisms and vice versa; and whether there are age and environmental dependencies when looking at the scaling relationships. Targeted IFU studies of this kind are already well underway and as a larger collective sample is built we will be able to better compare these results with theoretical predictions of feedback and inform new models.

The reduced data cubes used in this study are available by request to facilitate further study beyond the scope of this project.

\acknowledgments

We thank the anonymous referee for their comments which helped improve the quality of this paper. The data presented herein were obtained at the W. M. Keck Observatory, which is operated as a scientific partnership among the California Institute of Technology, the University of California and the National Aeronautics and Space Administration. The Observatory was made possible by the generous financial support of the W. M. Keck Foundation. The authors wish to recognize and acknowledge the very significant cultural role and reverence that the summit of Maunakea has always had within the indigenous Hawaiian community.  We are most fortunate to have the opportunity to conduct observations from this mountain. 

\vspace{5mm}
\facilities{Keck:II(KCWI)}

\software{astrodendro \citep{astrodendro},
          Astropy \citep{astropy:2013, astropy:2018}, 
          IPython \citep{ipython}
          Matplotlib \citep{Hunter2007},
          NumPy \citep{numpy},
          pandas \citep{pandas, pandas2},
          photutils \citep{photutils},
          PyStan \citep{Pystan},
          reproject \citep{reproject}
          SHAPE \citep{shape}
          }

\clearpage
\appendix

\section{\hii Region Naming Convention}\label{app:region_naming}
The previous naming of IC\,10 \hii regions was developed by \citet{HL1990} where regions are assigned a number in order of increasing RA. If that region breaks up into smaller knots then a letter is added after the complex number (e.g., 111a). Throughout the literature this region identifier is typically preceded by either ``HL90" or simply ``HL" to indicate the origin of the identifier. This is a simple and clear way of tabulating the \hii regions found in this early study but there are some significant shortcomings of this system now. One difficulty is in quickly identifying regions in the \ha \, maps as regions are only numbered based on their RA while there can be a large spread in Dec from one region to the next in the sequence. The other more problematic issue is that with better resolution and sensitivity one would expect to identify new \hii regions and complexes breaking into more knots. When this occurs there is not a clear way in which to assign an identifier to these new regions. If the next number in the sequence is assigned to each new region there would no longer be a clear ordering based on RA, and reassigning numbers to each region with every new identification would make comparison between studies exceedingly difficult.

We have therefore proposed a new naming convention that we believe addresses these issues for our study and allows extension to future studies with even wider fields using the grid described in Section \ref{sec:naming}. This numbering scheme simplifies identification of nearby regions in both RA and Dec as well as extension to a larger FoV. This could be applied to the larger IC\,10 \hi envelope by increasing the numbering range in Dec. and extending the RA designation to double and/or negative lettering (e.g., AA or -A). Fainter \hii regions may be identified in already occupied grid squares, but the next trailing letter in the sequence can be added as these should have lower luminosity than what is identified here.

\setcounter{figure}{0} 
\renewcommand{\thefigure}{\Alph{section}\arabic{figure}}

\begin{figure*}[h!]
    \centering
    \includegraphics[width=0.55\textwidth]{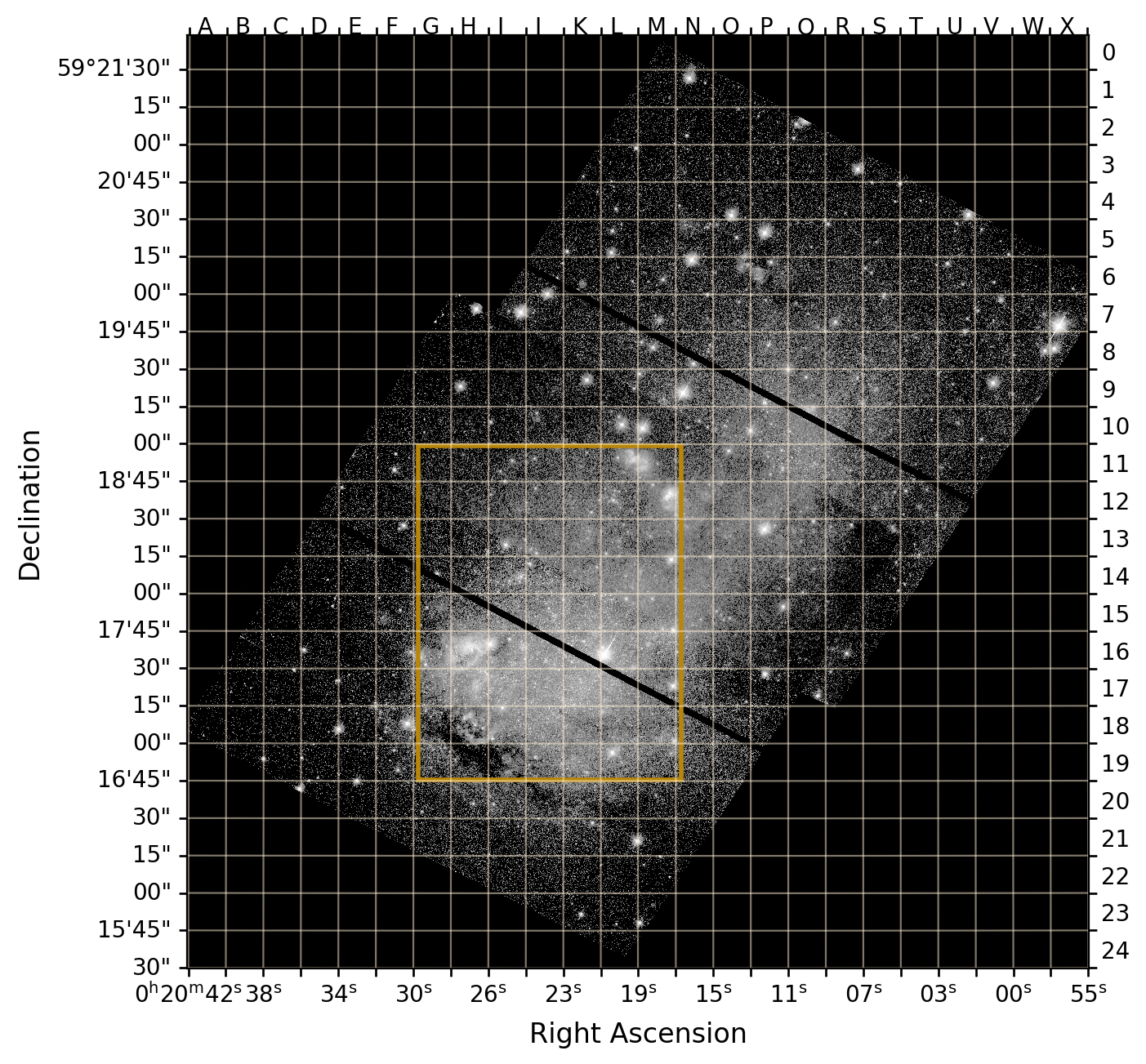}
    \caption{Coordinate grid for proposed \hii region naming scheme overlaid on an HST/ACS image showing the optical extent of the galaxy. \hii regions are named based on the lettered column and numbered row corresponding to their center. This could be extended to the larger \hi envelope of IC\,10 by going to larger positive and negative numbers as well as double and/or negative lettering (e.g., AA or -A). The gold rectangle outlines the region of IC 10 which our KCWI observations fall in. \label{fig:coord_grid}}
\end{figure*}

\section{Impact of Radius Definition} \label{app:radius_MCMC}
The choice of how to define the radius of an \hii region varies significantly between studies, and particularly between local and high-redshift studies.

In observational studies of local \hii regions there is wide variation in the methods used to define the size of the region, with this often not being a critical aim of the study. For example, the Green Bank Telescope \hii Region Discovery Survey (HRDS) measures the sizes of 441 \hii regions in the Milky Way by taking the mean of the FWHM of Gaussians fit to the RA and Dec components of continuum observations \citep{Anderson2011}. In the extensive CALIFA survey of over 26,000 extragalactic \hii regions, a custom procedure called \texttt{HIIEXPLORER} \citep{Sanchez2012b} \citep[and the Python version \texttt{PYHIIEXPLORER};][]{Espinosa-Ponce2020} is used to identify \hii regions. This procedure is similar in methodology to \texttt{astrodendro} except that the maximum expected extent of \hii regions is provided as an input constraint. The sizes of identified regions then tend towards a relatively uniform distribution \citep{Sanchez2012b}, and the authors note that extracting reliable sizes was not a main goal of their methodology partly due to the resolution of their observations. 

Studies of local GMCs typically use the second moments of the cloud structure to determine its properties \citep{Heyer2015}. Often the geometric mean of the second moments of the cloud structure (in the direction of greatest elongation and perpendicular to that) is used to describe the RMS extent of a cloud, $\sigma_{\rm r}$. An empirical factor is then used to determine the radius of a spherical cloud, $R=\eta\sigma_{\rm r}$. This factor was first determined empirically in \citet{Solomon1987} to be $\eta=1.91$ for converting their rectangular regions to spherical clouds and is used throughout the literature to convert the second moments of a variety of structures to a spherical radius. \citet{Rosolowsky2006} go through the derivation of this factor of $\eta$ for a spherical cloud with a density profile of $\rho\propto r^{-1}$ and determine a theoretical value of $\sqrt{6}\approx2.45$. They suggest that deviation may be due partly to the use of CO data to trace the GMC density which is shallower than the actual density profile due to saturation in dense regions and lack of detection in low density regions. This would mean that the ``true" value of $\eta$ would lie somewhere between the empirical value of 1.91 from \citet{Solomon1987} and the value of 2.45 determined from their toy model of a GMC. However, they recommend continued use of the radius definition from \citet{Solomon1987} in order to remain consistent with this data set. Since our observations are of ionized rather than molecular gas we do not use this same factor of $\eta$, and instead assume a Gaussian profile in our determination of $r_{1/2}^*$ from the second moments.

\citet{Zaragoza-Cardiel2017} combine CO and \ha \, observations in local LIRGs with regions identified using \texttt{astrodenro}. They define the radius, $r$, using the second moments of the structures in both cases, with the factor of 1.91 from \citet{Solomon1987}. They compare the ratio of this radius to that derived from the area of the full structure, $r_{\rm eff}$, and find an average $\frac{r}{r_{\rm eff}}=0.86\pm0.14$. In another sample of local LIRGs with star-forming regions identified by \texttt{astrodendro}, \citet{Larson2020} use $r_{\rm eff}$ to define the size of regions. \citet{Larson2020} also performs a comparison of the \texttt{astrodendro} and \texttt{CLUMPFIND} identification routines finding similar average radii and SFR, but a narrower range of fluxes for a given radius of star-forming region due to a lack of local background subtraction.

In studies of high-redshift star-forming clumps, differences in the method used to determine region sizes are also present, although perhaps less significant due to the decrease in resolution. \citet{Wisnioski2012} compares the size determined from isophotes of constant flux and from fitting a Gaussian to the radial surface profile of \hii regions. They find $r_{\rm eff}$ from the isophotal method to be systematically larger than $r_{1/2}$ determined via Gaussian fitting, but the luminosities to be consistent between the two methods. This comes from the emission being dominated by the higher intensity cores of the \hii regions. They argue that $r_{1/2}$ from Gaussian fitting is a better defined observational parameter as it is less likely to be contaminated by diffuse emission that could be significant at high-redshift. \citet{Livermore2012} also compare clump sizes determined by $r_{\rm eff}$ of \texttt{CLUMPFIND} isophotal structures and from fitting a 2D elliptical Gaussian profile to emission peaks. They find that using \texttt{CLUMPFIND} for their sample gives 25\% larger estimates of size than from the FWHM of the 2D elliptical Gaussian, but include error bars encompassing both measures. They note a smaller deviation between their \texttt{CLUMPFIND} radii and the FWHM of fitted Gaussians than in \citet{Wisnioski2012} due in part to the use of multiple isophote levels in \texttt{CLUMPFIND} that does not need to be tuned in the same way as other single isophote methods. 

Which of the many possible radius definitions is used has a significant impact on the typical size of star-forming regions and on the scaling relationships determined from those properties. To quantify just how much impact the choice of using the pseudo half-light radius ($r_{1/2}^*$) definition for the \hii region radii has in this study we perform MCMC fitting with different choices of radius definition and identification constraint for our sample. We fit the relationship between region size and luminosity for (i) the IC\,10 \hii regions alone, for (ii) local samples only, and for (iii) the full sample of local and high-redshift \hii regions and clumps. These results are shown in Table \ref{tbl:radmeth_results}.

\begin{deluxetable}{lcccccccc} [h]
	 \tablecaption{Radius Definition Implications} \label{tbl:radmeth_results} 
	 \tablehead{\colhead{\texttt{astrodendro} Constraint} & \colhead{Radius Definition} & \colhead{Filtering} & \colhead{$\rm N_{regions}$} & \colhead{$\rm R_{avg}$} & \multicolumn{3}{c}{Slopes} \\
	 & & & & (pc) & \colhead{IC\,10} & \colhead{Local} & \colhead{All Data}} 
\startdata 
\multirow{5}{6em}{2r $>$ FWHM} & $r_{1/2}^*$ & \nodata & 46 & 4.0 & $3.601^{+0.471}_{-0.385}$ & $3.049^{+0.029}_{-0.025}$ & $3.002^{+0.023}_{-0.024}$ \\ 
 & $r_{1/2}^*$ & r$>$FWHM &  20 & 5.6 & $2.972^{+0.793}_{-0.483}$ & $3.147^{+0.032}_{-0.035}$ & $3.067^{+0.032}_{-0.021}$ \\ 
 & $r_{\rm eff}$ & \nodata & 45 & 6.8 & $2.979^{+0.426}_{-0.344}$ & $3.045^{+0.025}_{-0.026}$ & $2.987^{+0.020}_{-0.028}$ \\ 
 & $r_{\rm maj}$ & \nodata & 46 & 5.1 & $3.176^{+0.451}_{-0.351}$ & $2.999^{+0.036}_{-0.026}$ & $2.951^{+0.031}_{-0.025}$ \\ 
\tableline 
 r $>$ FWHM & $r_{1/2}^*$ & \nodata & 23 & 6.1 & $2.888^{+0.558}_{-0.388}$ & $3.161^{+0.026}_{-0.025}$ & $3.043^{+0.025}_{-0.024}$ \\ 
\enddata 
\tablecomments{Results of model slope determined from MCMC fitting of the size-luminosity relationship for IC\,10 \hii regions and the local and high-redshift comparison sample based on different ways of defining and constraining the radius of IC\,10 \hii regions. \\
Col 1: constraint used in \texttt{astrodendro} to define an independent structure. The top section uses the more relaxed constraint that the diameter of the region must be larger than the FWHM determined from standard star observations (this always uses the definition of $r_{1/2}^*$ to match the Gaussian fit to the stars PSF). In the bottom row we require this diameter to be twice the standard star FWHM. \\
Col 2: radius definition used to determine \hii region size; $r_{1/2}^*$, and $r_{\rm eff}$ used as described in Section \ref{sec:radius}, and $r_{\rm maj}$ is determined by converting the second moment in the direction of greatest elongation to the HWHM of a Gaussian. \\
Col 3: additional filtering applied to exclude regions from MCMC fits beyond the manual filtering of bad regions as described in Section \ref{sec:astrodendro} \\
Col 4: Number of \hii regions found in IC\,10 based on these constraints and filtering. \\
Col 5: Average radius of IC\,10 \hii regions. \\
Col 6 - 8: slope and uncertainty determined from MCMC fitting of the size-luminosity relationship for IC\,10 regions only (7), local regions only (8) and all local and high-redshift data (9).} 
\end{deluxetable}

Regardless of our chosen definition for region size, the slope of the local and full sample is consistent within $<5\%$. However, there is a larger difference in the slope of the IC\,10 \hii region sample alone. Using the most relaxed resolution constraint in \texttt{astrodendro} and $r_{1/2}^*$ produces a slope for which the nominal value deviates by $18\%$ from either a stricter resolution constraint (bottom row) or the definitions of $r_{\rm eff}$ and $r_{\rm maj}$ for the region size. Fitting only the IC\,10 \hii regions, however, leads to significantly larger uncertainties so the nominal slopes for each radius constraint are still consistent within the $1\sigma$ uncertainties.

\newpage
\section{Spectra Thumbnails}\label{app:spec_thumbs}
\setcounter{figure}{0} 

\begin{figure}[h]
    \centering
    \includegraphics[width=\textwidth]{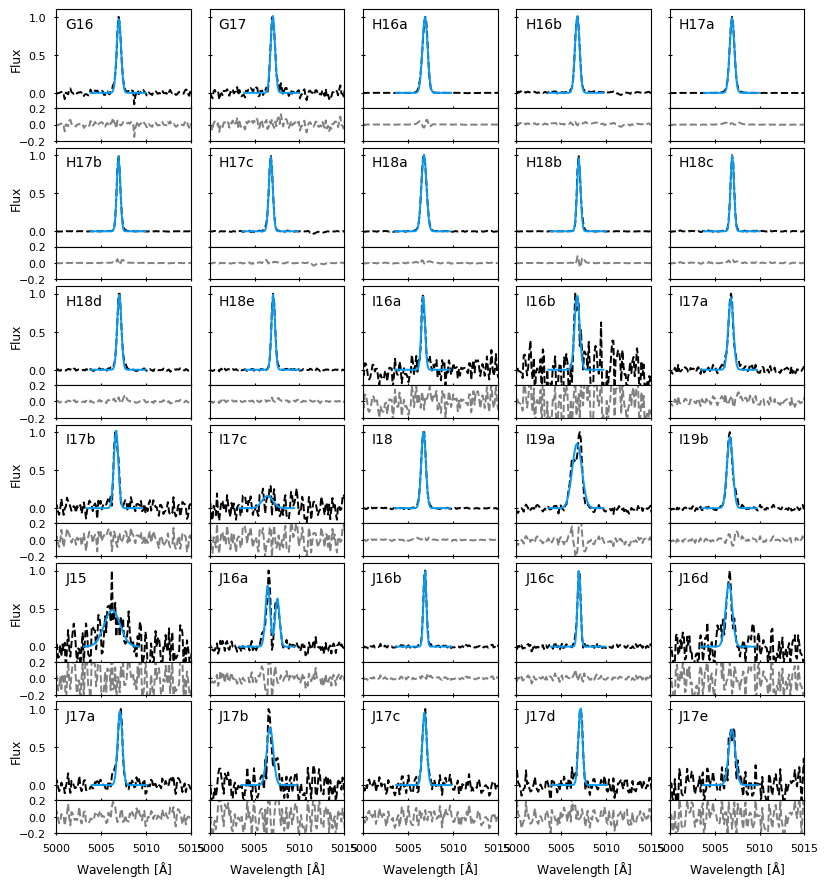}
    \caption{Spectral thumbnails of \oiii5007\ang \, from all \hii regions detected in the small slicer, R$\sim$18,000 observing mode. These lines are fit by a single Gaussian profile (except for J16a with a double Gaussian profile) shown in cyan with the residuals shown in grey below the associated spectrum. The spectra have been normalized to the peak of the \oiii5007\ang \, line.}
\label{fig:BH3_spectra}
\end{figure}

\begin{figure}[h]
    \centering
    \includegraphics[width=\textwidth]{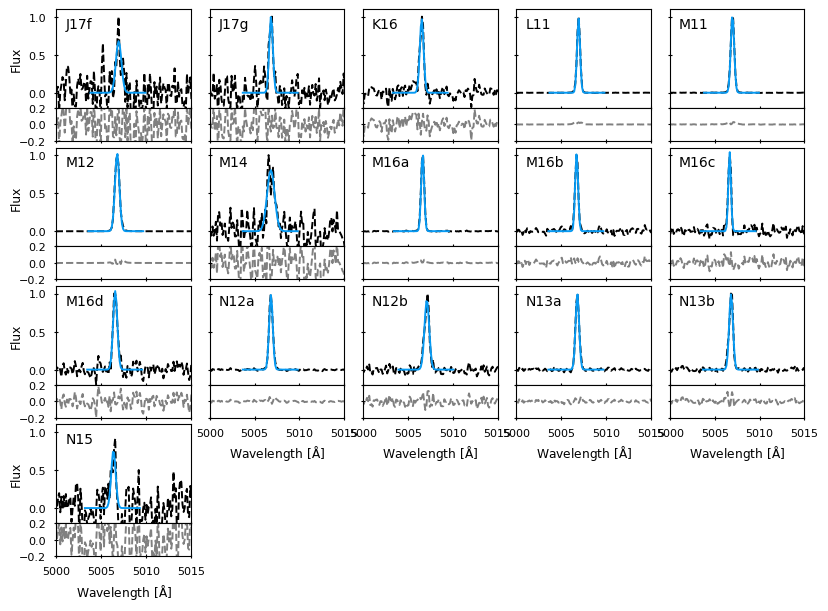}
    \caption{Figure \ref{fig:BH3_spectra} cont'd.}
\label{fig:BH3_spectra_b}
\end{figure}

\clearpage
\section{\hii Region Maps}\label{app:map_thumbs}
\setcounter{figure}{0}
\begin{figure}[h]
\centering
\includegraphics[width=.92\textwidth]{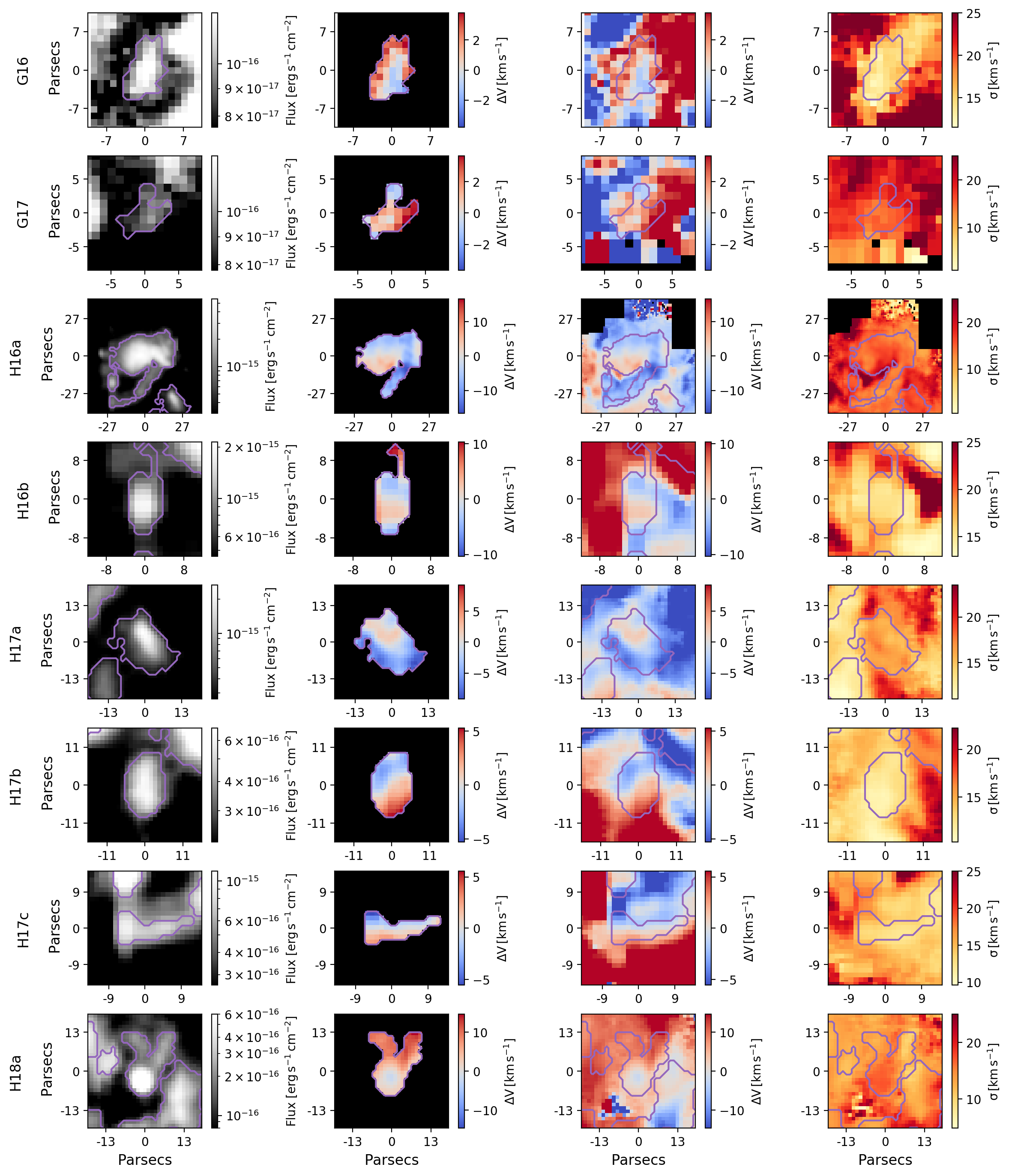}
\caption{Thumbnail maps of \hii regions identified in our KCWI observations. (Left): Flux maps of the surrounding area. (Center Left): Velocity shift of spaxels within the \hii region relative to the systemic velocity of the region. (Center Right): Velocity shift of the \hii region and the surrounding gas. (Left): Velocity dispersion within the \hii region and the surrounding gas. Regions of elevated velocity dispersion may be indicative of outflowing gas, particularly when correlated with a velocity shift relative to the surrounding gas.}
\label{fig:BH3_mapthumbs_app}
\end{figure}

\begin{figure}[h]
\centering
\includegraphics[width=.92\textwidth]{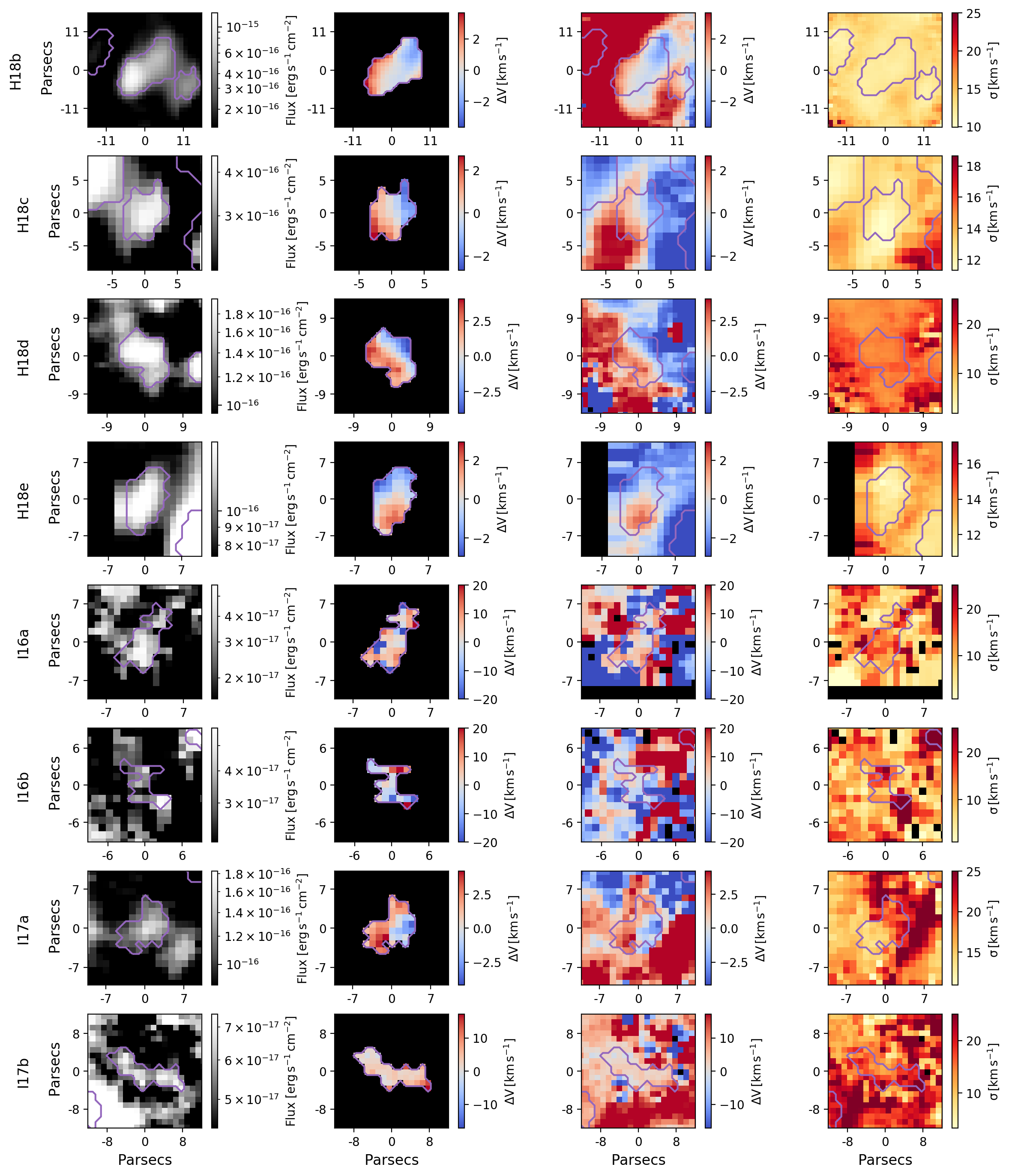}
\caption{Figure \ref{fig:BH3_mapthumbs_app} cont'd.}
\label{fig:BH3_mapthumbs_b}
\end{figure}

\begin{figure}[h]
\centering
\includegraphics[width=.92\textwidth]{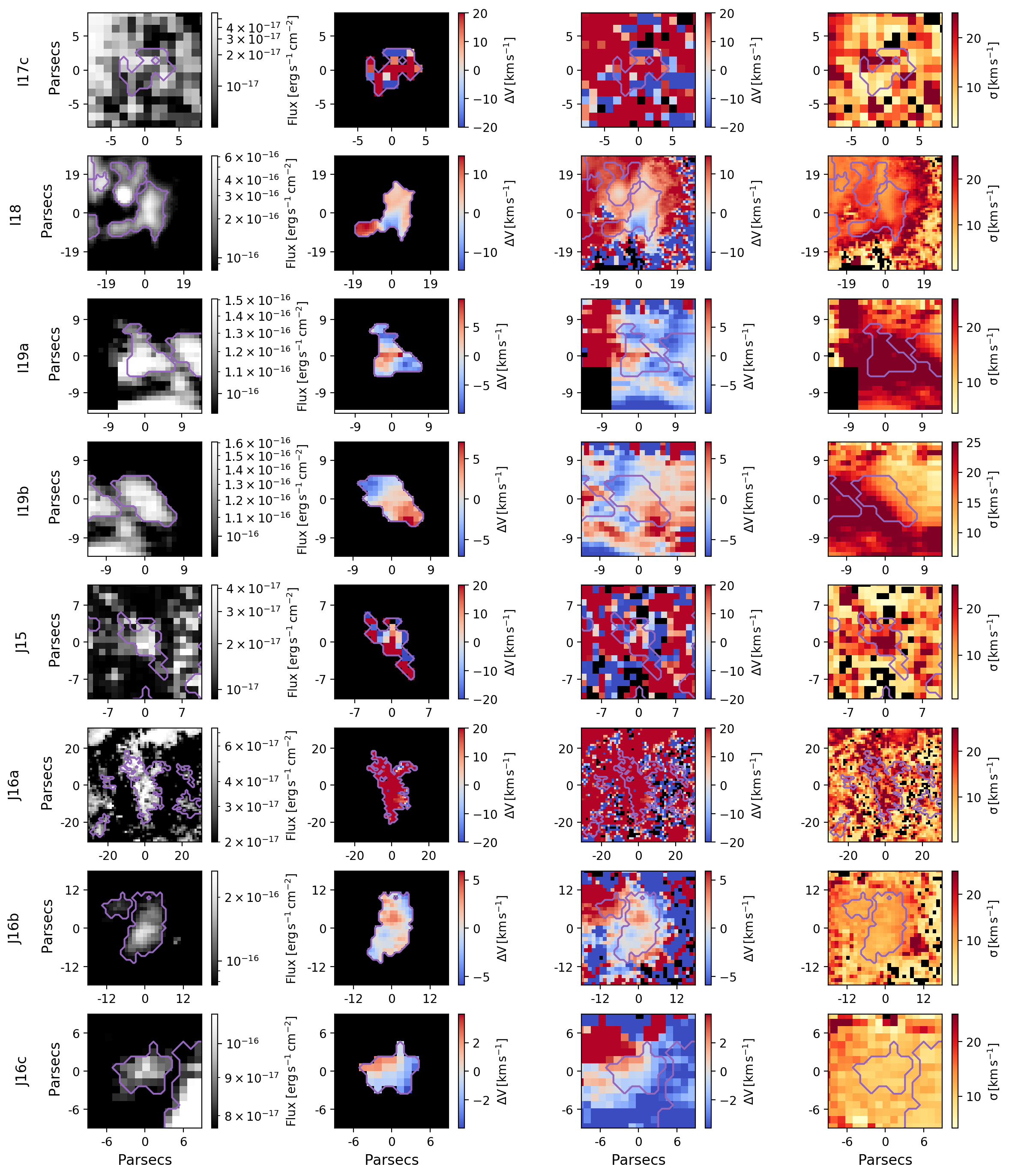}
\caption{Figure \ref{fig:BH3_mapthumbs_app} cont'd.}
\label{fig:BH3_mapthumbs_c}
\end{figure}

\begin{figure}[h]
\centering
\includegraphics[width=.92\textwidth]{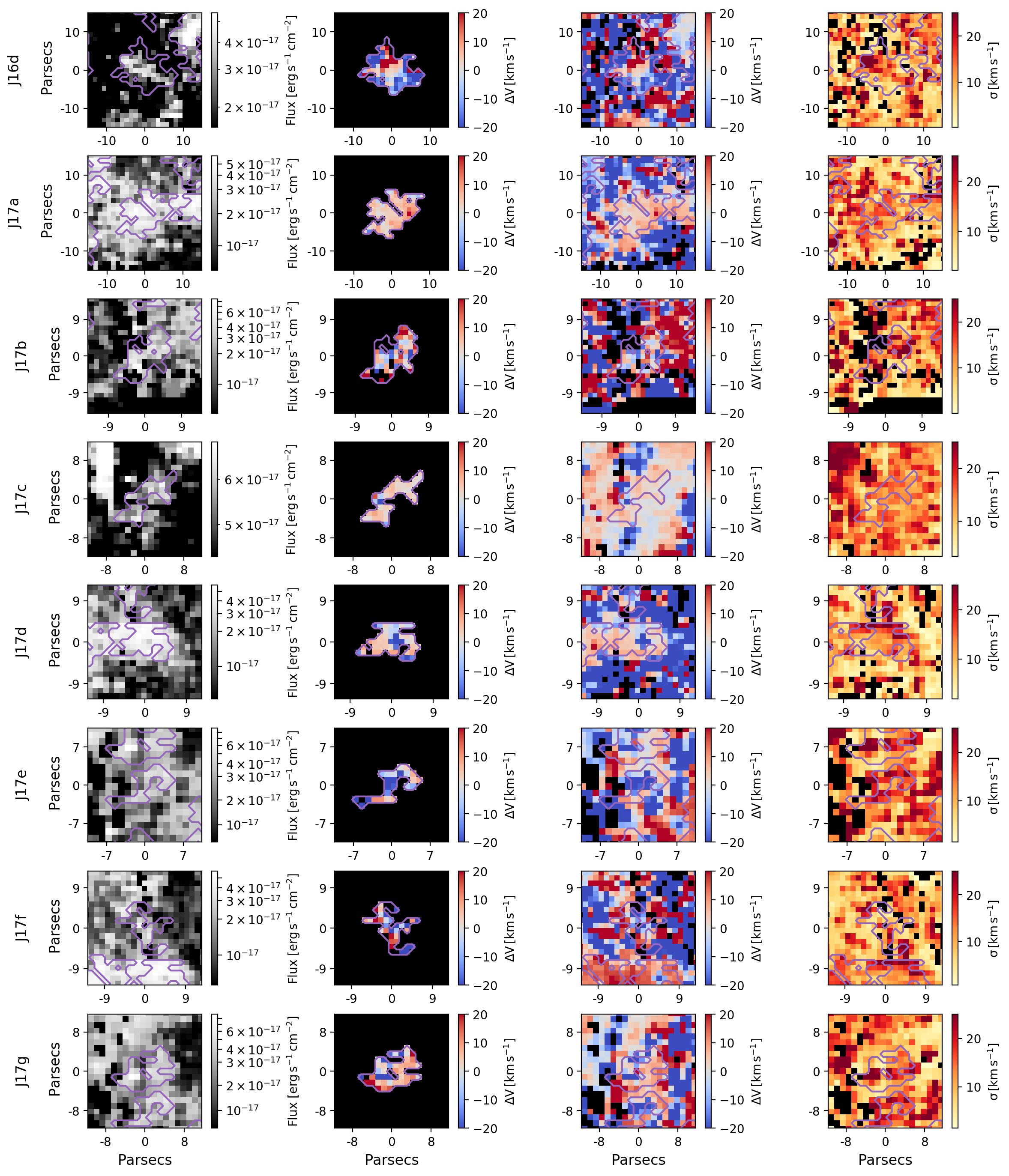}
\caption{Figure \ref{fig:BH3_mapthumbs_app} cont'd.}
\label{fig:BH3_mapthumbs_d}
\end{figure}

\begin{figure}[h]
\centering
\includegraphics[width=.92\textwidth]{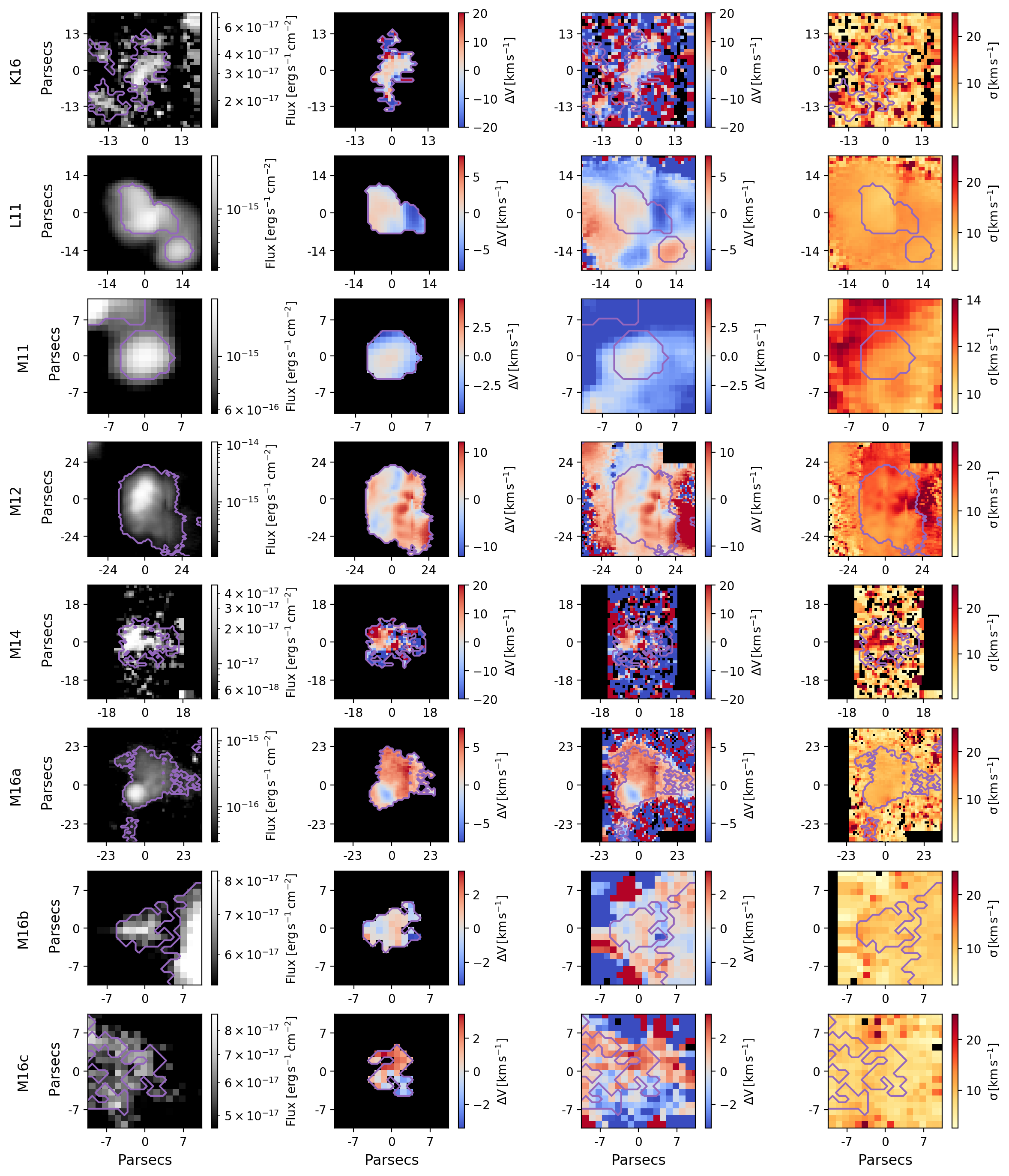}
\caption{Figure \ref{fig:BH3_mapthumbs_app} cont'd.}
\label{fig:BH3_mapthumbs_e}
\end{figure}

\begin{figure}[h]
\centering
\includegraphics[width=.92\textwidth]{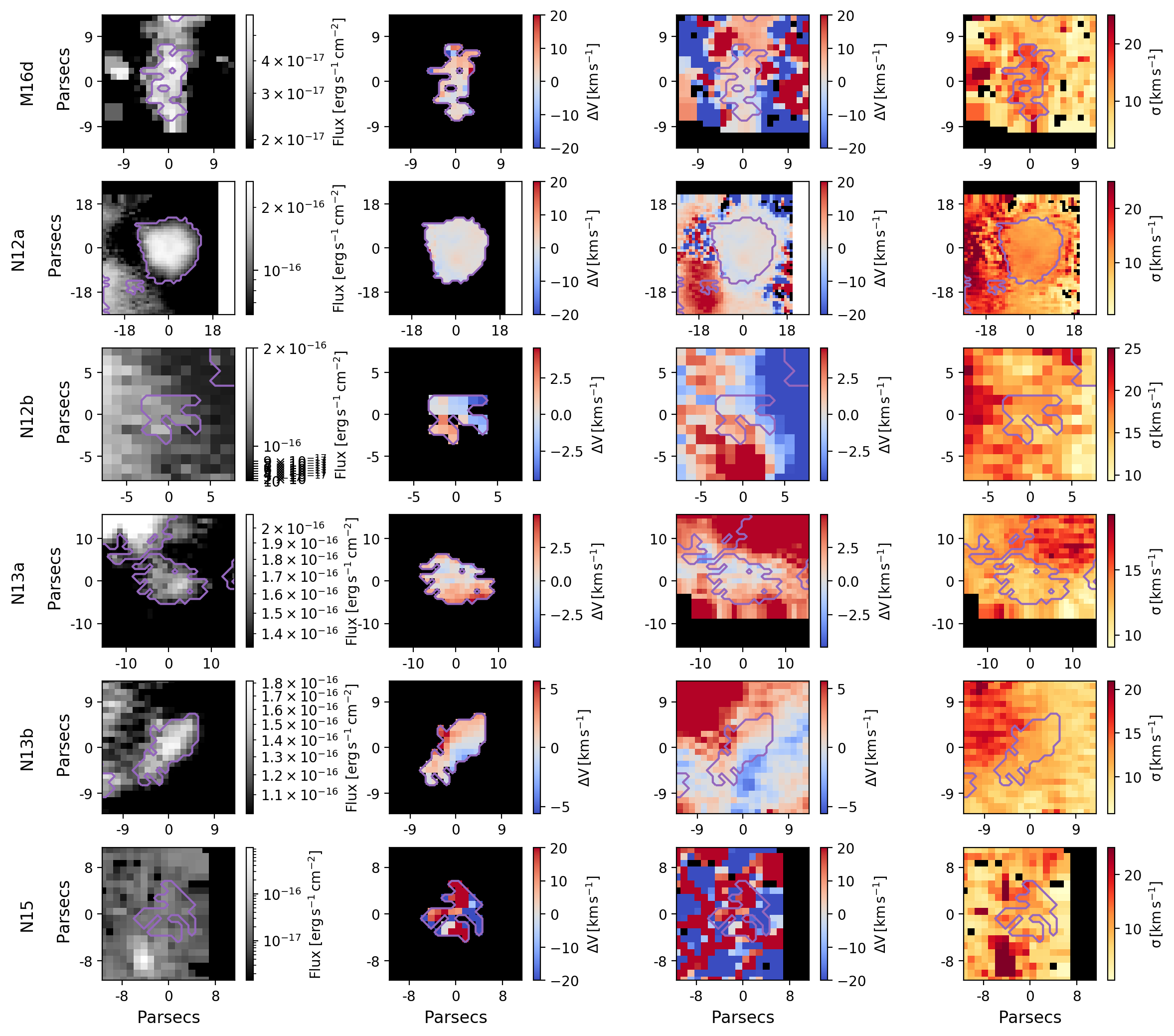}
\caption{Figure \ref{fig:BH3_mapthumbs_app} cont'd.}
\label{fig:BH3_mapthumbs_f}
\end{figure}
\clearpage

\bibliography{ic10}{}
\bibliographystyle{aasjournal}

\end{document}